\documentclass[mnsc,nonblindrev]{informs3}

\OneAndAHalfSpacedXI
\usepackage{graphicx}
\usepackage[letterpaper,margin=1in]{geometry} %
\usepackage{booktabs} %
\usepackage[dvipsnames]{xcolor}
\usepackage[ruled]{algorithm2e} %

\SetAlFnt{\small}
\SetAlCapFnt{\small}
\SetAlCapNameFnt{\small}
\SetAlCapHSkip{0pt}
\IncMargin{-\parindent}

\usepackage{mathrsfs} %

\usepackage[framemethod=tikz]{mdframed}

\usepackage{nicefrac}

\usepackage{pifont}
\usepackage{cancel}
\definecolor{BrickRed}{rgb}{0.8,0.25,0.33}
\usepackage{booktabs}
\usepackage{multirow}
\usepackage{dsfont}
\usepackage{bigstrut}
\usepackage{caption}
\usepackage{subcaption}
\usepackage{tcolorbox}
\usepackage[noend]{algorithmic}
\usepackage{amsmath,cases}
\usepackage{float}
\usepackage{tabulary,booktabs,multirow,mathtools}
\usepackage{tikz}
\usepackage{pgfplots}
\pgfplotsset{compat=1.18}
\usetikzlibrary{chains,arrows.meta,quotes,bending}
\usetikzlibrary{decorations.pathreplacing}

\usepackage{xcolor}
\usepackage{booktabs,colortbl}

\definecolor{BrickRed}{rgb}{0.8,0.25,0.33}
\definecolor{darkgreen}{HTML}{06402B}
\definecolor{navyblue}{HTML}{000080}

\usepackage[pagebackref,colorlinks,citecolor=blue,linkcolor=BrickRed]{hyperref}

\usepackage{mathtools}
\usepackage{cancel}
\usepackage{mwe}
\usepackage[shortlabels]{enumitem}
\usepackage{bbm}
\usepackage{cases}
\usepackage{soul}
\usepackage{tabularx}
\usepackage{amssymb}
\usepackage{booktabs}
\usepackage{url}
\usepackage{hyperref}
\usepackage{thm-restate}
\usepackage{cleveref}
\DeclareRobustCommand{\PrimalDualImprovement}{\texorpdfstring{\hyperref[proc:primal-dual-improvement]{\textcolor{BrickRed}{Primal-Dual Improvement}}}{Primal-Dual Improvement}}

\usepackage{natbib}
 \bibpunct[, ]{(}{)}{,}{a}{}{,}%
 \def\bibfont{\small}%

\newcommand{\qedsymb}{\hfill{\rule{2mm}{2mm}}}
\newcommand{\newreptheorem}[2]{%
\newenvironment{rep#1}[1]{%
 \def\rep@title{#2 \ref{##1}}%
 \begin{rep@theorem}}%
 {\end{rep@theorem}}}
\makeatother
\newreptheorem{theorem}{Theorem}
\newreptheorem{lemma}{Lemma}
\newreptheorem{proposition}{Proposition}
\newreptheorem{corollary}{Corollary}
\newreptheorem{definition}{Definition}
\newreptheorem{fact}{Fact}

\crefname{thm}{Theorem}{theorems}
\crefname{cla}{Claim}{claims}
\crefname{lem}{Lemma}{lemmas}
\crefname{fact}{Fact}{facts}
\crefname{exm}{Example}{examples}

\Crefname{thm}{Theorem}{theorems}
\Crefname{cla}{Claim}{claims}
\Crefname{lem}{Lemma}{lemmas}
\Crefname{fact}{Fact}{facts}
\Crefname{exm}{Example}{examples}

\hypersetup{
    colorlinks=true,
    urlcolor=black,  %
}

\newcommand{\appendix}{
    \par
    \setcounter{section}{0}%
    \setcounter{subsection}{0}%
    \renewcommand\thesection{\Alph{section}}%
    \renewcommand\thesubsection{\thesection.\arabic{subsection}}%
    \renewcommand{\sectionname}{Appendix}%
}
\renewcommand{\AppendixFontSize}{\normalsize}

\usepackage{natbib}
 \bibpunct[, ]{(}{)}{,}{a}{}{,}%
 \def\bibfont{\small}%

\TheoremsNumberedThrough   
\ECRepeatTheorems
\EquationsNumberedThrough 

\theoremstyle{plain}
\renewenvironment{proof}[1][]{\begin{trivlist} \item[\hspace{\labelsep}{\bf \noindent Proof#1.\/}] }{\qedsymb\end{trivlist}}%

\newcommand{\bs}[1]{\boldsymbol{#1}}
\newcommand{\pr}[1]{{\rm Pr} \left[ #1 \right]}
\newcommand{\prpar}[1]{{\rm Pr} [ #1 ]}

\newcommand{\ex}[1]{{\mathbb E} \left[ #1 \right]}
\newcommand{\expar}[1]{{\mathbb E} [ #1 ]}
\newcommand{\prtwo}[2]{{\rm Pr}_{#1} \left[ #2 \right]}

\newcommand{\expartwo}[2]{{\mathbb E}_{#1} [ #2 ]}
\newcommand{\extwo}[2]{{\mathbb E}_{#1} \left[ #2 \right]}

\newcommand{\eps}{\varepsilon}
\newcommand{\bb}{\mathbb}

\newcommand\restr[2]{{%
  \left.\kern-\nulldelimiterspace %
  #1 %
  \littletaller %
  \right|_{#2} %
  }}

\usepackage{xspace}

\newcommand{\iid}{\textsf{IID}\xspace}
\newcommand{\noniid}{\textsf{NON-IID}\xspace}
\newcommand{\gen}{\textsf{CORREL}\xspace}

\newcommand{\thr}{\tau^{ \ast }}

\makeatletter
\newcommand{\vastmin}{\bBigg@{3}}
\newcommand{\vast}{\bBigg@{4}}
\newcommand{\Vast}{\bBigg@{7}}
\newcommand{\boldvast}{\bBigg@{9}}
\makeatother

\newcommand{\poly}{\mathrm{poly}}

\newcommand{\ind}[1]{\mathbbm{I}\left [ #1 \right] }

\newcommand{\obl}{${\cal A}_{\rm base}$\xspace}
\newcommand{\mart}{${\cal A}_{\rm thr}$\xspace}

\newcommand{\ton}{\tau^\star}

\newcommand{\zz}{Z_t^{(0)}}

\newcounter{mysubequations}

\newcommand{\off}{{\textsf{OFF}}}
\newcommand{\on}{{\textsf{ON}}}

\newcommand{\qedsymbol}{\hfill{\rule{2mm}{2mm}}}
\allowdisplaybreaks

\tcbset{
    theorembox/.style={
        colback=white,
        colframe=black,
        width=\textwidth,
        arc=5mm,
        auto outer arc,
        boxsep=5pt, %
        before skip=10pt, %
        after skip=10pt,  %
        left=6pt, %
        right=6pt, %
        top=6pt,   %
        bottom=6pt %
    }
}

\usepackage{color}         
\usepackage{xcolor}

\definecolor{ForestGreen}{RGB}{0,180,0} %
\makeatletter
\renewcommand{\HOOKb}{%
  \@thanks
  \global\let\@thanks\@empty
}
\makeatother
\begin{document}
\RUNTITLE{Bayesian Optimal Stopping: \\ 
Invincible Prophets and Achievable Philosophers}
\TITLE{\fontsize{19}{23}\selectfont
Bayesian Optimal Stopping: \\ 
Invincible Prophets and Achievable Philosophers}

\ARTICLEAUTHORS{%
\AUTHOR{Alireza AmaniHamedani\thanks{Corresponding author: a.amanihamedani@imperial.ac.uk}}
\AFF{Imperial College London, \EMAIL{a.amanihamedani@imperial.ac.uk}}
\AUTHOR{Ali Aouad} 
\AFF{Massachusetts Institute of Technology, \EMAIL{maouad@mit.edu}} %
\AUTHOR{Senem I\c{s}\i{}k} 
\AFF{Stanford University, \EMAIL{senemi@stanford.edu}} %
\AUTHOR{Thomas Kesselheim} 
\AFF{University of Bonn, \EMAIL{thomas.kesselheim@uni-bonn.de}} 
\AUTHOR{Amin Saberi}
\AFF{Stanford University, \EMAIL{saberi@stanford.edu}}
}
\ABSTRACT{
We study a model of Bayesian optimal stopping, where a ground truth world $H$ is drawn from $K$ worlds according to a known prior but is hidden to the decision-maker. The world $H$ then determines the joint distribution of $n$ observations $(X_1, \cdots, X_n)$. The decision-maker selects one observation irrevocably, without knowing the future observations or $H$, to maximize its value. Unlike the classical adversarial or fully-known distributional models, here the decision-maker can learn about the world as observations arrive, and the optimal policy must balance this learning against the value of waiting. We compare online algorithms against two benchmarks: the Bayes-optimal online policy (the philosopher), given by a dynamic program over exponentially many histories, and the offline optimum (the prophet).

Our results show that the philosopher is achievable in that we find a $(1-\eps)$-approximation of the Bayes-optimal policy that runs in time $\poly(|{\cal I}|^{\log^* |{\cal I}| + 1/\eps} \cdot (1/\eps)^{1/\eps^2})$, where $|\cal I|$ is the size of the input. By contrast, the prophet is invincible and no algorithm can obtain a competitive ratio that substantially exceeds the worst-case guarantees. When the observations are conditionally independent and identically distributed given $H$, the secretary's rule is $1/e$-competitive. We show that no algorithm can guarantee a competitive ratio better than $1/e + 1/\poly(K)$ and this bound is attainable. 

Our approximation scheme is built on a {\em primal improvement} subroutine that connects the {prophet} to the {philosopher}: given a baseline algorithm that is $\alpha$-competitive (against the prophet), we construct an improved $\alpha^+$-approximation algorithm with a {strictly} better approximation factor (against the philosopher). This step is done via randomizing between the baseline algorithm and another algorithm that leverages the problem's Bayesian structure to construct \emph{dual martingales,} setting dynamic thresholds for accepting new observations. We show that this improvement step yields a new, tighter dual martingale, which allows the step to be applied repeatedly. By sequential primal-dual updates, we can iteratively find better approximation algorithms and tighter dual martingales. The martingales are approximated in polynomial time through importance-weighted Monte Carlo sampling. Such dual martingales may be useful technical tools for the design and analysis of Bayesian algorithms in more general Bayesian online {optimization} problems. {For instance, we also use them to establish the aforementioned upper bound on the competitive ratio}.
}

\maketitle
\tableofcontents
\clearpage
\section{Introduction}\label{sec:intro}

Online algorithms are traditionally analyzed under either adversarial or stochastic independent and identically distributed (IID) inputs; the former model can be overly pessimistic while the latter is too optimistic. 
To bridge this gap, researchers have developed several approaches that go \emph{beyond worst-case} (see \cite{roughgarden} for a survey). 

We propose a Bayesian framework for online algorithm design in which a prior over \emph{world} models induces a mixture distribution over IID inputs. Through the observations in the online phase, the algorithm learns about the world while making decisions. This model interpolates between adversarial and IID models, and captures the stochastic correlations in the input. Within this Bayesian framework, we develop a theory of optimal stopping---a canonical online selection problem---and use it to design and analyze new algorithms.

\paragraph{Bayesian setup.} In our model, the ground truth world $H$ is drawn from $K$ worlds according to a known prior $\pi$ but is hidden to the decision-maker. The world $H$ then determines the distribution of $n$ IID observations $X_1, \cdots, X_n$. The decision-maker selects one observation irrevocably without knowing the future observations or $H$, and its reward is the value of the selected observation. This model interpolates smoothly between classical settings: the known distribution prophet inequalities model of \cite{samuel1984comparison} corresponds to our setup when the prior has a single component and the observations are conditionally independent given the world; conversely, by Yao's minimax principle, the unknown IID {} model corresponds to arbitrary mixtures with any number of components, i.e., $K=\infty$.\footnote{A minimax argument shows, informally speaking, that in more general online decision-making problems, the competitive ratio of the optimum online algorithm on the worst-case Bayesian problem is equal to the best possible competitive ratio in the adversarial model (see \citet[Theorem A.1]{posterior-matching} for a formal statement).} Our setup can then capture the spectrum between these two extremes, providing the flexibility to model various levels of uncertainty for decision-making. In fact, our results are for more general models, where (i) the observations are conditionally independent given $H$ but may have different distributions, or (ii) even conditional on $H$, the observations may be correlated.

A key advantage of this Bayesian model is that it provides a framework for designing and analyzing online algorithms that learn while making decisions. The two extreme models discussed above often obscure the value of online learning: when the distribution is known, there is nothing to learn; when it is unknown, an elegant result of \cite{Correa2} shows that no amount of additional online learning can overcome the worst-case barrier of $1/e$, which the classical secretary rule \citep{dynkin1963} already attains. In our setup, by contrast, the decision-maker learns about the world as it decides. Each observation updates the posterior over $H$ and, with it, the decision-maker’s beliefs about future observations. Waiting is therefore valuable for two reasons: a later observation may be larger, and earlier observations may reveal which world generated the sequence. Indeed, the next example shows that in the intermediate regime $1<K<\infty$, an algorithm that ignores what it learns can be far from optimal. %

\vspace{0.3cm}
\begin{example}
    {\it Suppose $K=2$, $n=3$, and $0<\varepsilon<1$. Conditional on the world, the observations are IID, with the prior and conditional probabilities given below.}
    \vspace{0.2cm}

    \begin{center}
        \begin{minipage}{0.85\linewidth}
            \centering
            \begingroup
            \normalfont
            \normalsize
            \renewcommand{\arraystretch}{1.08}
            \setlength{\tabcolsep}{8pt}
            \setlength{\arrayrulewidth}{0.4pt}
            \arrayrulecolor{navyblue!55}

            \resizebox{\linewidth}{!}{%
            \begin{tabular}{c|c|*{3}{c}}
                \rowcolor{navyblue!6}
                \rule[-0.75ex]{0pt}{3.6ex}%
                \textcolor{navyblue}{World $h$}
                & \textcolor{navyblue}{Prior $\pi_h$}
                & \textcolor{navyblue}{$\Pr[X_t=0 \mid H=h]$}
                & \textcolor{navyblue}{$\Pr[X_t=\varepsilon^{-1} \mid H=h]$}
                & \textcolor{navyblue}{$\Pr[X_t=\varepsilon^{-3} \mid H=h]$} \\
                \hline
                \rule[-0.75ex]{0pt}{3.6ex}%
                $1$
                & $1-\varepsilon^2$
                & $1-\varepsilon$
                & $\varepsilon$
                & $0$ \\

                $2$
                & $\varepsilon^2$
                & $0$
                & $1-\varepsilon$
                & $\varepsilon$
            \end{tabular}%
            }

            \arrayrulecolor{black}
            \endgroup
        \end{minipage}
    \end{center}
    \vspace{0.2cm}

    {\it Observing a 0 completely identifies $H$ and is essential to obtain near-optimal performance. To see this, consider a ``history-oblivious'' algorithm that does not use the observations before time $t$ to decide whether or not to accept $X_t$. It is obvious that a reasonable algorithm would never accept 0 and would always accept $1/\eps^3$. Thus, it is easy to see that there are only two reasonable history-oblivious algorithms: one that accepts $X_1 = 1/\eps$, and one that would reject $X_1 = 1/\eps$ but would accept $X_2 = 1/\eps$. A simple calculation shows that the latter performs better and its expected reward is $4-O(\eps)$. By contrast, the optimal dynamic programming algorithm {} obtains a reward of $5-O(\eps)$. Consequently, letting $\eps \to 0$ demonstrates that an algorithm that does not learn from its observations---in this case, checking whether $X_1 = 0$---has a suboptimal performance. In fact,  in Appendix~\ref{app:oblivious-algos}, we extend this construction to show that the gap between history-oblivious and optimal algorithms can be arbitrarily large.}
\end{example}

\vspace{0.2cm}

This example shows that a good Bayesian algorithm must adapt its decisions to what it learns from past observations. Such adaptivity has been largely absent from prior work on algorithms with partial information.\footnote{For instance, in the context of optimal stopping with distributional advice, \cite{bai2025optimal} note: ``Our framework is limited in that it does not adjust thresholds adaptively based on the observed values. Intuitively, a \textit{learning} algorithm may adjust thresholds toward [the robust algorithm] if the observed values are unlikely to be from the predicted prior, and conversely, toward the [consistent] algorithm [\ldots] if the observed values are consistent with the predicted prior.''} The reason is that while adaptivity enables better algorithms, finding good adaptive policies is, in general, computationally challenging. 

\paragraph{Research questions.} The Bayes-optimal policy is the best adaptive algorithm: it solves a dynamic program {(DP)} over all reachable histories (roughly corresponding to posteriors) and thereby accounts for both the value of learning and the value of waiting. It is thus the natural benchmark for Bayesian algorithms, and it is tighter than the offline prophet, i.e., the benchmark that knows the input sequence.
However, the state space of this dynamic program contains every reachable history and posterior, which can be exponentially large. This raises our first research question  about approximating the Bayes-optimal algorithms, i.e., \emph{philosopher inequalities} \citep{papadimitriou2021online, braverman2024new} in Bayesian optimal stopping:

\vspace{0.2cm}
\begin{center}
    {\it Can we design polynomial-time approximations of the Bayesian optimal stopping algorithms? Is there a gap between polynomial-time algorithms and the Bayes-optimal one?}
\end{center}
\vspace{0.2cm}

The above {question} concerns \emph{computational} barriers, i.e., how well a polynomial-time algorithm can approximate the Bayes-optimal policy. Our second question concerns \emph{information-theoretic} barriers. {We would like to derive {\em prophet inequalities} that bound the worst-case ratio between online algorithms and the offline prophet that knows the input sequence upfront.}
At both extremes of our model, this question is well understood. When $K=1$ and the observations are independent, \cite{krengel1977semiamarts} and \cite{samuel1984comparison} showed that the best competitive ratio is $1/2$, which led to many extensions (e.g., \cite{kleinberg2012matroid,Rubinstein}); when the observations are also identically distributed, the best competitive ratio improves to $\approx 0.745$ \citep{hill1982comparisons,Correa3}. When $K=\infty$, the minimax principle reduces the problem to the setting where the distribution is unknown: for conditionally IID observations, the secretary algorithm of \cite{dynkin1963} is $1/e$-competitive, and \cite{Correa2} showed that no algorithm does better. For $1<K<\infty$, however, the best-possible competitive ratio is not known.{}

\vspace{0.2cm}
\begin{center}
    {\it How does the best competitive ratio depend on the number of worlds $K$? Is there a separation between the computational and information-theoretic barriers?}
\end{center}
\vspace{0.2cm}

\subsection{Preview of Our Results}

Our results answer both questions. To state them, we distinguish between three formulations of the Bayesian problem: (i) in the \iid model, the observations are conditionally IID given $H$; (ii) in the \noniid model, they are conditionally independent given $H$ but may have different distributions; and (iii) in the \gen model, they may be correlated even conditional on $H$. We compare online algorithms against two benchmarks: the Bayes-optimal policy (\on) and the offline prophet (\off), who observes the whole sequence upfront. The next theorem summarizes our findings; the formal statements are Theorems~\ref{thm:main}--\ref{thm:iid-comp} and \Cref{prop:noniid-cr}. Below, $|\cal I|$ denotes the input size.
\begin{theorem}[Informal]{}{}
    For the Bayesian optimal stopping problem, there exists a $(1-\eps)$-approximation of \on{} with runtime $\poly(|{\cal I}|^{\log^* |{\cal I}| + 1/\eps} \cdot (1/\eps)^{1/\eps^2})$, whereas no online algorithm is substantially more competitive against \off{} than in the worst case. In particular:
    \begin{itemize}
        \item For the \iid problem, there is a polynomial-time approximation scheme against \on. Against \off, we obtain a $(1/e + 1/\poly(K))$-competitive algorithm, and no algorithm is better than $(1/e + 1/\poly(K))$-competitive.
        \item For the \noniid problem, there is an approximation scheme against \on. Against \off, we obtain a $\nicefrac 1 {(2K)}$-competitive algorithm, and no algorithm is better than $\nicefrac 1 K$-competitive.
        \item For the \gen problem, there is an approximation scheme against \on. Against \off, the best possible competitive ratio is $\nicefrac 1 n$.
    \end{itemize}
\end{theorem}
 Our approximation scheme runs in time $\poly(|{\cal I}|^{\log^* |{\cal I}| + 1/\eps} \cdot (1/\eps)^{1/\eps^2})$, where $|\cal I|$ is the input size (defined formally in \Cref{ssec:model}), $\log^*(x)$ is the number of times the $\log$ function must be applied to $x$ before the result is at most 1, and $\eps \in (0,1]$ is the accuracy parameter. While this is not, strictly speaking, a polynomial-time approximation scheme (PTAS), its runtime is far below the typical $\poly(|\cal I|)^{\log |\cal I|}$ runtime of a quasi-PTAS. In the \iid model, and in the \noniid model with $K=O(1)$, the same scheme is a genuine PTAS (\Cref{thm:main}).\footnote{When $K=O(1)$, the \iid and \noniid models also admit an FPTAS by discretizing the space of posteriors and solving an approximate dynamic program; since this approach is fairly standard, we did not include it in the paper.}

The theorem shows a sharp separation between philosopher and prophet inequalities in the Bayesian set-up. Against \on, the approximation ratio can be made arbitrarily close to 1 in all three models. Against \off, the competitive ratio degrades with $K$ toward its worst-case value: in the \iid model, the advantage over the $1/e$ guarantee for unknown distributions vanishes inversely polynomially in $K$, and in the \noniid model the competitive ratio is $\Theta(1/K)$. In other words, the prophet remains out of reach for online algorithms, whereas the Bayes-optimal policy can be approximated efficiently. This result resolves, in part, the open questions posed by \cite{correa2020unknowniidprophetsbetter} for the competitive ratio in our \iid setting with finite $K$, saying ``[the $1/e$ asymptotic bound] leaves open the possibility of better
guarantees when the number of candidate distributions is small, a setting we believe is of
strong interest to applications.'' {} {}

\subsection{Our Techniques and Contributions}
Beyond the modeling framework, our technical contributions are as follows.

\subsubsection*{A primal-dual iterative approach.} Our approximation scheme is built around a \emph{primal improvement} step that converts a competitive algorithm into one with a strictly better approximation ratio, i.e., transforming the guarantee against \off{} to a better guarantee against \on{} through an improved algorithm (\Cref{prop:one-step-improvement}). We generalize the notion of competitiveness to be measured against a \emph{dual martingale\/}, a term we use to refer to a martingale $(Z_t)_t$ that satisfies $Z_t \ge X_t$ almost surely for every $t$.  Because $(Z_t)_t$ is a martingale and $Z_t \ge X_t$, we infer $\expar{Z_\tau} \ge \expar{X_{\tau}}$ for any (bounded) stopping time $\tau$. Letting $\tau = \tau^*$ be the stopping time of the Bayes-optimal policy, Doob's optional sampling theorem implies that no online algorithm can obtain more than $\expar{Z_1} = \expar{Z_{\tau}}$, so $\expar{Z_1} \ge \on$; we say that an algorithm is $\alpha$-competitive against $(Z_t)_t$ if it earns at least $\alpha \cdot \expar{Z_1}$. A natural starting point is the Doob martingale of the prophet's reward, $Z_t = \expar{\max_{s \in [n]} X_s \mid \cal F_t}$, against which competitiveness is the usual competitiveness against \off.

The improvement step, \Cref{alg:approx}, takes as input a dual martingale $(Z_t)_{t \in [n]}$ and an algorithm $\cal A$ that is $\alpha$-competitive against it, and returns an improved algorithm that is a $\alpha^+$-approximation (against \on), where $\alpha^+ = (1+\ln(1/\alpha))^{-1} > \alpha$. To do so, \Cref{alg:approx} randomizes between two subroutines. With some probability, it runs the $\alpha$-competitive baseline $\cal A$. With the remaining probability, it runs a \emph{dual-threshold} subalgorithm \mart, which draws a (random) cutoff $Q \in [\alpha, 1]$ and accepts the first $X_t$ with $X_t \ge Q \cdot Z_t$. The two subroutines complement each other. On sample paths where some observation is large relative to the dual martingale, i.e., $X_t / Z_t \ge q$ for some $t$, \mart with cutoff $q$ stops and collects at least a $q$ fraction of the dual value. Conversely, the baseline algorithm covers other sample paths on which every ratio $X_t/Z_t$ is small, i.e., the gap between the observations and the dual martingale is large. \Cref{prop:one-step-improvement} shows that this mixture is an $\alpha^+$-approximation of \on, where $\alpha^+ > \alpha$ for every $\alpha \in (0,1)$; for instance, starting from the $1/e$-competitive secretary algorithm in the \iid model gives a $1/2$-approximation by one step improvement.

Since $\alpha^+ > \alpha$, it is natural to apply the improvement step again, with the improved algorithm as the new baseline. However, the improvement step cannot be applied directly to its own output. It requires a baseline that is competitive against a dual martingale, whereas \Cref{prop:one-step-improvement} only shows that the output approximates \on. We resolve this by showing that the improvement step also produces a new, tighter dual martingale against which its output is competitive, which turns the primal improvement into a \emph{primal-dual} one. Specifically, let $M = Z_n$ be the terminal value of $(Z_t)_t$ and define $\rho = \max_{t} X_t / Z_t \le 1$ on each sample path. We show that the output of \Cref{alg:approx} is $\alpha^+$-competitive against the Doob martingale of the \emph{shrunken} certificate $M \cdot \max\{\rho, \alpha\}$, which still dominates $(X_t)_t$ and is tighter than $M$. \Cref{alg:iteration} pairs the primal improvement with this \emph{dual multiplicative shrinkage}, producing pairs $(\cal A_j, Z^{(j)})$ in which $\cal A_j$ is $\alpha_j$-competitive against $Z^{(j)}$, with
\[
    \alpha_0 \le \alpha_1 \le \cdots < 1 \ , \qquad \expar{Z_1^{(0)}} \ge \expar{Z_1^{(1)}} \ge \cdots \ge \on \ , \qquad \alpha_{j+1} = \left(1 + \ln \nicefrac 1 {\alpha_j}\right)^{-1} \ .
\]
Thus, our construction approaches \on{} simultaneously from below, through improving algorithms, and from above, through shrinking dual certificates. The recursion reaches a constant after $O(\log^* \nicefrac 1 {\alpha_0})$ steps and reaches $1-\eps$ after $O(1/\eps)$ further steps. To start this process, the initial dual martingale is the Doob martingale from the prophet's reward, i.e, $Z_t^{(0)} = \expar{\max_{s \in [n]} X_s\mid \cal F_t}$. As baselines, we use the secretary algorithm in the \iid model ($\alpha_0 = 1/e$), the $\nicefrac 1 2$-competitive threshold algorithm of the world $h^\star$ with the largest contribution to \off{} in the \noniid model ($\alpha_0 = \nicefrac 1 {2K}$), and accepting the time step $t^\star$ with the largest contribution to \off{} in the \gen model ($\alpha_0 = \nicefrac 1 n$). Putting these together, \Cref{alg:iteration} returns the desired approximation scheme for the Bayesian optimal stopping problem.  

\subsubsection*{Efficient approximation of dual martingales.} The main obstacle to implementing \Cref{alg:iteration} is that the martingales $Z^{(j)}$ cannot be computed exactly. They are defined recursively: evaluating $Z_t^{(j)}$ at a history $h_t$ requires averaging the certificate $M_j = Z_n^{(j)}$ over all continuations of $h_t$, and $M_j$ on each continuation depends on $Z^{(j-1)}$ at all of its prefixes. What makes approximation possible is that our dual shrinkage is local: it rescales the certificate on each sample path using only quantities along that path. Hence, $Z_t^{(j)}(h_t)$ can be estimated by Monte Carlo, sampling continuations of $h_t$ and recursively estimating $Z^{(j-1)}$ along each of them. By contrast, the optimal dual martingale, like the Snell envelope, is determined by backward induction over the entire scenario tree. To obtain estimates that are accurate within a multiplicative factor, we sample from a change of measure that reweights each path by its prophet reward, so that rare paths with large rewards are not missed. The resulting estimator runs in time $\poly(2^{j^2} |\cal I|^{j+1})$, which gives an approximation scheme in general and a PTAS whenever $\alpha_0$ is a constant (the \iid model, and the \noniid model with $K = O(1)$).

\subsubsection*{Hardness of prophet inequalities.} To understand the information-theoretic barriers in this problem, we construct an \iid instance in which no online algorithm is better than $(1/e + O(1/\sqrt K))$-competitive. A minimax argument shows that $1/e$ is the best constant as $K \to \infty$ \cite[Theorem~5.1]{correa2020unknowniidprophetsbetter}; our construction is explicit and quantifies the dependence  on $K$. The instance must balance two requirements. On the one hand, observations should not quickly reveal the world; otherwise an online algorithm could identify it and tailor its thresholds. On the other hand, the worlds must differ enough that their prophet rewards diverge. We achieve both with a set of distributions that have nested supports and such that observed samples are minimally informative, making the posterior change only mildly. Analyzing the Bayesian DP directly on this instance is difficult because the posterior still evolves and its analysis is intricate. Instead, we bound the reward of every online algorithm by constructing an explicit dual martingale and by then comparing that with \off. An interesting takeaway of our paper is that dual martingales can serve both for algorithm design and for impossibility results in Bayesian models.

We complement this hardness result with a $(1/e + \Omega(1/K^2))$-competitive algorithm, which adapts the robustness--consistency algorithm of \cite{bai2025optimal} by committing to the world with the largest contribution to \off{} and setting its parameters in terms of $K$. Moreover, in the \noniid model, we show that the $\nicefrac 1 {2K}$-competitive baseline above is optimal up to a factor of 2. Finally, by constructing a martingale-based hard instance, we show that in the \gen model, $1/n$ is the optimal competitive ratio. 

\subsection{Related Literature} {}

\subsubsection*{Prophet and philosopher inequalities.}
Prophet inequalities originate with \cite{krengel1977semiamarts} and \cite{samuel1984comparison}, who showed that the best competitive ratio for independent observations is $1/2$; for IID observations, it is $\approx 0.745$ \citep{hill1982comparisons,Correa3}. These results have been extended in many directions, e.g., to matroids \citep{kleinberg2012matroid} and posted pricing \citep{feldman2014combinatorial,dutting2020prophet}; see \cite{lucier} for a survey. Closest to our work are models with limited distributional knowledge, such as unknown distributions \citep{Correa2,correa2020unknowniidprophetsbetter}, samples \citep{Kaplan,Correa,Rubinstein}, inaccurate priors \citep{Kesselheim}, and predictions \citep{bai2025optimal,Kehne}. While this line of work has mostly focused on the offline benchmark, the online benchmark is an equally natural one, studied through the recently introduced notion of philosopher inequalities \citep{papadimitriou2021online}. For online stochastic matching, approximating the optimal online policy within some constant factor is PSPACE-hard \citep{papadimitriou2021online}, and constant-factor approximations are known \citep{papadimitriou2021online,braverman2022max,braverman2024new}; see also \cite{anari2019nearly,dutting2023prophet,fu2018ptas}. More related to our work, \cite{posterior-matching} recently used the minimax principle to reduce the worst-case competitive analysis of online combinatorial problems to Bayesian design under arbitrarily correlated distributions. Through this reduction, they aim to provide a unifying framework for designing competitive algorithms for online combinatorial problems, i.e., to study the information-theoretic barrier. Their main technical idea is to use the posterior of the offline optimum to guide online decisions. Our focus, by contrast, is on Bayesian optimal stopping, which we study across the whole spectrum of uncertainty, parametrized by the prior size $K$ and in different models (\iid, \noniid, \gen). We study both the computational and information-theoretic barriers, and by devising an approximation scheme against \on, we show a separation between the two.

\subsubsection*{Stochastic control.}
Our problem is a partially observable Markov decision process with a stopping action \citep{Rieder}, and computing optimal policies for such processes is PSPACE-hard in general \citep{Papadimitriou}. In applied probability, a rich literature on pricing American options approximates optimal stopping by simulation \citep{longstaff2001valuing}, together with dual upper bounds given by martingales \citep{rogers2002monte,haugh2004pricing,chen2007additive,andersen2004primal}; \cite{brown} extend this duality to information relaxations of general stochastic dynamic programs. Our dual martingales are closely related to these representations. These methods typically focus more on heuristics than on approximation algorithms with worst-case guarantees. An exception is \cite{chen-goldberg2026}, who give an additive $\eps$-approximation that runs in polynomial time for every fixed $\eps$, for optimal stopping problems with a bounded rewards assumption. When the offline benchmark has a bounded coefficient of variation, their algorithm is a multiplicative approximation. Their main technical tool is an expansion of the value function that is related to our dual martingales through additive updates. However, we show that under their information-access model which is based on sample oracles, obtaining an approximation ratio better than 6/7 is not possible (see \Cref{ssec:model}). We instead assume access to a probability and value oracle and obtain an approximation scheme with no assumption on reward ranges or coefficients of variance---which are typically the hard instances for optimal stopping approximation algorithms.

\section{Problem Formulation and Preliminaries}

\subsection{Bayesian Optimal Stopping}\label{ssec:model}
 We consider an optimal stopping problem over a horizon $[n] = \{1,\cdots, n\}$. There are $K$ possible worlds $\{1,\cdots, K\}$.  Each world $k$ corresponds to a sequence of distributions $(D_t^{(k)})_{t \in [n]}$ on non-negative reward values $v_1, \cdots, v_m \in {\bb R}_+$. We assume that $v_1, \cdots, v_m$ are given parts of the input and the overall bit complexity of the input includes the encoding of these values. The ground truth world $H$ is drawn randomly from $\{1,\cdots, K\}$ according to the prior distribution $\pi = (\pi_1, \cdots, \pi_K)$. Then, there is a sequence of $n$ random rewards $(X_1, \cdots, X_n)$, drawn from the joint distribution $D^{(H)}$ conditional on world $H$. 
 While the algorithm designer knows the prior distribution $\pi$ and value distributions $(D^{(h)})_{h \in [K]}$---formally explained in the next paragraph, it does not observe $H$. When for each $h\in [K]$, the distrbutions $D_t^{(h)}$ are independent and identical across time steps, we call our model the \iid case; when they are only independent but not identical; we call it \noniid; and when they are not even independent, we call it \gen. If $K=1$, the \noniid case is the classic prophet inequality problem of \cite{samuel1984comparison}, whereas the \iid case is simply the IID prophet inequality problem of \cite{Correa3}. 
 
 {Our goal is to devise a polynomial-time algorithm ${\cal A}$
that selects one item from $X_1, \cdots, X_n$ of maximal expected value, only knowing the prior over $K$ worlds and their corresponding distributions. We measure running time by the number of arithmetic operations and calls to the information oracles defined below, counting each as one unit of time. Our bounds are expressed in terms of the number of worlds $K$, the support size $m$, the horizon $n$, and the accuracy $\eps$.} We denote the reward of our algorithm by $r({\cal A})$. 

\subsubsection*{Information access model.} We assume access to two information oracles: (i) a probability oracle that, given any history $h_t$ and value $v_q$, returns $\prpar{X_{t+1} = v_q \mid \cal F_t}$, where $\cal F_t$ is the natural filtration from the history $h_t$; and (ii) a value oracle that returns $\expar{\max_{t' \in [n]} X_{t'} \mid \cal F_t}$ which is the conditional expected reward of \off{} given history $h_t$. In the \iid and \noniid models, both oracles can be efficiently constructed from the probability mass function (PMF) of each distribution $D^{(h)}$, where we note that having the PMF is a standard assumption in the prophet inequality literature (see Appendix~\ref{sssec:exact-off} for the construction). In this case, the given PMFs also allow us to compute the posterior over worlds exactly by Bayes’ rule. For the \gen model, however, these oracles are additional computational assumptions, as the input distributions may not have compact representations. 

One potential alternative is to assume access to a sample oracle that, given any history $h_t$, outputs a fresh sample for the continuation of $h_t$ from the conditional distribution $\prpar{\cdot \mid \cal F_t}$. This is the information model used by \cite{chen-goldberg2026}. However, under such model, we show in Appendix~\ref{app:conditional-sampling-hardness} that even the classical problem with $K=1$ admits no approximation scheme and no polynomial-time algorithm can guarantee an approximation ratio better than 6/7. This result shows that we need stronger information access models than sampling oracles to obtain near-optimal performance.

\subsubsection*{Benchmarks.} We consider two benchmarks. The first one is the offline benchmark that knows $M = \max \{X_1, \cdots, X_n \}$ upfront. Thus, the expected reward of this benchmark is $\ex{M}$, where the expectation is with respect to the true world $H$ and the randomness in $X_1, \cdots, X_n$ conditional on $H$. With a slight abuse of notation, we denote the benchmark and its expected reward by \off. Consequently, we define the \emph{competitive ratio} by
\(
    \text{CR}({\cal A}) = \frac{\ex{r({\cal A})}}{\off}.
\)

The second benchmark we consider is the online benchmark that has the same information as us, but it has unlimited computational power. Indeed, this algorithm and its expected reward correspond to an (exponentially-sized) Bayesian DP. We denote both by \on. Then, the \emph{approximation ratio} is defined as
\(
    \text{AR}({\cal A}) = \frac{\ex{r({\cal A})}}{\on}.
\)

The Bayesian DP has two actions: accept $X_t$ and terminate, or reject it and continue. Its value-to-go, after observing history $h_t = (X_1, \cdots, X_t)$ and before making the decision at time $t$, is the \emph{Snell envelope} $S_t=V_t(h_t)$, defined by backward induction:
\begin{align}
    S_n=X_n, \qquad
    S_t=\max\left\{X_t,\expar{S_{t+1}\mid\cal F_t}\right\}
    \quad (t<n). \label{eq:snell}
\end{align}
Thus, $S_t$ compares the reward from stopping immediately with the expected reward from continuing optimally, with the conditional expectation incorporating the posterior over worlds. It is the smallest supermartingale dominating $(X_t)_{t \in [n]}$, and thus, $\expar{S_1}=\on$. %

\subsubsection*{Notation and assumption.} For a sample path $\omega$ of all $n$ observations and a time $t$, let $\omega_t$ be the history until time $t$ under $\omega$, i.e., the $t$-prefix of $\omega$. The state space is $\Omega = \{v_1, \cdots, v_m \}^n$. Throughout the paper, we assume that
$\expar{\max_{t'\in[n]} X_{t'} \mid h_t}>0$
for every reachable history $h_t$.
This assumption is only for notational convenience and
can be removed without affecting our results by simply making the algorithm ignore sample paths that violate this assumption.

\subsection{Martingale Preliminaries}\label{ssec:martingales}

\begin{definition}
An integrable process $(Z_t)_{t\in[n]}$ is a martingale with respect to filtration $({\cal F}_t)_{t\in[n]}$ if for every $t < n$,  $Z_t$ is ${\cal F}_t$-measurable, and
\(\expar{Z_{t+1}\mid\cal F_t}=Z_t \) almost surely.  Replacing the equality by ``$\le$'' defines a supermartingale. 
\end{definition}

Any $\cal F_n$-measurable random variable $M$ is called a \emph{terminal} random variable. Given a terminal random variable $M$, the corresponding Doob's martingale is defined as $Z_t = \expar{M \mid {\cal F}_t}$.\footnote{The process $(Z_t)_t$ is a martingale because $\expar{Z_{t+1} \mid \cal F_t}
    =\expar{\expar{M\mid\cal F_{t+1}}\mid\cal F_t}
    =\expar{M\mid\cal F_t}
    =Z_t$ by the tower property.} Conversely, any martingale $(Z_t)_t$ can be expressed in terms of its terminal value by setting $M = Z_n$ and noting $Z_t = \expar{Z_n \mid \cal F_t} = \expar{M \mid \cal F_t}$. For any history
$h_t=(x_1,\ldots,x_t)$, we write $Z_t(h_t)$ as the value of the martingale at time $t$,
conditional on having observed $h_t$. Formally, we mean $Z_t(h_t) = \expar{M \mid X_1=x_1, \cdots, X_t=x_t}$.

The main property of a martingale is that its (unconditional) expectation remains the same, so that $\expar{Z_1} = \cdots = \expar{Z_n}$.
We now recall Doob's optional sampling/stopping theorem---a key result from martingale theory---that concerns the expected value of a martingale 
at stopping times. A random variable $\tau \in [n] \cup \infty$ is a \emph{stopping time} if $\{\tau \le t\}$ is $\cal F_t$-measurable for every $t \in [n]$. A stopping time $\tau$ is bounded if $\tau \le n$ almost surely.

\begin{theorem}[Doob's Optional Sampling Theorem, cf. \cite{williams}]\label{thm:optional-sampling}
Let $(Z_t)_{t\in[n]}$ be an integrable martingale, and let $\tau$ be a bounded stopping time. Then, we have $\expar{Z_\tau} = \expar{Z_1}$. 
\end{theorem}

Next, we define the notion of \emph{dual martingales} which is central to our paper. %
\begin{definition}\label{def:dual-martingale}
    A martingale $(Z_t)_t$ is a dual martingale for the sequence $(X_t)_t$ if, for every $t \in [n]$,
    we have $Z_t \ge X_t$ almost surely. Moreover, the corresponding terminal random variable $Z_n$ is called a \textit{dual certificate}. 
\end{definition}
A dual martingale provides an upper bound on the performance of the optimal algorithm. 
Indeed, for any bounded stopping time $\tau$, Doob's optional sampling (\Cref{thm:optional-sampling}) implies $\expar{X_\tau} \le \expar{Z_{\tau}} = \expar{Z_n}$. Letting $\tau = \tau^*$ be the stopping time corresponding
to the optimal algorithm shows the claim. Consequently, we can define the notion of \emph{competitiveness against a dual martingale}: An algorithm $\cal A$ is $\alpha$-competitive against $(Z_t)$, if $r(\cal A) \ge \alpha \cdot \expar{Z_1}$. If this holds, because $\expar{Z_1} \ge \on$, we immediately infer that $\cal A$ is an $\alpha$-approximation to the optimum online algorithm as well.

The term \emph{dual} refers to the following pair of optimization problems:
\[
    \on
    = \sup_{\substack{\tau\text{ stopping time}\\1\le\tau\le n}}
      \ex{X_\tau}
    \quad \le \quad
    \inf_{\substack{(Z_t)_t\text{ martingale}\\
                    Z_t\ge X_t\text{ a.s. for every }t}}
      \ex{Z_n} \ .
\]
The primal problem chooses a stopping policy to maximize expected reward,
whereas the dual problem chooses a dominating martingale to minimize
its expected value. The inequality is weak duality, established above
by optional sampling. In fact, equality holds: this is strong duality,
which our analysis in \Cref{ssec:primal-dual} also establishes.

\section{Approximation Algorithms}\label{sec:approximation}
In this section, we present our main result, which is a near-polynomial time approximation scheme for the Bayesian optimal stopping problem. In the following theorem, $\log^*$ denotes the iterated logarithm function. 
\begin{theorem}\label{thm:main}
{For every $0<\eps\le1$, there is an algorithm $\cal A$ such that for any instance of the Bayesian optimal stopping problem in the \gen model, $r(\cal A) \ge (1-\eps)\textup{\on}$ and ${\cal A}$ runs in time $\poly(K m n^{1/\eps + \log^* n} \cdot ({1/\eps})^{1/\eps^2})$.} In the special case of the \noniid model, the exponent of $n$ in the running time is $1/\eps + \log^*(\min \{n, 2K\})$, and for the \iid model, this exponent is $1/\eps$, making this algorithm a polynomial-time approximation scheme.
\end{theorem}

Notice that if $K=O(1)$ so that the number of different worlds is bounded, our algorithm for the \noniid model is in fact a genuine PTAS. 

\begin{remark}[Extension to exchangeable observations]
The PTAS guarantee for the \iid model in \Cref{thm:main} extends to the more general setting where the observations are exchangeable conditional on $H$. Indeed, conditional exchangeability implies unconditional exchangeability, under which the classical secretary algorithm remains $1/e$-competitive against \off. As we show in this section, the runtime of our approximation scheme depends on the competitive ratio of a baseline algorithm, while the remainder of our analysis applies to arbitrarily correlated observations. Thus, the existence of a $1/e$-competitive baseline yields a PTAS for conditionally exchangeable observations as well, assuming polynomial-time access to the probability and value oracles of \Cref{ssec:model}. This extends the PTAS guarantee established for the prophet secretary problem by \cite{dutting2023prophet} to the Bayesian optimal stopping problem.
\end{remark}

The rest of this section establishes \Cref{thm:main}. In \Cref{ssec:primal-improvement}, we present \Cref{alg:approx} that can transform any competitive algorithm into an approximation algorithm with a strictly better approximation factor; we call this process ``Primal Improvement.'' \Cref{ssec:primal-dual} develops the \PrimalDualImprovement{} subroutine which leads to \Cref{alg:iteration} that repeatedly uses \Cref{alg:approx} and is the main construction underlying \Cref{thm:main}. 
Next, \Cref{ssec:implementation} explains how to implement \Cref{alg:iteration} in polynomial time. Finally, \Cref{ssec:proof} analyzes our final algorithm, thereby establishing \Cref{thm:main}.

\subsection{Warm-up: Approximation Algorithms via Primal Improvement}\label{ssec:primal-improvement}
In this subsection, we present \Cref{alg:approx}, a central piece of our overall approximation scheme. As inputs to \Cref{alg:approx}, we assume having access to a dual martingale $(Z_t)_t$ and a baseline algorithm $\cal A_{\rm base}$ that is known to be $\alpha$-competitive against $(Z_t)$. Then, \Cref{alg:approx} converts these two inputs into a $g(\alpha)$-approximation of $\on$, where $g(x) = 1/(1+\ln(1/x)) > x$, and provides a performance guarantee strictly better than $\alpha$.
\begin{proposition}\label{prop:one-step-improvement} Consider a Bayesian optimal stopping problem instance. Given a dual martingale $(Z_t)_t$ and a polynomial-time algorithm $\cal A_{\rm base}$ that is $\alpha$-competitive against $Z_t$ (with $\alpha \in (0,1))$, Algorithm~\ref{alg:approx} returns a $1/(1+\ln(1/\alpha))$-approximation against \textup{\on}.
\end{proposition}
While the base algorithm \obl is itself an $\alpha$-approximation, we can leverage it to obtain a better approximation factor; the multiplicative improvement grows as the baseline performance factor $\alpha$ decreases (see \Cref{fig:boost} in Appendix~\ref{app:figures}). 

Before presenting the algorithm, we give three immediate implications of Proposition~\ref{prop:one-step-improvement} for our models of \iid, \noniid, and \gen. To use \Cref{prop:one-step-improvement}, we must specify a dual martingale and a corresponding $\alpha$-competitive baseline algorithm. Let $M = \max_{t \in [n]} X_t$ be the random variable corresponding to the offline optimum, and let $Z_t = \expar{M \mid \cal F_t}$ be its corresponding Doob's martingale. We use $(Z_t)_t$ as our dual martingale. 

We now specify a baseline competitive algorithm in each model. For the \iid model, we use the classical secretary algorithm of \cite{dynkin1963}, which is $1/e$-competitive against $M$. For the \noniid model, we can determine the world $h^* = \arg\max_{h \in [K]} \pi_h \expar{\off \mid H = h}$ that has the highest contribution to \off. Then, our baseline algorithm is one that assumes $H = h^*$, and uses the corresponding {threshold-based prophet} $1/2$-competitive algorithm for world $h^*$; it follows that this baseline algorithm is $1/(2K)$-competitive for the \noniid Bayesian problem. Finally, for the {\gen} model, we pick  a timestep $t^*$ uniformly at random from $\{1,\cdots,n\}$, and accept the observation at time $t^*$; this baseline algorithm chooses the maximum observation with probability $1/n$ and is $(1/n)$-competitive. 

Applying \Cref{prop:one-step-improvement} by specifying the above baseline algorithms and the dual martingale $(Z_t)$, the following corollary ensues:
\begin{corollary}\label{cor:baseline-approximations}
The Bayesian optimal stopping problem admits a polynomial-time
algorithm with approximation factor
\[
\begin{cases}
    \dfrac{1}{2},
        & \text{in the \iid{} model}; \\[6pt]
    \dfrac{1}{1+\ln(2K)},
        & \text{in the \noniid{} model}; \\[6pt]
    \dfrac{1}{1+\ln n},
        & \text{in the \gen{} model}.
\end{cases}
\]
\end{corollary}

\subsubsection{Algorithm.}\label{sssec:approx-algo}
We now present the details of \Cref{alg:approx}. In addition to the baseline algorithm $\cal A_{\rm base}$, we also use another ``dual-threshold'' subroutine called $\cal A_{\rm thr}$. Consequently, with probability $\gamma = (1+\ln(1/\alpha))^{-1}$, we run the baseline algorithm and with the remaining probability, we run $\cal A_{\rm thr}$. The subroutine $\cal A_{\rm thr}$, explained in Line~\ref{line:thr-beg}-\ref{line:thr-end}, uses the dual martingale $(Z_t)$ as thresholds. Concretely, before any observations, \mart draws a random cutoff $Q \in [\alpha,1]$ with density $f_Q(q) \propto 1/q$. Next, the subroutine accepts the first arriving item $X_t$ that satisfies  $X_t \ge Q Z_t$. In other words, $Q$ is the threshold on $X_t / Z_t$ for accepting a newly arriving item.

Notice that our current definition for $Z_t$ is based on the hindsight optimal $M$, which makes the martingale easily computable for every history $h_t$ using the value oracle assumed in \Cref{ssec:model}. Nevertheless, we present \Cref{alg:approx} in a more general form in which $Z_t$ may not be possible to evaluate exactly (in polynomial time) and we must approximate it. This more general form will be used in \Cref{ssec:implementation} to develop approximation schemes.
 
\begin{algorithm}[H]
    \caption{Primal improvement}
    \label{alg:approx}
    \begin{algorithmic}[1]
    {
        \STATE {\bf Input:} (i) A dual martingale $(Z_t)_t$; (ii) An $\alpha$-competitive algorithm $\cal A$ against $(Z_t)_t$
        \STATE Set $\gamma = \frac{1}{1 + \ln(1/\alpha)}$
        \STATE With probability $\gamma$, run $\cal A$
        \STATE With the remaining probability $1-\gamma$, run the following algorithm \mart:
        \STATE \qquad Draw $Q \in [\alpha, 1]$ with density $f(q) = \frac{1}{q \ln(1/\alpha)}$ \label{line:thr-beg}
        \STATE \qquad \textbf{for all} times $t = 1, \cdots, n$ \textbf{do}
        \STATE \qquad \qquad Let $h_t$ be the history at time $t$
        \STATE \qquad \qquad Using fresh randomness, compute $L_t(h_t)$ such that $\frac{L_t(h_t)}{Z_t(h_t)} \in [1-\eps, 1]$  \\ \qquad \qquad with probability at least $1-\eps/n$. \label{line:approximate-z}\quad \CommentSty{\color{red} // A Monte Carlo estimate, as in \Cref{lem:accuracy-analysis}}
        \STATE \qquad \qquad If $X_t \ge QL_t$, accept $X_t$ and terminate \label{line:thr-end}
        }
    \end{algorithmic}
\end{algorithm}

Intuitively, the two subalgorithms \obl and \mart extract complementary parts of the optimal reward $X_{\thr}$, where $\thr$ is the stopping time corresponding to the optimal online algorithm.
Writing the classical base value and surplus decomposition of the rewards \citep{lucier}:
\[
    X_{\thr} \le
    \underbrace{\alpha Z_{\thr}}_{\text{base value}}
    +
    \underbrace{\{ X_{\thr} - \alpha Z_{\thr} \}_+}_{\text{surplus}}
    \ ,
\]
the subroutine \obl captures a reward of $\alpha Z_{\thr}$, whereas, as will be shown, \mart roughly captures a fraction of $\{ X_{\thr} - \alpha Z_{\thr} \}_+$. The challenge, however, is that it is not a priori known which component is larger, which is why we randomize between the two subalgorithms. 

While \obl may not necessarily use any Bayesian information, \mart crucially leverages the Bayesian structure of our problem. The dual martingale $Z_t$ is an approximation (upper bound) to the Snell envelope \eqref{eq:snell}. Suppose that at some history $h_t$, we have $Z_t(h_t) = S_t(h_t)$. Then it is clear that we should accept $X_t$ only if it exceeds $Z_t(h_t)$---corresponding to deterministically setting $Q = 1$. 
However, because the actual value of $Z_t$ may differ from $S_t$, we use a more permissive threshold $Q \in [0,1]$ on $X_t / Z_t$. 

The specific randomization of $Q$ warrants some explanation. To simplify the notation, consider some fixed $x = X_{\thr}$ and $z = Z_{\thr}$. Our goal is to obtain a reward close to $x$, which can be written in the layer cake representation as 
\(
    x = \int_0^1 z \cdot \ind{x \ge qz} dq
\). When \mart accepts an item $X_t$, it is guaranteed that $X_t \ge q Z_t$, so we can only be sure of obtaining a reward of at least $qZ_t$; thus, our collected reward
can be lower bounded by $qZ_t \cdot \ind{X_t \ge qZ_t}$. Assume, for the moment, that $t = \thr$ and $Z_t = z$. Then, choosing $Q \sim f(q)$ would yield a reward of $\int_0^1 qf(q)z \cdot \ind{x \ge qz}$. For this reward to be a constant-factor approximation of $x$, we thus set $f(q) \propto 1/q$; with appropriate renormalization, this intuition drives our approximation factor. 
Given that $(Z_t)_t$ is a martingale, its expectation remains constant, implying $\expar{Z_t} = z$ in general. The analysis in the next section formalizes this argument and proves that $Z_t$ can be replaced with $Z_{\thr}$ to obtain an approximation factor analysis.

\subsubsection{Analysis of the Primal Algorithm.}

We will next prove Proposition~\ref{prop:one-step-improvement}, where we recall that we assume $Z_t$ itself is used instead of the approximation $L_t$ in \Cref{alg:approx}. The argument follows from a careful analysis of \mart, which relies on the evolution of the process $(Z_t)_t$. For a fixed $q\in[0,1]$, let $r_q$ be the expected reward of {\mart} when $Q=q$ and $L_t=Z_t$ at every time $t \in [n]$, and recall that $\tau^\star$ is the stopping time of \on. We can now lower bound $r_q$ as follows.

\begin{lemma}\label{lem:posterior-prophet}
For every fixed $q\in [0,1]$, we have $r_q \ge q \cdot \ex{Z_{\tau^\star} \cdot \ind{X_{\tau^\star} \ge qZ_{\tau^\star}} }$.
\end{lemma}

Before proving this lemma, we leverage it to establish Proposition~\ref{prop:one-step-improvement}.

\begin{proof}[ of Proposition~\ref{prop:one-step-improvement}]
Clearly, $\tau^\star$ is always at most $n$, and thus it is bounded almost surely. Consequently, Doob's Optional Sampling Theorem implies $\expar{Z_{\thr}} = \expar{Z_{1}}$. Thus, {\obl} has an expected reward of at least $\alpha \cdot {\expar{Z_{1}}} = \alpha \cdot \expar{Z_{\thr}}$. 

Recall that Algorithm~\ref{alg:approx} chooses {\mart} using a parameter $Q \in [\alpha, 1]$ with density $f(q) = \frac{1}{q \ln(1/\alpha)}$. We can thus lower bound the expected reward of Algorithm~\ref{alg:approx} by
    \begin{align}
        &\gamma \alpha \cdot \ex{Z_{\thr}} + (1-\gamma)  \int_{\alpha}^1 r_{q}  f(q) \ dq \notag
        \\ &=  \gamma  \alpha \cdot \ex{Z_{\ton}} + \frac{(1-\gamma)}{\ln(1/\alpha)}  \cdot \int_{\alpha}^1 r_{q}/q \ dq \notag
        \\ & \overset{(i)}{\ge} \gamma  \alpha \cdot \ex{Z_{\ton}} + \frac{(1-\gamma)}{\ln(1/\alpha)} \int_{\alpha}^1 \ex{Z_{\tau^\star} \cdot \ind{X_{\tau^\star} \ge qZ_{\tau^\star}} } \ dq \notag
        \\ 
        & = \gamma   \alpha \cdot \ex{Z_{\ton}} + \gamma \int_{\alpha}^1 \ex{Z_{\tau^\star} \cdot \ind{X_{\tau^\star} \ge qZ_{\tau^\star}} } \ dq \notag
        \\ & \overset{(ii)}{=} \ex{ \gamma  \alpha \cdot {Z_{\tau^\star}} + \gamma  \int_{\alpha}^1 {Z_{\tau^\star} \cdot \ind{X_{\tau^\star} \ge qZ_{\tau^\star}}} \ dq} \ , \label{eq:r-lb}
    \end{align}
    where $(i)$ uses Lemma~\ref{lem:posterior-prophet}, and $(ii)$ uses linearity of expectation.

    Now, notice that for values $0 \le x \le z$, a simple integral derivation gives the layer cake representation 
    \[
        z\int_{\alpha}^1 \ind { x \ge qz} \ dq = \{ x-\alpha z \}_+  \ .
    \] Thus, because $Z_{\tau^\star} \ge X_{\tau^\star}$ almost surely, \eqref{eq:r-lb} is equal to
    \begin{align}
        \ex{\gamma  \alpha {Z_{\tau^\star}} + \gamma  \left \{ X_{\tau^\star} - \alpha {Z_{\tau^\star}} \right \}_+} & =  \ex{\max \left \{ \gamma X_{\tau^\star}, \gamma  \alpha {Z_{\tau^\star}} \right \} } \label{eq:reward} \\ & \ge \gamma  \ex{X_{\tau^\star}} = \gamma  \cdot \on  = \frac{1}{1+\ln(1/\alpha)} \cdot \on \ , \notag
    \end{align}
    proving that Algorithm~\ref{alg:approx} is a $\frac{1}{1+\ln(1/\alpha)}$-approximation and the proposition is established. \hfill
\end{proof}

It remains to prove Lemma~\ref{lem:posterior-prophet}. 

\begin{proof}[ of Lemma~\ref{lem:posterior-prophet}]
For every stopping time $\sigma\le n$, Doob's Optional Sampling  implies
\(
\ex{Z_\sigma}
= \ex{Z_1}
\).
Consider a coupling of the optimal stopping rule $\tau^\star$ and {\mart} with parameter $Q=q$, in which $H$ and the values $X_1,\ldots,X_n$ are shared.
Define $\tau_q = \inf \{t: X_t \ge q Z_t\}$ to be the stopping time of \mart, where $\inf\emptyset=+\infty$, and set $X_{\tau_q}=Z_{\tau_q}=0$ on $\{\tau_q=\infty\}$. Let ${\cal E}=\{\tau_q\le\tau^\star\}$ be the event that {\mart} stops no later than \on. Now, apply Doob's Optional Sampling Theorem to the bounded stopping time $\tau=\tau_q\wedge\tau^\star$:
\begin{align}
\ex{Z_\tau}=\ex{Z_{\tau}\ind{{\cal E}}}+\ex{Z_{\tau}\ind{{\cal E}^c}}
=\ex{Z_{\tau_q}\ind{{\cal E}}}+\ex{Z_{\tau^\star}\ind{{\cal E}^c}} \ .  \label{eq:tau-doob}
\end{align}
Similarly, we can use the Optional Sampling Theorem for the bounded stopping time $\tau^\star$:
\begin{align}
    \ex{Z_{\tau^\star}}=\ex{Z_{\tau^\star}\ind{{\cal E}}}+\ex{Z_{\tau^\star}\ind{{\cal E}^c}} \ . \label{eq:tau-star-doob}
\end{align}
Observing $\expar{Z_\tau} = \expar{Z_{\thr}} = \expar{Z_1}$ and combining \eqref{eq:tau-doob} and \eqref{eq:tau-star-doob} gives 
\begin{align}
    \ex{Z_{\tau^\star}\ind{{\cal E}}}=\ex{Z_{\tau_q}\ind{{\cal E}}} \ . \label{eq:z-star-ta}
\end{align}
On ${\cal E}$, we have $\tau_q\le n$ and therefore
\begin{align}
    X_{\tau_q}\ind{{\cal E}}\ge qZ_{\tau_q}\ind{{\cal E}} \quad \text{almost surely}. \label{ineq:ta-A}
\end{align}
On ${\cal E}^c$, the reward of {\mart} is nonnegative. Therefore, \eqref{eq:z-star-ta} and \eqref{ineq:ta-A} imply
\[
\begin{aligned}
r_q\ge\ex{X_{\tau_q}\ind{{\cal E}}}
\ge q\ex{Z_{\tau_q}\ind{{\cal E}}}=q\ex{Z_{\tau^\star}\ind{{\cal E}}}
\ge q\ex{Z_{\tau^\star}\ind{X_{\tau^\star}\ge qZ_{\tau^\star}}} \ .
\end{aligned}
\]
The last inequality follows from $\{X_{\tau^\star}\ge qZ_{\tau^\star}\}\subseteq{\cal E}$. Indeed, if the threshold condition holds at $\tau^\star$, then {\mart} stops no later than $\tau^\star$. The proof is now complete.
\end{proof}

\subsection{Approximation Scheme via Iterative Primal-Dual Improvement}\label{ssec:primal-dual}

We now turn to the main construction behind Theorem~\ref{thm:main}. The construction builds on a key observation from Algorithm~\ref{alg:approx}: the martingale $Z_t$ need not be the Doob martingale generated by the offline benchmark $\max_{t\in [n]}X_t$. Rather, it suffices to have some dual martingale $(Z_t)_t$ and a corresponding $\alpha$-competitive baseline algorithm $\cal A$.

While \Cref{alg:approx} improves the approximation ratio against \on, one cannot simply run it iteratively. 
The reason is that the next invocation requires an improved baseline competitive algorithm against a dual martingale, but the algorithm itself does not provide such martingale. We therefore need to improve the primal algorithm in tandem with shrinking its dual benchmark, preserving the invariant needed for the next iteration.
The \PrimalDualImprovement{} subroutine does exactly this. Given a dual certificate $M$ and an algorithm $\cal A$ known to be $\alpha$-competitive against it, \PrimalDualImprovement{} yields a tighter dual certificate $M^+$ and a new primal algorithm $\cal A^+$ that is $\alpha^+$-competitive against it, with a strictly better competitiveness factor $\alpha^+ > \alpha$. 

\begin{tcolorbox}[
    title={\phantomsection\label{proc:primal-dual-improvement}Primal-Dual Improvement},
    colback=white,
    colframe=black!60,
    colbacktitle=black!8,
    coltitle=BrickRed,
    fonttitle=\bfseries,
    boxrule=0.6pt,
    arc=2pt,
    left=8pt, right=8pt, top=6pt, bottom=6pt
]
\textbf{Input.} A dual certificate $M$ and its Doob's martingale $Z_t=\expar{M\mid\cal F_t}$; an algorithm $\cal A$ that is $\alpha$-competitive against $(Z_t)_t$

\begin{enumerate}[leftmargin=*, itemsep=4pt, topsep=6pt]
    \item \textbf{Primal update.} Apply \Cref{alg:approx} to $\alpha$, $\cal A$, and $(Z_t)_t$, to obtain $\cal A^+$. Set
    \(
        \alpha^+=\frac{1}{1+\ln(1/\alpha)}.
    \)
    \item \textbf{Dual update.} Define, on each sample path,
    \[
        \rho=\max_{t\in[n]}\frac{X_t}{Z_t},
        \qquad M^+=M\max\{\rho,\alpha\},
        \qquad Z_t^+=\expar{M^+\mid\cal F_t} \ .
    \]
\end{enumerate}

\textbf{Output.} The triple $(M^+,\cal A^+,\alpha^+)$
\end{tcolorbox}
Note that our dual update is valid:
\begin{lemma}\label{lem:dual-martingale}
    The process $(Z_t^+)_t$ is a dual martingale.
\end{lemma}
{\bf Proof.} We have
\begin{align*}
    Z_t^+
    = \ex{M^+ \mid \cal F_t} \ge \ex{\left. M \max_{t'\in[n]}\frac{X_{t'}}{Z_{t'}} \right | \cal F_t} \ge \ex{M\frac{X_t}{Z_t} \mathrel{\Big|} \cal F_t} = \frac{X_t}{Z_t}\expar{M\mid\cal F_t}
     = \frac{X_t}{Z_t}Z_t
     = X_t \ .  \tag*{\rule{2mm}{2mm}}
\end{align*}

While \Cref{prop:one-step-improvement} implies that algorithm $\cal A^+$ is an $\alpha^+$-approximation, we can show a stronger result:
\begin{lemma}\label{lem:exact-pd-improve}
    The martingale $(Z_t^+)_t$ is a dual martingale, and the output algorithm $\cal A^+$ is $\alpha^+$-competitive against $(Z_t^+)_t$. 
\end{lemma}
Since a stronger version of this lemma will be established in the proof of \Cref{thm:main}, we do not repeat it here. A consequence of this lemma is that we can repeatedly invoke the \PrimalDualImprovement{} subroutine, as \Cref{alg:iteration} formalizes.

\begin{algorithm}[H]
    \caption{Iterative primal-dual improvement}
    \label{alg:iteration}
    \begin{algorithmic}[1]
    {
        \STATE {\bf Input:} An $\alpha$-competitive algorithm ${\cal A}$ with known $\alpha \in (0, 1)$; an iteration count $J$
        \STATE Let $M_{0} \gets \max_{t \in [n]} X_t$ be the terminal random variable for the hindsight optimal \off.
        \STATE Initialize ${\cal A}_0 \gets {\cal A}$, $\alpha_{0} \gets \alpha$, and $W_0 \gets 1$.
        \FORALL{$j = 0, \cdots, J-1$}
            \STATE Let $(M^+, \cal A^+, \alpha^+)$ be the output of \PrimalDualImprovement{} with input $(M_j, \cal A_j, \alpha_j)$ \label{line:subroutine-call}
            \STATE Set $M_{j+1} \gets M^+, \cal A_{j+1} \gets \cal A^+, \alpha_{j+1} \gets \alpha^+$
        \ENDFOR
        \STATE Return $\cal A_J$
    }
    \end{algorithmic}
\end{algorithm}

\paragraph{The approximation scheme of \Cref{thm:main} from \Cref{alg:iteration}.} The output $\mathcal A_J$ is our final online algorithm for \Cref{thm:main}.
The recursive construction defines a mixture of the baseline policy
and dual-threshold policies. Concretely, unrolling the recursion provides the following alternative description of $\cal A_J$: We first obtain dual certificates $M_0, M_1, \cdots, M_{J-1}$. Then, we randomly pick one of these dual certificates to use for our stopping decisions. We select $M_j$ (for $0 \le j < J$) with probability $(1-\alpha_{j+1})\prod_{\ell = j+2}^J \alpha_\ell$ and we choose the baseline with the remaining probability $\prod_{\ell = 1}^J \alpha_\ell$. We then implement the subroutine \mart (in \Cref{alg:approx}) with $Q \in [\alpha_{j},1]$ and dual certificate $M_{j}$. Otherwise if the baseline is chosen, we simply use the baseline algorithm $\cal A_0$. It is straightforward to see that this algorithm is equivalent to $\cal A_J$ as the output of \Cref{alg:iteration}. 

For now, we are not concerned with the runtime of \Cref{alg:iteration}. The next subsection explains how this algorithm can be (approximately) implemented efficiently. In the rest of this subsection, we provide the intuition behind \PrimalDualImprovement{} and \Cref{alg:iteration}. 

\paragraph{Intuition behind \Cref{alg:iteration}.} The central invariant behind our design is the dual identity
\begin{equation}\tag{Dual-Identity}\label{eq:dual-identity}
    \ex{M_{j+1}} = \alpha_j \cdot \ex{M_{j}} + \int_{\alpha_j}^1 \ex{M_j \cdot \ind{\rho_j \ge q}} \, dq \ ,
\end{equation}
which will be proved in the proof of \Cref{thm:main} in \Cref{ssec:proof}. 

The reward of our updated primal algorithm is closely tied to \eqref{eq:dual-identity}: On the one hand, when \Cref{alg:approx} is ran on $\cal A_j$ and $M_j$---more precisely, on its associated Doob's martingale $(Z_t^{(j)})_t$---the baseline algorithm $\cal A_j$ has a reward of $\alpha_j \cdot \expar{M_j}$. On the other hand, the dual-threshold algorithm \mart with $Q=q$ obtains a reward $r_q$ related to the second term in \eqref{eq:dual-identity}. Recall that $\tau_q$ is the stopping time of \mart with $Q = q$. 
Suppose that $Z_t^{(j)}$ can be exactly evaluated at every time. We thus have:
\[
\begin{aligned}
r_q
&=\sum_{t=1}^n \ex{X_t \cdot \ind{\tau_q=t}}\\
&\ge q \cdot \sum_{t=1}^n \ex{Z_t^{(j)} \cdot \ind{\tau_q=t}}\\
&= q \cdot \sum_{t=1}^n \ex{\ex{M_j \mid \cal F_t} \cdot \ind{\tau_q=t}}\\
&=q \cdot \sum_{t=1}^n \ex{M_j \cdot \ind{\tau_q=t}}\\
&=q \cdot \ex{M_j \cdot \ind{\tau_q\le n}} = q \cdot \ex{M_j \cdot \ind{\rho_j\ge q}}.
\end{aligned}
\]
Consequently, choosing $Q$ from $[\alpha_j,1]$ with density $f(q) \propto 1/q$ would cancel out the $q$ factors, yielding a reward proportional to the second term in \eqref{eq:dual-identity}. 

Because $\rho_j \le 1$,\footnote{This follows from the fact that $Z_t^{(j)}$ is a dual martingale, and thus, $Z_t^{(j)} \ge X_t$ for every $t$ almost surely.} by iteratively using \PrimalDualImprovement, we obtain a sequence of dual certificate $M_0 \ge M_1 \ge M_2, \cdots$. We claim that this sequence converges to the 
\emph{optimal dual martingale}
\begin{align}
    M_* = \argmin_{\substack{\text{terminal random} \\\text{variable } M}} \left \{\ex{M}: \ex{M \mid \cal F_t} \ge X_t \; \text{almost surely for every } \; t \in [n]\right \} \ . \label{eq:optimal-dual}
\end{align}
As we will formally prove $r(\cal A_{j}) \ge \alpha_j \cdot \expar{M_j} \ge \alpha_j \cdot \on$ for every $j$, and because $\alpha_j \to 1$ as $j \to \infty$, we infer that $\expar{M_*} = \lim_{j \to \infty} \expar{M_j} = \on$. In other words, the optimal dual martingale has the same expected value
as the optimal online algorithm's reward, which is the \emph{strong duality} claimed in \Cref{ssec:martingales}. This form of {strong duality} has been similarly established in the context of ``information relaxation'' methods for stochastic dynamic programming \citep{brown}.

Beyond the close connection of our primal guarantee and \eqref{eq:dual-identity}, a main motivation for our particular form of dual update is to be able to approximate the dual martingales efficiently. Indeed, finding the optimal dual certificate $M_*$ in \eqref{eq:optimal-dual} is as hard as finding the optimum policy.
Thus, we start with a feasible solution of the optimization problem in \eqref{eq:optimal-dual}---which is $M_0$---and progressively refine it to get closer to $M_*$. 

\begin{figure}
    \centering
    \begin{tikzpicture}[
        x=1cm, y=1cm,
        font=\footnotesize,
        scenario node/.style={circle, draw=black!55, fill=white, inner sep=0pt, minimum size=4pt},
        scenario edge/.style={draw=black!35, line width=0.5pt},
        scenario path/.style={draw=BrickRed, line width=1.2pt},
        history label/.style={fill=white, inner sep=1pt, anchor=west}
    ]
        \foreach \panel in {0,1} {
            \begin{scope}[xshift=8.1cm*\panel]
                \begin{scope}[xshift=3cm, x=0.65cm]
                \fill[navyblue!6, rounded corners=3pt]
                    (1.6,2.42) -- (3.3,-0.23) -- (-0.1,-0.23) -- cycle;
                \fill[navyblue!16, rounded corners=3pt]
                    (0.8,1.38) -- (1.7,-0.18) -- (-0.1,-0.18) -- cycle;
                \coordinate (root) at (3.2,3.2);
                \coordinate (left) at (1.6,2.15);
                \coordinate (right) at (4.8,2.15);
                \coordinate (leftleft) at (0.8,1.1);
                \coordinate (leftright) at (2.4,1.1);
                \coordinate (rightleft) at (4,1.1);
                \coordinate (rightright) at (5.6,1.1);
                \draw[scenario edge] (leftleft) -- (left) -- (root) -- (right) -- (rightleft);
                \draw[scenario edge] (leftright) -- (left) (right) -- (rightright);
                \foreach \parent/\firstleaf/\secondleaf in {
                    leftleft/0.4/1.2, leftright/2/2.8,
                    rightleft/3.6/4.4, rightright/5.2/6
                } {
                    \draw[scenario edge] (\firstleaf,0) -- (\parent) -- (\secondleaf,0);
                }
                \ifnum\panel=0
                    \draw[scenario path] (root) -- (left) -- (leftleft) -- (1.2,0);
                \else
                    \draw[scenario path, dashed] (root) -- (left) -- (leftleft) -- (1.2,0);
                \fi
                \foreach \vertex in {root,left,right,leftleft,leftright,rightleft,rightright} {
                    \node[scenario node] at (\vertex) {};
                }
                \foreach \leafposition in {0.4,1.2,2,2.8,3.6,4.4,5.2,6} {
                    \node[scenario node] at (\leafposition,0) {};
                }
                \node[history label] at (1.75,2.15) {$h_1$};
                \node[history label] at (1,1.1) {$h_2$};
                \coordinate (highlightedleaf) at (1.2,0);
                \node[scenario node, draw=BrickRed, fill=BrickRed] at (1.2,0) {};
                \ifnum\panel=0
                    \node[anchor=east, xshift=-3mm, font=\scriptsize, text=navyblue]
                        at (left) {$Z_1^{(j)}(h_1)=\expar{M_j\mid h_1}$};
                    \node[anchor=east, xshift=-3mm, font=\scriptsize, text=navyblue]
                        at (leftleft) {$Z_2^{(j)}(h_2)=\expar{M_j\mid h_2}$};
                \else
                    \node[anchor=east, xshift=-3mm, font=\scriptsize, text=navyblue]
                        at (left) {$Z_1^{(j+1)}(h_1)=\expar{M_{j+1}\mid h_1}$};
                    \node[anchor=east, xshift=-3mm, font=\scriptsize, text=navyblue]
                        at (leftleft) {$Z_2^{(j+1)}(h_2)=\expar{M_{j+1}\mid h_2}$};
                \fi
                \end{scope}
                \ifnum\panel=0
                    \node[font=\small\bfseries] at (3.6,4) {(a) Read one sample path};
                    \draw[BrickRed, line width=0.8pt] (leftleft) circle[radius=0.12cm];
                    \node[anchor=north, yshift=-3mm, text=BrickRed] at (highlightedleaf) {$M_j(\omega)$};
                    \node at (3.6,-1.2) {$\displaystyle \rho_j(\omega)=\max_{t\in[n]}\frac{X_t(\omega)}{Z_t^{(j)}(\omega)}$};
                \else
                    \node[font=\small\bfseries] at (3.6,4) {(b) Update that sample path};
                    \node[anchor=north, yshift=-3mm, text=BrickRed] at (highlightedleaf) {$M_{j+1}(\omega)$};
                    \node at (3.6,-1.2) {$M_{j+1}(\omega)=M_j(\omega)\max\{\alpha_j,\rho_j(\omega)\}$};
                \fi
            \end{scope}
        }
    \end{tikzpicture}
    \caption{The pathwise dual update on the scenario tree}
    \label{fig:scenario-tree}
    \vspace{0.2cm}
    \begin{minipage}{\linewidth}
    \footnotesize
    Each node shows a certain history $h_t$, with the leaves being complete sample paths. The figure shows updating the dual value $M_j(\omega)$ for the highlighted path $\omega$, while leaving the value of $M_j$ the same for other sample paths; note that the algorithm defines the update rule for all sample paths simultaneously, in contrast with the figure. 
    The update scales $M_j(\omega)$ by $\max\{\alpha_j,\rho_j(\omega)\}$, where $\rho_j(\omega)=\max_{t \in [n]} {X_t(\omega)}/{Z_t^{(j)}(\omega)}$ and the circled node in panel (a) attains this maximum ratio.
     The shaded subtrees contain the continuations of histories $h_1$ and $h_2$: their leaf sets are nested, illustrating the laminar structure of histories (any two such sets are nested or disjoint). The labels beside $h_1$ and $h_2$ give the corresponding conditional expectations: each is an average of the leaf values in its shaded subtree, weighted by their probabilities conditional on that history. In panel (b), only the highlighted leaf is updated; consequently, only internal values along its path to the root can change. These values are recomputed as conditional expectations of the updated certificate. The pathwise ratio bound ensures dual feasibility at these ancestors, while all other node values remain unchanged.
    \end{minipage}
\end{figure}
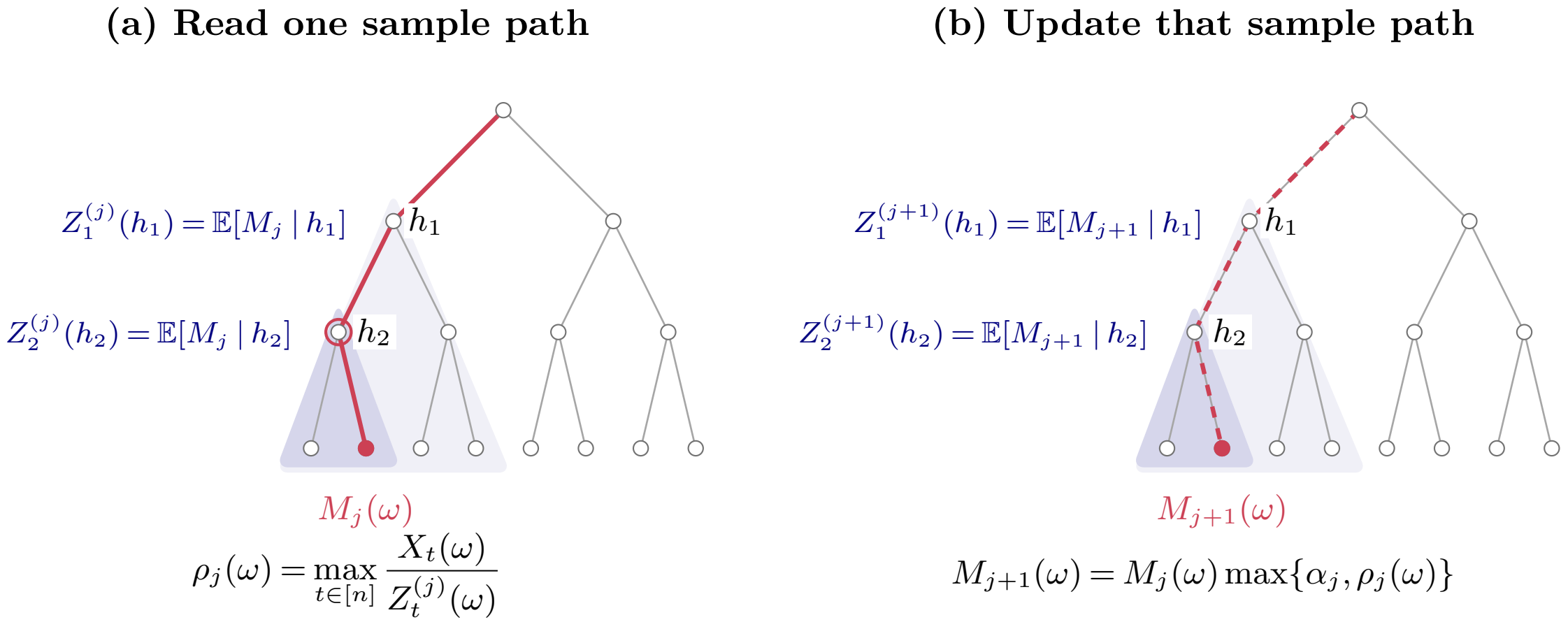

The update dual has a simple pathwise interpretation. Fix a time $t$. Since $\rho_j \ge X_t/Z_t^{(j)}$ on every sample path, and $X_t/Z_t^{(j)}$ is $\cal F_t$-measurable, we have
\[
    Z_t^{(j+1)} = \expar{M_j \max\{\rho_j,\alpha_j\} \mid \cal F_t}
    \;\ge\; \frac{X_t}{Z_t^{(j)}} \cdot \expar{M_j \mid \cal F_t} = X_t \ .
\]
In other words, the dual feasibility constraints $\expar{M \mid \cal F_t} \ge X_t$---which couple all sample paths sharing the same history---are guaranteed if, for each sample path $\omega$ of the $n$-step trajectory, $M(\omega)$ is scaled by a factor of at least $\max_{t} X_t(\omega)/Z_t^{(j)}(\omega)$, a condition that can be checked on each path in isolation. The factor $\rho_j(\omega)$ is the smallest one satisfying this condition, and since $\rho_j \le 1$, the scaling can only shrink $M_j$. Because the condition is checked separately on each path, we may shrink $M_j(\omega)$ on any subset of paths, or on all of them at once, without violating dual feasibility (see also \Cref{fig:scenario-tree} for this intuition on the scenario tree).

This pathwise structure is also what makes the update computable. Unrolling the recursion gives $M_j = M_0 \prod_{\ell < j} \max\{\rho_\ell, \alpha_\ell\}$, so $M_j(\omega)$ depends only on the path $\omega$ itself and on the earlier martingales $Z^{(\ell)}$, $\ell < j$, evaluated at the $n$ prefixes of $\omega$. Hence, $Z_t^{(j)}(h_t) = \expar{M_j \mid h_t}$ can be estimated by Monte Carlo: simulate continuations of $h_t$, and evaluate $M_j$ on each simulated path, which in turn calls for estimates of $Z^{(j-1)}$ along that path. For a fixed $j$, this recursion requires only $n^{O(j)}$ simulated paths, rather than the exponentially many trajectories of the full scenario tree. By contrast, the optimal dual $M_*$ in \eqref{eq:optimal-dual}, like the Snell envelope, is determined by backward induction over the entire tree and admits no such local description. Finally, we note that the floor at \(\alpha_j\) for defining $\rho_j$ prevents the dual certificate from shrinking too aggressively, which is needed for efficient estimation via Monte carlo simulation. {Indeed, the normalized multiplier
$W_j=\prod_{\ell<j}\max\{\rho_\ell,\alpha_\ell\}$
lies between $\prod_{\ell<j}\alpha_\ell$ and $1$ on every sample path.
Under the importance-sampling measure introduced in the next subsection, we estimate
the expectation of this bounded variable; the lower bound on its
mean controls the number of samples needed for its relative accuracy.}

We note that related dual shrinkage updates have been proposed in the applied probability literature on optimal stopping \citep{chen-goldberg2026} and information relaxations \citep{chen2007additive,brown} that have different forms (such as additive updates) or not been implemented in polynomial time.

\begin{remark}
    In light of the alternative description of $\cal A_J$ provided after \Cref{alg:iteration}, one may attempt to only use the final dual martingale $M_J$ because it is the tightest martingale, i.e., $M_J \le \cdots, M_0$ almost surely. Interestingly, it seems non-trivial to obtain a similar performance only using $M_J$. One challenge is how to use $M_J$ to make decisions; \Cref{alg:approx} used the dual martingales via the threshold policy \mart. We now show that using $M_J$ with \mart, our analysis leads to a worse approximation ratio than $\alpha_J$, which is proved for $\cal A_J$. Suppose that \mart draws $Q \in [a,1]$ with density $f_Q(s) \propto 1/s$ and accepts $X_t$ if and only if $X_t / \expar{M_J \mid \cal F_t} \ge Q$, where $a$ is a parameter of choice. Let this algorithm be called $\cal A'$. 

    Ignore the approximation error in computing the martingales and suppose that we can implement \PrimalDualImprovement{} exactly. \Cref{alg:iteration} gives $\cal A_J$ with an approximation ratio of $\alpha_J$. Conversely, similar to the proof of \Cref{prop:one-step-improvement}, we can show
    \[
        \frac{r(\cal A')}{\on} \ge \frac{\ex{(X_{\thr} - a Z_{\thr})_+}}{\ln (1/a) \cdot \on} \ge \frac{\on - a \ex{M_J}}{\ln (1/a) \cdot \on} \ .
    \]
    Let $\beta_J = \on / \expar{M_J} $ be the gap between the dual certificate $\expar{M_J}$ and \on, which makes the right-hand side above equal to $\frac{1-a/\beta_J}{\ln(1/a)}$. For the sake of this remark, we even allow $\cal A'$ to choose \(a\) using the unknown ratio \(\beta_J\). Optimizing it for $a$ gives $a^* = -\frac{\beta_J}{W_{-1}(-\beta_J/e)}$, where $W$ is the Lambert function. This choice of $a = a^*$ yields $$r(\cal A') \ge \gamma(\beta_J) \cdot \on = -\frac{1}{W_{-1}(-\beta_J/e)} \cdot \on \ .$$
    However, we unfortunately only know that $\beta_J \ge \alpha_J$ and not its exact value.\footnote{There exist instances in which $\frac{1-\beta_J}{1-\alpha_J} \to 1/2$ as $J \to \infty$, which means that one cannot prove a lower bound for $\frac{r(\cal A')}{\on}$ better than $1-\sqrt{\eps} + O(\eps)$. In other words, the duality-gap-based performance analysis cannot yield a faster convergence rate merely by establishing tighter universal bounds on \(\beta_J\). A concrete example is a \noniid instance with $K=1$, $n=2$, $X_1=1$, and $X_2 = 2 \text{Bern}(1/2)$. For this instance, it can be shown that $1-\beta_J \sim 1/J$ while our analysis in \Cref{lem:alpha-convergence} implies $1-\alpha_J \sim 2/J$.} Consequently, replacing $\beta_J$ with $\alpha_J$ and letting $\alpha_J =1-\eps$ for a small $\eps \ll 1$, we obtain 
    \begin{align}
        \frac{r(\cal A')}{\on} \ge (1-\sqrt{2\eps}) + O(\eps) \ , \label{ineq:aprime-lb}
    \end{align}
    which is smaller than the $1-\eps$ approximation ratio of $\cal A_J$. Thus, with the same number of iterations $J$, our analysis would prove a better approximation ratio for \Cref{alg:iteration} than only using the final dual certificate. Moreover, whereas \Cref{alg:iteration} needs to set $J=O(\log^* n + 1/\eps)$ to obtain a $(1-\eps)$ approximation ratio, this analysis requires $J = O(\log^* n + 1/\eps^2)$ for the $\cal A'$ to achieve the same performance. Thus, under this analysis, certifying a \((1-\eps)\)-approximation using only the final dual martingale requires an additional factor of \(1/\eps\) in the number of iterations.

    This comparison illustrates a potential advantage of our iterative primal–dual construction. Rather than relying exclusively on the tightest available dual martingale, the algorithm maintains primal policies associated with earlier, looser dual certificates. Our analysis exploits the complementary contributions of these policies to obtain a faster convergence guarantee. Whether a policy using only the final dual martingale can achieve the same guarantee remains an interesting open question.
\end{remark}

\subsection{Efficient Implementation of \Cref{alg:iteration}}\label{ssec:implementation}
We now discuss how to efficiently implement $\cal A_J$, the output of \Cref{alg:iteration}. Recall that to define $\cal A_J$, \Cref{alg:iteration} repeatedly uses \PrimalDualImprovement{} and then invokes \Cref{alg:approx} using dual martingales $Z_t^{(j)} = \expar{M_j \mid \cal F_t}$. Given that we cannot exactly compute martingales $Z_t^{(j)}$ (for superconstant values of $j$), we need to approximate it. Thus, in order to run $\cal A_J$ in polynomial time, the invocation of \PrimalDualImprovement{} in Line~\ref{line:subroutine-call} must use \Cref{alg:approx} with a $(1-\eps)$-approximation $L_t^{(j)}$ of $Z_t^{(j)}$. 
Except for this approximation, the rest of Algorithms~\ref{alg:approx} and \ref{alg:iteration} can be readily implemented in polynomial time. Therefore, we prove the following lemma (proof in Appendix~\ref{prf:accuracy-analysis}) that establishes our approximation can be done in a computationally efficient manner.

\begin{lemma}\label{lem:accuracy-analysis}
    Suppose $0 < \alpha_0 < 1$ and consider accuracy parameters $0<\eps,\delta<1$. For each $j \ge 0$, time $t \in [n]$, and history ${h}_t$, we can compute $L_t^{(j)}(h_t)$ such that
    \[
        \frac{L_t^{(j)}(h_t)}{Z_t^{(j)}(h_t)} \in [1-\eps, 1] \quad \text{with probability at least $1-\delta$ in time }  \poly\!\left(2^{j^2}\left(\frac{Kmn}{\alpha_0\eps\delta}\right)^{j+1}\right). 
    \]
    More precisely, the runtime is
    \[
        O\!\left(
    \zeta^{j^2}K m\, n^{j+1}(\alpha_0\eps)^{-(2j+2)}
    \left(j+\ln\frac{n}{\alpha_0\delta\eps}\right)^{j+1}
    \right) \quad \text{where $\zeta=2^{19}$.}
    \]
\end{lemma}

To compute $Z_t^{(j)}(h_t)$ for a fixed history $h_t$, one needs to consider all continuations from $h_t$ until time $n$, compute $M_j$ for each of them, and then average them by their probabilities. Evaluating $M_j(\omega)$ for a single sample path $\omega$ is itself challenging because it requires the computation of $Z_t^{(j)}(h_t)$ for $n$ prefixes $h_t$ of $\omega$; each of them require this process recursively. 

The estimation process underlying \Cref{lem:accuracy-analysis} leverages the pathwise structure of our dual updates to approximate $Z_t^{(j)}(h_t)$ by sampling continuations of $h_t$, and evaluating $M_j(\omega)$ for each $\omega$ of such continuations. As said in the previous paragraph, this requires further (recursive) computations of dual martingales which are again done via Monte carlo simulations. Then, $(M_\ell)_{\ell \le j}$ and $(\rho_\ell)_{\ell \le j}$
values are estimated using these samples. 
 
One challenge for Monte carlo simulations is that an unlikely sample path $\omega$ can have a large value of $M_j(\omega)$, which would be missed if $\omega$ is not sampled. To resolve this issue, 
use an importance-sampling change of measure to draw our Monte carlo samples. 

Define the biased probability measure 
\begin{align}
    \prtwo{*}{\omega \mid {\cal F}_t} = \pr{\omega \mid \cal F_t} \cdot \frac{M_0(\omega)}{Z_t^{(0)}} \ , \label{expr:biased-prob}
\end{align} 
where $\omega$ is a sample path compatible with the history until time $t$ denoted by $h_t$.\footnote{The biased distribution is a valid probability measure because $\sum_{\omega} \prtwo{*}{\omega \mid \cal F_t} = \sum_{\omega} \pr{\omega \mid \cal F_t} \cdot \frac{M_0(\omega)}{Z_t^{(0)}} = \frac{\ex{M_0 \mid \cal F_t}}{Z_t^{(0)}} = 1$ using the martingale property.} See Appendix~\ref{app:star-sampling} for a quick description of how to exactly sample from $\prtwo{*}{\cdot}$ in polynomial time. Let $\extwo{*}{\cdot}$ be the expectation operator with respect to $\prtwo{*}{\cdot}$. The biased distribution has the following useful property (proof in Appendix~\ref{prf:z-alter-def}):
\begin{claim}\label{clm:z-alter-def} Let $W_j = \prod_{\ell = 0}^{j-1} \max \{ \alpha_\ell, \rho_\ell \}$.
For every $j \ge 0$, $t \in [n]$, and a history $h_t$, we have
\( 
    Z_t^{(j)}(h_t) = \ex{M_0 W_j \mid h_t} = Z_t^{(0)}(h_t) \cdot \extwo{*}{W_j \mid h_t}
\). 
\end{claim}

\Cref{clm:z-alter-def} is useful because $Z_t^{(0)}(h_t) = \expar{M_0 \mid h_t}$ can be analytically computed exactly in polynomial time. Consequently, to estimate $Z_t^{(j)}$, we draw sufficiently many samples $\omega^{(1)}, \omega^{(2)}, \cdots$ for continuations $X_{t+1}, \cdots, X_n$ from $\prtwo{*}{\cdot \mid \cal F_t}$, each sharing the same history $h_t$. For each sample path $\omega^{(i)}$, we approximate $W_j(\omega^{(i)})$. By Chernoff bound, with sufficiently many samples, the average value from these samples is close to the true expected value $\extwo{*}{W_j \mid h_t}$ with high probability; \Cref{clm:z-alter-def} then provides the estimation of $Z_t^{(j)}(h_t)$. 
The formal description of approximating $W_j(\omega)$ is in \Cref{alg:w-oracle}, with its analysis and the proof of \Cref{lem:accuracy-analysis} deferred to Appendix~\ref{prf:accuracy-analysis}.

\begin{algorithm}[H]
    \caption{Recursive estimator $\cal W_j(\omega, \eps, \delta)$}
    \label{alg:w-oracle}
    \begin{algorithmic}[1]
    {
        \STATE {\bf Input:} Recursion level $j$; sample path $\omega$; accuracy parameters $0<\eps,\delta<1$ 
        \STATE {\bf Output:} A randomized output $\widehat W_j(\omega)$ such that $\widehat W_j(\omega)/W_j(\omega) \in [(1-\eps), (1-\eps)^{-1}]$ w.p. $1-\delta$ 
            \IF{$j = 0$}
                \STATE Return 1
            \ENDIF
            \STATE $\eps' \gets 1-(1-\eps)^{1/3}$ and $G_{j-1} \gets \prod_{\ell = 0}^{j-1} \alpha_\ell$ \qquad \qquad \CommentSty{\color{red} // with $(\alpha_\ell)_\ell$ as defined in \Cref{alg:iteration}}  
            \STATE $I_j \gets \lceil \frac{3}{\eps'^2G_{j-1}}\ln\frac{4n}{\delta}\rceil$ and $\delta' \gets \frac{\delta}{2(nI_j + 1)}$
            \FORALL{time step $t=1,\ldots,n$}
                \STATE Let $h_t$ be the history of first $t$ steps in $\omega$. 
                    \FORALL{$i=1,\ldots,I_j$}
                        \STATE Independently sample a fresh continuation $\tilde h^{(t,i)}=(\tilde X_{t+1}^{(t,i)},\ldots,\tilde X_n^{(t,i)})$ from $\prtwo{*}{\cdot\mid h_t}$
                        \STATE Let $\tilde \omega^{(t, i)}$ be the sample path formed by concatenating $h_t$ and $\tilde h^{(t,i)}$
                        \STATE Recursively set $\widehat W^{(t,i)}_{j-1} \gets \cal W_{j-1}(\tilde \omega^{(t,i)}, \eps', \delta')$  \label{line:sample-w}
                    \ENDFOR
                    \STATE $\widehat Z_t^{(j-1)} \gets Z_t^{(0)}(h_t) \cdot \frac{1}{I_j} \sum_{i=1}^{I_j}  \widehat W_{j-1}^{(t,i)}$ \label{line:prev-z}
            \ENDFOR
            \STATE $\widehat \rho_{j-1} \gets  \max_{t \in [n]} {X_t(\omega)}/{\widehat Z_t^{(j-1)}}$
        \STATE $\widehat W_{j-1}(\omega) \gets \cal W_{j-1}(\omega, \eps', \delta')$ \label{line:self-w}
        \STATE Return $\widehat W_j(\omega) \gets \widehat W_{j-1}(\omega) \cdot \max \{\alpha_{j-1},\widehat \rho_{j-1} \}$
    }
    \end{algorithmic}
\end{algorithm}

\subsection{Completing the Proof of \Cref{thm:main}: Analysis of \Cref{alg:iteration}}\label{ssec:proof}

\color{black}

In this subsection, we analyze \Cref{alg:iteration} and prove that it satisfies the properties needed for \Cref{thm:main}. 

The main technical analysis is done via the following lemma---which is a stronger version of \Cref{lem:exact-pd-improve} that accounts for martingale approximations.
Recall that given an $\alpha$-competitive algorithm ${\cal A}$, the sequence of approximation ratios $(\alpha_j)_{j\ge0}$ is defined recursively by $\alpha_j = (1 + \ln(1/\alpha_{j-1}))^{-1}$ with $\alpha_0 = \alpha$. We then have the following lemma (proved in \Cref{sssec:shrink-performance}):
\begin{lemma}\label{lem:shrink-performance}
Suppose $0<\beta<1$ and we are given an $\alpha$-competitive algorithm ${\cal A}$ with $0<\alpha\le1$. If Algorithm~\ref{alg:iteration} starts with the baseline algorithm $\cal A$ and $\alpha_0 = \alpha$, and uses estimation accuracy $\beta$ when calling \Cref{alg:approx}, then every constructed level $j\ge0$ satisfies $$r({\cal A}_j)\ge \alpha_j(1-\beta)^2\ex{M_j} \ge \alpha_j(1-\beta)^2 \textup{\on} \ . $$
\end{lemma}

Before proving \Cref{thm:main}, we determine how many iterations are required for $\alpha_j$ to approach $1$. The following lemma gives the needed bound (proof in Appendix~\ref{app:alpha-convergence}). 

\begin{lemma}\label{lem:alpha-convergence}
Suppose $0<\eps\le1$, $0<\alpha_0\le1$, and $\alpha_{j+1}=1/(1+\ln(1/\alpha_j))$ for every $j\ge0$. There exists an integer
\(
    J=O(\log^*\!\frac1{\alpha_0}+\frac1\eps)
\)
such that $\alpha_J\ge1-\eps/2$.
\end{lemma}

\subsubsection{Proof of \Cref{thm:main}.} We may assume $n\ge2$ (the case $n=1$ is trivial). Set the internal estimation accuracy in Algorithm~\ref{alg:iteration} (when calling \Cref{alg:approx}) to $\beta = \eps/4$. By Lemma~\ref{lem:alpha-convergence}, there exists an integer $J=O(\log^*\!\frac1{\alpha_0}+\frac1\eps)$ with $J\ge2$ such that $\alpha_{J}\ge1-\eps/2$. 
Consequently, by Lemma~\ref{lem:shrink-performance}, the algorithm ${\cal A}_J$ satisfies \[
\begin{aligned}
    r({\cal A}_J)
    &\ge\alpha_J(1-\beta)^2\on\\
    &\ge\left(1-\frac\eps2\right)
       \left(1-\frac\eps4\right)^2\on \\
    & \ge(1-\eps)\on \ ,
\end{aligned}
\] as desired.

It remains to account for the running time. At each time, Algorithm~\ref{alg:approx} invokes Lemma~\ref{lem:accuracy-analysis} with accuracy $\eps/4$ and failure probability $\delta = \eps/(4n)$. During a run of ${\cal A}_J$, the algorithm either runs ${\cal A}_0$ or enters {\mart} at some level $j\in\{0,\ldots,J-1\}$. Thus, the algorithm makes at most $n$ calls to compute $L_t(h_t)$ for $t \in [n]$, all at a single level. This dominates all the other operations in that level in terms of runtime. The runtime bound in Lemma~\ref{lem:accuracy-analysis} for level $j$ is increasing in $j$, so the largest cost occurs at $j=J-1$. Thus, the total running time is
\[
O\!\left(
\zeta^{J^2}Km\,n^{J+1}(\alpha_0\eps)^{-2J}
\left(J+\ln\frac{n}{\alpha_0\eps}\right)^J
\right).
\]
Because $\alpha_0 \ge 1/n$ (by accepting at a time chosen uniformly at random) and $J=O(\log^*\!\frac1{\alpha_0}+\frac1\eps)$, the overall runtime of our algorithm is $\poly(K m n^{1/\eps + \log^* n} \cdot ({1/\eps})^{1/\eps^2})$, which establishes the runtime of our approximation scheme. For the \noniid model, we have $\alpha_0 \ge \max\{\frac1{2K}, n\}$ because we can also choose the algorithm that commits to the world that has the largest contribution to \off{} and adopt a 1/2-competitive algorithm for that world, which would be $\frac1{2K}$-competitive. 
Finally, in the special case of \iid, we have $\alpha_0 \ge 1/e$ (by using the secretary algorithm), which makes our algorithm a geniune PTAS. The proof is now complete. 
\hfill \qedsymbol

\subsubsection{Proof of \Cref{lem:shrink-performance}.}\label{sssec:shrink-performance}

    We prove the result by induction on $j$. The base $j=0$ is given by the lemma's hypothesis. For the inductive step, fix $j\ge1$, and note that $\alpha_j < 1$. Let $L_t$ be the approximation of $Z_t^{(j-1)}$, as in Line~\ref{line:approximate-z} of Algorithm~\ref{alg:approx}. For every $q \in [0,1]$, define
    \[
        \tau_q = \inf \{ t: X_t \ge q \cdot Z_t^{(j-1)} \} \quad \text{and} \quad \widehat \tau_q = \inf \{ t: X_t \ge q \cdot L_t \} 
    \]
    with the convention that $\inf \emptyset = \infty$, and with all quantities indexed by an infinite stopping time set to zero. Furthermore, define $ \widehat A_q = \expar{X_{\widehat \tau_q} \cdot \ind{\widehat \tau_q \le n}}$. The proof proceeds in six steps as follows. %
    
     \textit{Step 1.} 
    We first show \eqref{eq:dual-identity}, which is equivalent to
    \[
        \ex{M_{j}} = \alpha_{j-1} \cdot \ex{M_{j-1}} + \int_{\alpha_{j-1}}^1 \ex{M_{j-1} \ind{\tau_q \le n}} \ dq \ . 
    \]
    Recall the definition $M_j = M_{j-1} \cdot \max \{\rho_{j-1}, \alpha_{j-1} \}$.
    Because $(Z_t^{(j-1)})_t$ is a dual martingale (from \Cref{lem:dual-martingale}), we have $0\le\rho_{j-1}\le1$. This gives
    \[
      \max \{\rho_{j-1}, \alpha_{j-1} \} = \alpha_{j-1} + \int_{\alpha_{j-1}}^1 \ind{\rho_{j-1} \ge q} \ dq \ \; \; \text{almost surely.} 
    \]
    By the definition of $\rho_{j-1}  = \max_{t \in [n]} \frac{X_t}{Z_t^{(j-1)}}$ and $\tau_q  = \inf \{ t: X_t \ge q \cdot Z_t^{(j-1)} \}$, we have $$M_{j-1}\ind{\tau_q\le n}=M_{j-1}\ind{\rho_{j-1}\ge q} \quad \text{almost surely for all }q.$$ This implies
    \[
      M_j = \alpha_{j-1}  M_{j-1} + \int_{\alpha_{j-1}}^1 M_{j-1}  \ind{\tau_q \le n} \ dq \ \; \; \text{almost surely.} 
    \]
    Taking expectations establishes Step~1.
     
    \textit{Step 2.} We next obtain the following lower bound for the reward of ${\cal A}_j$:
    \[
        r(\cal A_j) \ge \gamma  \left ( \alpha_{j-1}(1-\beta)^2 \cdot \ex{M_{j-1}} + \int_{\alpha_{j-1}}^1 \frac{\widehat A_q}{q} \ dq \right ) \ .
    \]
    When \Cref{alg:approx} runs \mart with $Q=q$, its reward is exactly $\widehat A_q$. Thus, by construction, with $\gamma=1/(1+\ln(1/\alpha_{j-1}))$, we have
    \begin{align*}
        r({\cal A}_{j}) &= \gamma  r({\cal A}_{j-1}) +  \frac{1-\gamma}{\ln(1/\alpha_{j-1})}  \int_{\alpha_{j-1}}^1 \frac{\widehat A_q}{q} \ dq 
        \\ & = \gamma r({\cal A}_{j-1}) + \gamma \int_{\alpha_{j-1}}^1 \frac{\widehat A_q}{q} \ dq 
        \\ & \ge \gamma  \left ( \alpha_{j-1}(1-\beta)^{2} \cdot \ex{M_{j-1}} + \int_{\alpha_{j-1}}^1 \frac{\widehat A_q}{q} \ dq \right ) \ .
    \end{align*}
    where the final inequality follows from the induction hypothesis.
    
    \textit{Step 3.} To analyze the estimation failures, we make the estimator's randomness explicit. Let $U= (U_1, \cdots, U_n)$ be auxiliary random seeds to compute $L_1, \cdots, L_n$; thus, $U$ is independent from the underlying sample path and natural filtration. Also, let ${\cal U}_t = \sigma(U_1, \cdots, U_t)$ be the corresponding filtration. Notice that $\cal F_t$ and $\cal U_t$ are independent, i.e., $\cal F_t \perp\!\!\!\! \perp \cal U_t$. 
    
    Define events $E_t=\{L_t\in[(1-\beta)Z_t^{(j-1)},Z_t^{(j-1)}]\}$ and ${\cal E}_t=\bigcap_{s=1}^tE_s$.
    At each realized history, we invoke Lemma~\ref{lem:accuracy-analysis} with accuracy $\eps = \beta$ and failure probability $\delta = \beta/n$, as stated in \Cref{alg:approx}. Because $U$ is independent of the full realized path and the estimator at time $t$ depends on that path only through its history $h_t$, the fixed-history guarantee gives $\prpar{E_t^c\mid\cal F_n}\le\beta/n$. A union bound therefore gives $\prpar{{\cal E}_t^c\mid\cal F_n}\le\beta$ for every $t\in[n]$.
    On $\{\widehat\tau_q=t\}\cap{\cal E}_t$, we have $$X_t\ge qL_t\ge q(1-\beta)Z_t^{(j-1)}.$$ This gives
    \begin{align}
        \widehat{A}_q \ge q(1-\beta) \cdot \ex{Z_{\widehat\tau_q}^{(j-1)} \cdot \ind{{\widehat\tau_q}\le n} \cdot \ind{{\cal E}_{\widehat\tau_q}}} \ . \label{ineq:step3}
    \end{align}
    
    \textit{Step 4.} Next, we prove
    \begin{align}
         \ex{Z_{\widehat\tau_q}^{(j-1)}\cdot \ind{{\widehat\tau_q}\le n}\cdot \ind{{\cal E}_{\widehat\tau_q}}} = \ex{M_{j-1}\cdot \ind{{\widehat\tau_q}\le n} \cdot \ind{{\cal E}_{\widehat\tau_q}}} \ . \label{eq:step4}
    \end{align}
    Using the tower property on the enlarged filtration $\cal G_t = \cal F_t \lor \cal U_t$, for every $t\in[n]$ we have
      \begin{align*}
        \ex{M_{j-1} \cdot \ind{\widehat \tau_q = t} \cdot \ind{\cal E_t}} &= \ex{\ex{M_{j-1} \cdot \ind{\widehat \tau_q = t}  \cdot \ind{\cal E_t} \mid \cal G_t}}
        \\ & \overset{(i)}{=} \ex{\ex{M_{j-1}  \mid \cal G_t} \cdot \ind{\widehat \tau_q = t} \cdot \ind{\cal E_t}}
        \\ & \overset{(ii)}{=} \ex{\ex{M_{j-1}  \mid {\cal F}_t} \cdot \ind{\widehat \tau_q = t} \cdot \ind{\cal E_t}}
        \\ & = \ex{Z_t^{(j-1)} \cdot \ind{\widehat \tau_q = t} \cdot \ind{\cal E_t}} \,
    \end{align*}
    where $(i)$ uses the fact that $\{\widehat \tau_q = t\}$ and $\ind{\cal E_t}$ are $\cal G_t$-measurable, and $(ii)$ follows from $\cal F_n$-measurability of $M_{j-1}$ and that $\cal F_n \perp\!\!\!\! \perp \cal U_n$. The final equality simply follows from the definition. Now, summing over $t \in [n]$ establishes \eqref{eq:step4}. \\

    \textit{Step 5.} We now show
    \begin{align}
        \ex{M_{j-1}\cdot \ind{{\widehat\tau_q}\le n}\cdot \ind{{\cal E}_{\widehat\tau_q}}} \ge \ex{M_{j-1}\cdot\ind{{\tau_q}\le n} \cdot \ind{{\cal E}_{\tau_q}}}  \ . \label{ineq:step5}
    \end{align}
    Suppose $\{\tau_q \le n\} \cap \cal E_{\tau_q}$ occurs. Then $L_s \le Z_s^{(j-1)}$ for every $s \le \tau_q$, and at time $\tau_q$ we have $X_{\tau_q}\ge qZ_{\tau_q}^{(j-1)}\ge qL_{\tau_q}$. Hence, the actual algorithm stops at some $\widehat \tau_q \le \tau_q \le n$. Since the events $\cal E_t$ are nested, $\cal E_{\widehat \tau_q}$ holds too. These observations establish the following event inclusion:
    \[
        \{ \tau_q \le n \} \cap \cal E_{ \tau_q} \subseteq \{\widehat \tau_q \le n \} \cap \cal E_{\widehat \tau_q} \ .
    \]
    Because $M_{j-1} \ge 0$, we can multiply both sides of this expression by $M_{j-1}$ pathwise to get
    \[
        {M_{j-1} \cdot \ind{{\widehat\tau_q} \le n} \cdot \ind{\cal E_{\widehat \tau_q}}} \ge {M_{j-1} \cdot \ind{{\tau_q} \le n} \cdot \ind{\cal E_{\tau_q}}} \quad \text{almost surely.}
    \]
    Taking an expectation proves \eqref{ineq:step5}. 
    
    \textit{Step 6.} At the final step, we demonstrate that
    \begin{align}
        \ex{M_{j-1} \cdot \ind{{\tau_q} \le n} \cdot \ind{\cal E_{\tau_q}}}  \ge (1-\beta) \cdot \ex{M_{j-1} \cdot \ind{{\tau_q} \le n}}  \ .  \label{ineq:step6}
    \end{align}
    To see this, note that
     \begin{align*}
         \ex{M_{j-1} \cdot \ind{\tau_q \le n} \cdot \ind{\cal E_{\tau_q}^c}} &= \ex{\ex{\left .M_{j-1} \cdot \ind{\tau_q \le n} \cdot \ind{\cal E_{\tau_q}^c} \right |  \cal F_n}}
         \\ & = \ex{ M_{j-1} \cdot \ind{\tau_q \le n} \cdot \ex{\left.  \ind{\cal E_{\tau_q}^c} \right | \cal F_n} }
         \\ & \le \ex{M_{j-1} \cdot \ind{\tau_q \le n} \cdot \beta }
         \\ & = \beta \cdot \ex{M_{j-1} \cdot \ind{\tau_q \le n} } \ ,
     \end{align*}
     where the inequality follows from the bound $\prpar{{\cal E}_t^c\mid\cal F_n}\le\beta$ in Step~3, since $\tau_q$ is $\cal F_n$-measurable.
     Noting 
     \[
        \ex{M_{j-1} \cdot \ind{\tau_q \le n}} = \ex{M_{j-1} \cdot \ind{\tau_q \le n} \cdot \ind{\cal E_{\tau_q}}} + \ex{M_{j-1} \cdot \ind{\tau_q \le n} \cdot \ind{\cal E_{\tau_q}^c}} 
     \]
     completes the proof of \eqref{ineq:step6}. 

    \textit{Completing the proof.} Now, we use Steps~1-6 to establish the lemma. Combining Steps~3-6 immediately gives
    \[
        \widehat A_q \ge q(1-\beta)^2  \cdot \ex{M_{j-1} \cdot \ind{{\tau_q} \le n}}  \ .
    \]
    Using Step 2, we obtain 
    \[
        r(\cal A_j) \ge \gamma  \left ( \alpha_{j-1}(1-\beta)^{2} \cdot \ex{M_{j-1}} + (1-\beta)^2 \cdot \int_{\alpha_{j-1}}^1 \ex{M_{j-1} \cdot \ind{{\tau_q} \le n}} \ dq \right )  \ .
    \]
    Comparing this lower bound with the equality for $\expar{M_j}$ in Step 1 immediately gives $r(\cal A_j) \ge \gamma(1-\beta)^2 \expar{M_j}=\alpha_j(1-\beta)^2\expar{M_j}$, since $\gamma=\alpha_j$. Thus, the induction step is established and the proof is complete. 
\hfill \qedsymbol

\section{Competitive Algorithms} \label{sec:competitive-algs}

In this section, we consider the problem of competing with the prophet benchmark. We uncover that this benchmark is significantly harder. In particular, in the \iid model, we find that there exists a polynomial-time algorithm with a $(1/e + 1/\poly(k))$ competitive ratio, but no online algorithm can guarantee a better competitive ratio (even without a runtime constraint). 

The main result of this section is as follows:

\begin{theorem}\label{thm:iid-comp}
    There exists a polynomial-time algorithm ${\cal A}$ for the \iid Bayesian optimal stopping problem that is $(1/e + c/K^2)$-competitive for some universal constant $c > 0$.
    Conversely, no algorithm ${\cal A}$ can guarantee a competitive ratio better than $1/e + c'/\sqrt{K}$ for some other universal constant $c' > 0$.
\end{theorem}

The algorithm in \Cref{thm:iid-comp} is an adaptation of \cite{bai2025optimal} where they consider an IID prophet inequality problem where a (potentially wrong) candidate $\widehat{D}$ for the distribution is given. For a consistency parameter $\alpha > 0$ and a robustness parameter $\beta > 0$, the authors devise an algorithm that is $\alpha$-competitive if $\widehat{D}$ is the true distribution, and is $\beta$-competitive otherwise. Indeed, this is done only for (a subset of) feasible pairs $(\alpha, \beta)$. By carefully adapting their results to our setup, we obtain our algorithm. 

Conversely, the hard instance is more challenging to construct, as it should imply hardness for any online algorithm, including the exponential-time Bayesian DP. While a uniform $1/e$ upper bound for the limit $K \to \infty$ has been previously established by \cite[Theorem~5.1]{correa2020unknowniidprophetsbetter} through a minimax argument, we instead provide a concrete construction which provides upper bounds on the competitive ratio in terms of $K$.

Extending \Cref{thm:iid-comp} to the \noniid case, we observe that it is not possible to obtain a competitive ratio substantially better than the immediate guarantee of $1/(2K)$. The proof of the following proposition is deferred to Appendix~\ref{prf:noniid-cr}.  
\begin{proposition}\label{prop:noniid-cr}
    There exists a polynomial-time algorithm ${\cal A}$ for the \noniid Bayesian optimal stopping problem that is $1/{(2K)}$-competitive. 
    Conversely, no algorithm ${\cal A}$ can guarantee a competitive ratio better than $1/K$. 
\end{proposition}

Finally, it is easy to see that $1/n$ is the optimal competitive ratio in the \gen case. We state it here for completeness (proof in Appendix~\ref{prf:gen-cr}):
\begin{proposition}\label{prop:gen-cr}
    There exists a polynomial-time algorithm ${\cal A}$ for the \gen Bayesian optimal stopping problem that is $1/n$-competitive and no algorithm can guarantee a better competitive ratio. 
\end{proposition}

The rest of this section is organized as follows. The hard instance is presented in \Cref{ssec:hard-instance} and \Cref{ssec:competitive-algo} discusses the algorithm. Putting together their analyses in Lemmas~\ref{lem:main-hardness} and~\ref{lem:iid-feasibility} proves \Cref{thm:iid-comp}. 

\subsection{Hard Instance of \Cref{thm:iid-comp}}\label{ssec:hard-instance}
In this subsection, we present and analyze an instance in which no online algorithm can obtain a competitive ratio better than $1/e + O(1/\sqrt{K})$. 
For almost all distributions in the support of the prior, we choose the per-distribution maximum to be sufficiently likely, overwhelmingly large, and unlikely to appear twice. Approximating the offline optimum therefore essentially requires selecting the sequence maximum, as in the secretary problem. Moreover, we use a nested family of truncated distributions. This structure greatly simplifies the analysis because the posterior depends only on the current time and running maximum. Thus, the nesting renders the additional information learned online minimal, creating an information-theoretic barrier. 

We remark here that, by a minimax argument between unknown IID and the Bayesian optimal stopping, \citet[Theorem~5.1]{correa2020unknowniidprophetsbetter} had previously established that $1/e$ is the best possible constant competitive ratio that one can guarantee. However, we construct a concrete hard instance which also yields an upper bound on the best possible competitive ratio in terms of $K$; letting $K \to \infty$ indeed corresponds to the $1/e$ bound. 

\paragraph{Construction.}
Fix $n\ge2$. For each world $h\in[K]$, let $D^{(h)}$ be supported on $\{1,n^3,\ldots,n^{3h}\}$, with
\[
    \pr{X_t=1\mid H=h}=\left(1-\frac1{n^2}\right)^h,
    \qquad
    \pr{X_t=n^{3r}\mid H=h}
    =\frac1{n^2}\left(1-\frac1{n^2}\right)^{h-r}
    \quad (r\in[h]).
\]
Conditional on $H=h$, the observations are drawn IID from $D^{(h)}$. We choose the prior
\[
    \pi_h=\frac{n^{-3(h-1)}}{c_K} \ ,
    \qquad
    c_K=\sum_{\ell=0}^{K-1}n^{-3\ell} \ .
\]
\Cref{tbl:hard-instance} visualizes this hard instance.

The largest value in each world appears with probability $1/n^2$ at each time and is a factor $n^3$ larger than every other value in that world. Thus, we have $\expar{\off \mid H = h} \approx n^{3h} \cdot p_n$, where $p_n = 1-(1-1/n^2)^n \approx 1/n$. In other words, a large $1-o(1)$ fraction of the reward of \off{} comes from the largest value in the support of $D^{(h)}$. The prior balances these contributions across worlds, so that the expected contribution of every world to \off{} is almost identical.

\begin{table}[h]
    \centering
    \captionsetup{
        font={small,normalfont},
        labelfont=normalfont,
        textfont=normalfont
    }

    \begingroup
    \footnotesize
    \renewcommand{\arraystretch}{1.08}
    \setlength{\tabcolsep}{6pt}
    \setlength{\arrayrulewidth}{0.4pt}
    \arrayrulecolor{navyblue!55}

    \resizebox{\linewidth}{!}{%
    \begin{tabular}{c|c|*{7}{c}}
        \multicolumn{2}{c}{}
        & \multicolumn{7}{c}{%
            \cellcolor{navyblue!6}\textcolor{navyblue}{Value}} \\
        \cline{3-9}
        \noalign{\vskip\arrayrulewidth}

        \rowcolor{navyblue!6}
        \rule[-0.75ex]{0pt}{3.6ex}%
        \textcolor{navyblue}{World $h$}
        & \textcolor{navyblue}{Prior $\pi_h$}
        & \textcolor{navyblue}{$1$}
        & \textcolor{navyblue}{$n^3$}
        & \textcolor{navyblue}{$n^6$}
        & \textcolor{navyblue}{$n^9$}
        & \textcolor{navyblue}{$\cdots$}
        & \textcolor{navyblue}{$n^{3(K-1)}$}
        & \textcolor{navyblue}{$n^{3K}$} \\
        \hline

        $1$
        & $c_K^{-1}$
        & $1-\frac{1}{n^2}$
        & $\frac{1}{n^2}$
        & & & & & \\

        $2$
        & $(c_K n^3)^{-1}$
        & $\left(1-\frac{1}{n^2}\right)^2$
        & $\left(1-\frac{1}{n^2}\right)\frac{1}{n^2}$
        & $\frac{1}{n^2}$
        & & & & \\

        $3$
        & $(c_K n^6)^{-1}$
        & $\left(1-\frac{1}{n^2}\right)^3$
        & $\left(1-\frac{1}{n^2}\right)^2\frac{1}{n^2}$
        & $\left(1-\frac{1}{n^2}\right)\frac{1}{n^2}$
        & $\frac{1}{n^2}$
        & & & \\

        $\vdots$ & $\vdots$ & & & & & & & \\

        $K-1$
        & $(c_K n^{3(K-2)})^{-1}$
        & $\left(1-\frac{1}{n^2}\right)^{K-1}$
        & $\left(1-\frac{1}{n^2}\right)^{K-2}\frac{1}{n^2}$
        & $\left(1-\frac{1}{n^2}\right)^{K-3}\frac{1}{n^2}$
        & $\cdots$
        & $\left(1-\frac{1}{n^2}\right)\frac{1}{n^2}$
        & $\frac{1}{n^2}$
        & \\

        $K$
        & $(c_K n^{3(K-1)})^{-1}$
        & $\left(1-\frac{1}{n^2}\right)^K$
        & $\left(1-\frac{1}{n^2}\right)^{K-1}\frac{1}{n^2}$
        & $\left(1-\frac{1}{n^2}\right)^{K-2}\frac{1}{n^2}$
        & $\cdots$
        & $\cdots$
        & $\left(1-\frac{1}{n^2}\right)\frac{1}{n^2}$
        & $\frac{1}{n^2}$ \\
    \end{tabular}%
    }

    \arrayrulecolor{black}
    \endgroup

    \vspace{10pt}
    \caption{The hard instance corresponding to \Cref{lem:main-hardness}. \\
    The entries under the value columns give the probability of sampling those values under the corresponding worlds.}
    \label{tbl:hard-instance}
\end{table}

\paragraph{Analysis.}
We prove a stronger upper bound for competitive ratio that yields the result of the theorem as a corollary: 

\begin{lemma}\label{lem:main-hardness}
    There exist absolute constants $c_0>0$ and $n_0\ge2$ such that, for every $K\ge1$ and $n\ge n_0$, every online algorithm $\mathcal A$ on the hard instance of \Cref{tbl:hard-instance} satisfies
    \[
        \text{CR}(\mathcal A)\le\frac1e+c_0\left(\frac1n+\frac nK\right) .
    \]
\end{lemma}
Taking $n=\lfloor\sqrt K\rfloor$ in this lemma then yields the hardness result in \Cref{thm:iid-comp}.

{\bf Proof of \Cref{lem:main-hardness}.} %

We prove this lemma by constructing a dual certificate $D$ and comparing it with \off, i.e., we show that
\begin{align}
    \frac{\ex{D}}{\ex{M}} \le \frac 1 e + O\left (\frac{1}{n} + \frac n K\right ) \ , \quad \text{where} \quad M = \max_{t\in[n]} X_t \ .\label{ineq:d-perf}
\end{align}
Since $D$ is a dual certificate, {by Doob's optional sampling theorem (\Cref{thm:optional-sampling}),} its expectation upper bounds the reward of any online algorithm, i.e., $r(\cal A) \le \expar{D}$; the lemma follows immediately. 

To construct $D$, we pick a suffix $X_{s}, X_{s+1}, \cdots, X_n$ and let $D = X_s + \cdots + X_n$. We show that if the suffix is chosen carefully, $D$ becomes a dual certificate. We choose this suffix by first drawing $T\in [n/e, n]$ with density $f_T(s) = 1/s$, and then letting the suffix be $X_{\lceil T \rceil}, \cdots X_n$. Finally, the dual certificate is 
\begin{align}
    D = \ex{ \widetilde D \,  \big | \cal F_n} \quad \text{for} \quad \widetilde D = \sum_{i=\lceil T \rceil}^n X_i + n^2 v_K \cdot \ind{X_n = v_K} \ ,  \label{eq:dual-def}    
\end{align}
where the second term in $\widetilde D$ is a correction term to control for finite values of $K$.

We first analyze $\expar{D} / \expar{M}$, and then show that $D$ is a dual certificate. 
A simple calculation shows that $\expar{T} = (1-1/e)n$. Since the suffix contains $n-\lceil T\rceil+1$ observations, we have
\[
    \ex{\sum_{s=\lceil T\rceil}^n X_s}
    = \ex{X_1}\cdot\ex{n-\lceil T\rceil+1}
    = \left(\frac ne+O(1)\right)\ex{X_1} \ ,
\]
where we used the fact that the observations are conditionally IID given $H$. 
To compute $\expar{X_1}$, note that in each world $h$, the largest value $v_h=n^{3h}$ arrives at any time $t$ with probability $1/n^2$, and every other value is at most $n^{3h-3}$. Thus, $\expar{X_1\mid H=h}=n^{3h-2}+O(n^{3h-3})$, and averaging over the prior gives
\begin{equation} \label{eq:hard-instance-off}
    \ex{X_1} = \sum_{h=1}^K \frac{1}{c_K n^{3h-3}} \left(n^{3h-2}+O(n^{3h-3})\right)
    = \frac{nK+O(K)}{c_K} \ .
\end{equation}

Similarly, the largest value appears among the $n$ observations with probability $1-(1-1/n^2)^n=1/n+O(1/n^2)$. Therefore,
\begin{align}
    \off
    &= \sum_{h=1}^K \frac{1}{c_K n^{3h-3}}
    \left(n^{3h}\left(1-\left(1-\frac1{n^2}\right)^n\right)+O(n^{3h-3})\right) \notag \\
    &= \frac{n^2K+O(nK)}{c_K} \ . \label{ineq:off-bound}
\end{align}
Finally, $X_n=v_K$ is possible only in world $K$, where it occurs with probability $1/n^2$. Hence, the correction term has expectation
\[
    \ex{n^2v_K\ind{X_n=v_K}}
    =v_K\pr{H=K}=\frac{n^3}{c_K} \ .
\]
Combining these, we obtain
\[
    \frac{\ex{D}}{\ex{M}}
    =\frac{n^2K/e+n^3+O(nK)}{n^2K+O(nK)}
    =\frac1e+O\left(\frac1n+\frac nK\right) \ ,
\]
where the expectation in the numerator is with respect to all the randomness of the instance as well as the choice of $T$. This is now the desired bound in the lemma. 

It remains to show that $D$ is a dual certificate, i.e., $\expar{D \mid \cal F_t} \ge X_t$ for every $t$ almost surely. 
Fix $t \ge 1$. If $T < t$, the suffix contains $X_t$ and the dual property clearly holds. We thus decompose the dual martingale as follows:
\begin{align}
    \ex{D \mid \cal F_t} &= \sum_{s=1}^n \ex{X_s \cdot \ind{T \le s} \mid \cal F_t} + \ex{n^2 v_K \cdot \ind{X_n = v_K} \mid \cal F_t} \notag \\
    & \ge \ex{X_t \mid \cal F_t} \cdot \pr{T \le t} + \sum_{s=t+1}^n \ex{X_s \mid \cal F_t} \cdot \pr{T \le s} + \ex{n^2 v_K \cdot \ind{X_n = v_K} \mid \cal F_t} \notag  \\
    & \ge X_t \cdot \pr{T \le t} + \sum_{s=t+1}^n \ex{X_s \mid \cal F_t} \cdot \pr{T \le s} + \ex{n^2 v_K \cdot \ind{X_n = v_K} \mid \cal F_t} \ ,  \label{ineq:d-lb}
\end{align}
where the first inequality uses the independence of $T$ and observations. In light of \eqref{ineq:d-lb}, we next focus on lower-bounding the second and third terms above. We argue
\begin{align*}
    \sum_{s = t+1} \ex{X_s \mid \cal F_t}   \cdot \pr{T \le s} 
    & = \sum_{s = t+1} \ex{\ex{X_s \mid H, \cal F_t} \mid \cal F_t } \cdot \pr{T \le s}\\ 
    & =  \ex{\mu_H \mid \cal F_t }  \cdot \sum_{s = t+1} \pr{T \le s} \ ,
\end{align*}
where the second equality uses the fact that the observations are conditionally {IID} given $H$. In other words, $\expar{X_s \mid \cal F_t}$ is identical for all $s > t$ and is equal to $\expar{\mu_H \mid \cal F_t}$, where $\mu_h$ is the expectation of $D^{(h)}$. We henceforth use the notation $\mu_t = \expar{\mu_H \mid \cal F_t}$. 

Now, to show that $D$ is a dual certificate, we analyze $\mu_t$ and $W_t = \sum_{s = t+1} \pr{T \le s}$. Let $M_t = \max_{s \le t}X_s$. To analyze $\mu_t$, suppose that $M_t = v_j = n^{3j}$ for some $j \in [K]$ and define $d = K - j$, where we note that if $M_t = 1$, we have $\expar{D \mid \cal F_t} \ge X_t =1$ trivially. 
To simplify the notation, let
\[
    p  = \frac{1}{n^2}, \quad q=1-p, \quad z = q^t, \quad a = \frac z {n^3}. 
\]
On the one hand, a simple calculation for the posterior shows that the posterior at time $t$ only depends on $M_t$ and, for $0 \le r \le d$, that
\[
    \pr{H = j + r \mid \cal F_t, M_t = v_j} = \pr{H = j + r \mid M_t = v_j} = \frac{a^r}{S_d} \ , \quad \text{where} \quad S_d = \sum_{r = 0}^d a^r = \frac{1-a^{d+1}}{1-a}.
\]
See Appendix~\ref{prf:posterior} for a quick proof. 

On the other hand, the contribution of world $H = j + r$ to $\mu_t$ comes mostly from the largest value in its support, which is $v_{j+r} = n^{3(j+r)}$ and condtional on $H = j + r$, it occurs at time $t+1$ with probability $p$. We thus have for every $j \in [K]$:
\begin{align}
    \mu_t \ge \sum_{r = 0}^d \pr{H = j + r \mid M_t = v_j} \cdot pv_{j+r}  = \sum_{r=0}^d \frac{a^r}{S_d} \cdot p n^{3r} v_j= \frac{p v_j}{S_d} \sum_{r=0}^d z^r \ . \label{ineq:mu-t-lb}
\end{align}
We define $$m_t^* = \frac{1-z}{p(1-a)} = \frac{1-q^t}{p(1-q^t/n^3)} \ .$$ 
\begin{claim}\label{clm:mstar}
    For all sufficiently large $n$ and $2 \le t \le n$, we have $m_t^* \le t$. 
\end{claim} 
Proof of this claim is in Appendix~\ref{prf:mstar}. 

Let $c(x) = \prpar{T \le x}$ and so $W_t = \sum_{s=t+1}^n c(s)$. The density of $T$ gives $c(x)=\max\{0,1-\ln(n/x)\}$ for $0<x\le n$. Since $c(x)$ is increasing, for every $t\in[n]$ we have
\[
    W_t = \sum_{s=t+1}^n c(s)
    \ge \int_t^n c(s)\,ds
    = \max\{t,n/e\}(1-c(t)) \ .
\]
For $2 \le t < n$, \Cref{clm:mstar} gives $m_t^*\le t$. At $t=1$, we instead have $m_1^*=1/(1-q/n^3)\le n/e$ for $n\ge3$. Thus, for every $t\in[n]$, we obtain
\begin{align}
    W_t \ge m_t^*(1-c(t)) \ . \label{ineq:w-bound}
\end{align}
Now, multiplying \eqref{ineq:mu-t-lb} by $m_t^* = (1-z)/(p(1-a))$ gives
\[
\begin{aligned}
m_t^*\mu_t \ge
\frac{v_j}{(1-a)S_d}
(1-z)\sum_{r=0}^dz^r =
v_j\frac{1-z^{d+1}}{1-a^{d+1}} \ge v_j(1-z^{d+1}).
\end{aligned}
\]
By \eqref{ineq:w-bound}, we have
\begin{align}
    \sum_{s = t+1} \ex{X_s \mid \cal F_t}   \cdot \pr{T \le s}  = W_t \mu_t  \ge m_t^* \mu_t (1-c(t)) \ge v_j(1-z^{d+1})(1-c(t)) \ , \label{ineq:d-second-lb}
\end{align}
which is the second term in \eqref{ineq:d-lb}. We now lower bound the third term in \eqref{ineq:d-lb}. Using the posterior formula, we have for $t < n$ and on the event $\{M_t = v_j \}$:
\begin{align}
    \ex{n^2 v_K \ind{X_n = v_K} \mid \cal F_t} &= v_K \cdot \pr{H = K \mid \cal F_t} \notag \\ 
    & = v_K \frac{a^d}{S_d} = \frac{v_j z^d}{S_d} \ge {v_j z^d (1-a)} \ge v_j z^{d+1} \ , \label{ineq:third-term-lb}
\end{align}
where the first inequality follows from $S_d = \sum_{r = 0}^d a^r \le 1/(1-a)$ and the second inequality follows from $1-z \ge p \ge a$.

Plugging in \eqref{ineq:d-second-lb} and \eqref{ineq:third-term-lb} in \eqref{ineq:d-lb} gives, for $t < n$ and on the event $\{M_t = v_j\}$, that
\begin{align}
    \ex{D \mid \cal F_t} &\ge c(t) X_t +  v_j(1-z^{d+1})(1-c(t)) + v_j z^{d+1} \notag \\
    & \ge c(t) X_t +  v_j(1-z^{d+1})(1-c(t)) + v_j z^{d+1}(1-c(t)) \notag \\
    & = c(t) X_t +  v_j(1-c(t))  \notag \\ 
    & \ge c(t) X_t + X_t(1-c(t)) \notag \\
    & = X_t \ . \notag
\end{align}
For $t=n$, $D$ includes $X_n$ as a summand and the dominance holds trivially. Hence, $D$ is indeed a dual certificate. Thus, we have $\ex{D} \ge \on$. Recalling \eqref{ineq:d-perf} completes the proof. 
\hfill \qedsymbol

\color{black}

\subsection{Algorithm of \Cref{thm:iid-comp}}\label{ssec:competitive-algo} Our algorithm proceeds as follows. First, we identify the world $h^\star$ that has the highest contribution to the expected reward of \off, which implies $\pi_{h^\star}\ex{\off | H = h^\star} \ge \off / K$. Next, we select $\alpha_K = 1/e + 1/K$ and $\beta_K = 1/e - 2/(eK^2)$. Finally, we use the algorithm of \cite{bai2025optimal} with parameters $\alpha_K, \beta_K$ and the candidate distribution $D_{h^\star}$. We recall that \cite{bai2025optimal} consider an \iid optimal stopping problem where $n$ observations are generated from a distribution $\cal D_0$. Given a candidate distribution $\cal D$ and feasible $(\alpha, \beta)$, the authors develop an algorithm that is $\alpha$-competitive if $\cal D_0 = \cal D$, and $\beta$-competitive otherwise. 

If $(\alpha_K, \beta_K)$ is feasible, a simple calculation shows that our algorithm has an expected reward of at least
\begin{align}
    & \alpha_K \cdot \ex{\off \cdot \ind{H = h} } + \beta_K \cdot \ex{\off \cdot \ind{H \neq h} } \notag \\ \; & = (\alpha_K - \beta_K) \cdot \ex{\off \cdot \ind{H = h} } + \beta_K \cdot \off  \notag
    \\ & \ge \left ( \frac{\alpha_K - \beta_K}{K} + \beta_K \right) \cdot \off  \notag
    \\ & \ge \left ( \frac{1}{e} + \frac{1-2/e}{K^2}\right ) \cdot \off \ . \label{eq:cr-final}
\end{align}

Thus, it suffices to show $(\alpha_K, \beta_K)$ is feasible. To this end, we leverage Theorem 1 of \cite{bai2025optimal}, that provides sufficient conditions for feasibility of $\alpha$ and $\beta$, to prove the following lemma:
\begin{lemma}\label{lem:iid-feasibility}
For every integer \(K \ge 1\), the pair $(\alpha_K, \beta_K)$ with
\[
\alpha_K=\frac1e+\frac1{\widehat K} \ ,
\qquad
\beta_K=\frac1e-\frac{2}{e{\widehat K}^2} \ , \qquad \widehat K = \max \{K, 39\}
\]
is feasible for \citet[Theorem 1]{bai2025optimal}. That is, for every horizon \(n\) and candidate
distribution \(\widehat{\cal D}\), there exists an online stopping
algorithm that is \(\alpha_K\)-competitive when the observations
are drawn IID from \(\widehat {\cal D}\), and
\(\beta_K\)-competitive when they are drawn IID from any other
distribution \({\cal D}\).
\end{lemma}
This lemma is proved in Appendix~\ref{prf:huang-suff}. Note that by this result, $(\alpha_K, \beta_K)$ is feasible for every $K \ge 39$. Therefore, \eqref{eq:cr-final} implies that our algorithm is $\big ({1}/{e} + {(1-2/e)}/{K^2}\big )$-competitive, as desired. For $K < 39$, we use the algorithm for $K=39$, i.e, we set $\alpha = 1/e + 1/39$ and $\beta_K = 1/e - 2/(e\cdot 39^2)$. Then, the bound in \eqref{eq:cr-final} becomes 
\[
    \left (\frac{1}{e} + \frac{1}{39K} - \frac{2(K-1)}{39eK^2}\right ) \cdot \off \ge \left (\frac{1}{e} + \frac{1}{39K^2} \right) \quad \text{for every } K \ge 1 \ , 
\]
which completes the proof of the lower bound in \Cref{thm:iid-comp} with $c' = \min \{\frac{1}{39}, 1-2/e \} = \frac1{39}$. 

\section{Conclusion and Open Questions}

In this work, we propose Bayesian models as a middle ground between pessimistic, adversarial models and optimistic, fully known stochastic models. We study this question in the context of Bayesian optimal stopping. In these models, the decision-maker learns about the hidden world while making decisions, and the Bayes-optimal policy accounts for both the value of learning and the value of waiting. Our results show a stark separation between philosopher and prophet inequalities. On the one hand, the Bayes-optimal policy admits efficient approximation schemes, despite being defined by an exponentially large dynamic program. On the other hand, no online algorithm can substantially surpass the worst-case competitive ratios against the prophet when the prior is broad enough, i.e., when $K$ is large.

The central idea of our approximation scheme is to start with a baseline algorithm and improve it iteratively. Each improvement step randomizes between the current algorithm and thresholds set relative to a dual martingale, which upper bounds the value function of the Bayesian DP. The same step also shrinks the dual martingale, so that the improved algorithm is competitive against a tighter one and the step can be repeated. Because the shrinkage acts on each sample path separately, the dual martingales can be approximated efficiently by Monte Carlo sampling. Together, these ingredients give an approximation scheme for Bayesian optimal stopping, which is a PTAS in the \iid model and in the \noniid model with $K = O(1)$.

For prophet inequalities, we construct an explicit \iid instance on which no online algorithm is better than $(1/e + O(1/\sqrt K))$-competitive. Rather than analyzing the Bayesian DP on this instance, we bound the reward of every online algorithm by an explicit dual martingale. Thus, dual martingales serve both for designing algorithms and for proving hardness results.

Several directions and open questions emerge from this paper.

\begin{itemize}
    \item {\it Information model.} In this work, we assume access to a probability oracle and a value oracle. This assumption is stronger than access to a sample oracle that returns fresh samples from the conditional distribution given the history. However, our results in Appendix~\ref{app:conditional-sampling-hardness} show that such sample oracles cannot yield approximation ratios better than a constant, even in the \iid model with $K=1$. It would therefore be interesting to explore information models in between, especially for the \gen model. Relatedly, identifying meaningful classes of problems within the \gen model but outside the \noniid model that admit compact representations and approximation schemes is another fruitful direction.
    \item {\it Better computational efficiency.} In this work, we obtain approximation schemes for the \gen and \noniid models and a PTAS for the \iid model. Are better runtimes possible, e.g., can the optimal algorithm be computed in polynomial time? Or is the \gen model, or even the \noniid model, APX-hard? The \noniid model is of particular interest: although the problem is well-structured, we are not aware of any genuinely polynomial-time approximation scheme for it, and our limited human- and AI-based efforts to find one have been unsuccessful even to obtain a constant-factor approximation.
    \item {\it Combinatorial problems.} We view this work as a stepping stone toward designing Bayesian models and algorithms for more general online decision-making problems. Can similar algorithms be developed for problems with a combinatorial feasible set, such as online matching? Our machinery readily extends to multi-unit selection when the budget is $O(1)$, but beyond that, we are not aware of straightforward extensions. What about other online problems with a different structure? 
\end{itemize}

\subsubsection*{AI usage disclosure.} The majority of ideas were from the authors, with ChatGPT models 5.3-6 used for brainstorming and honing human ideas. Notable  parts with input from AI were the randomization over $Q$ in \mart in \Cref{alg:approx} when given the algorithm with a deterministic $Q=q$, and the construction of the dual martingale for the hardness proof of competitive ratio in \Cref{lem:main-hardness}, which were then simplified and modified by the authors; the authors previously had a technical argument for the proof of \Cref{lem:main-hardness} that was based on an analysis of the Bayesian dynamic program. The paper was fully written by the authors, with those ChatGPT models and Claude Opus 5.5 used for minor editorial purposes to improve the exposition. The authors verified the correctness and originality of all content, including references.

\bibliographystyle{apalike}
\bibliography{refs}

\begin{APPENDICES}
\renewcommand{\theHsection}{appendix.\Alph{section}}

\section{Omitted Proofs from \Cref{sec:intro}}
{
\subsection{Performance of History-oblivious Algorithms}\label{app:oblivious-algos}
The next proposition shows that no \textit{history-oblivious} algorithm can guarantee a constant-factor competitive ratio. An algorithm is history-oblivious if its decision at time $t$ is independent of $X_1, \cdots, X_{t-1}$. The proposition relies on the construction and definitions in \Cref{ssec:hard-instance}.
\begin{proposition}\label{prop:history-hardness}
For every $\eps>0$, there exists an instance of the Bayesian optimal stopping problem in which every history-oblivious online algorithm $\mathcal A$ satisfies
\(
    r(\mathcal A)\le\eps\on,
\)
where \on{} is the Bayes-optimal algorithm defined in \eqref{eq:snell}.
\end{proposition}

\begin{proof}
We use the same hard instance from Section~\ref{ssec:hard-instance}, shown in Table~\ref{tbl:hard-instance}. Recall Equation \eqref{ineq:off-bound} in the proof of \Cref{lem:main-hardness} that gives
\begin{equation}
    \off=\frac{n^2K}{c_K}\left(1+O\!\left(\frac1n\right)\right),
    \label{eq:history-off}
\end{equation}
where $\off=\ex{\max_{t\in[n]}X_t}$ denotes the expected reward of the offline benchmark. We choose $K=n^2$ so that every supported value greater than $1$ has probability at least $1/(en^2)$, i.e.
\begin{equation}
    \begin{gathered}
        \pr{X_t=v_j\mid H=h}
        =\frac1{n^2}\left(1-\frac1{n^2}\right)^{h-j} \ge\frac1{n^2}\left(1-\frac1{n^2}\right)^{K-1}
        \ge\frac1{en^2}
    \end{gathered}
    \label{eq:history-mass}
\end{equation}
for every $j\le h\le K$, where the last inequality follows from $(1-1/n^2)^{K-1}=(1+1/(K-1))^{-(K-1)}\ge e^{-1}$.

Now, take an arbitrary algorithm $\mathcal A$ whose decision at time $t$ depends only on the Bayesian instance (prior $\pi$ and world distributions $D^{(h)}$), observation $X_t$, time $t$, and its internal randomness (represented by seed $U$). Let $\tau_{\mathcal A}$ be its stopping time. At each time $t$, let the algorithm's acceptance set be $$A_t = \{j\in [K]: \mathcal A \text{ accepts $v_j$ at time $t$}\}.$$ Fix any realization $u$ of $U$. Define the expected reward and probability of selecting per-distribution maxima as
\[
    r_u=\ex{X_{\tau_{\mathcal A}}} \quad \text{and} \quad
    P_h=\pr{X_{\tau_{\mathcal A}}=v_h\mid H=h},
\]
respectively. Every reward below $v_h$ is at most $v_h/n^3$. Thus, using $\pi_hv_h=n^3/c_K$ and \eqref{eq:history-off}, we obtain
\begin{equation}
    \frac{r_u}{\off}
    \le\left(1+O\!\left(\frac1n\right)\right)
    \left(\frac nK\sum_{h=1}^K P_h+\frac1{n^2}\right),
    \label{eq:history-reduction}
\end{equation}

To bound $\sum_hP_h$, we compare each accepted value with the smaller values the algorithm would also accept at the same time $t$. For an integer $1\le L\le K$, to be chosen below, let $A_t^-$ contain the $\min\{|A_t|,L\}$ smallest indices in $A_t$, and let $A_t^+=A_t\setminus A_t^-$.  
First, consider all pairs $(t,h)$ such that $h \in A_t^-$. The probability of selecting $v_h$ at time $t$ in world $h$ is at most $1/n^2$, since this is the probability that $v_h$ appears at that time. Then, taking the union bound over all times $t\in[n]$ with $h \in A_t^-$, these selections contributes at most
\begin{align}
    \sum_{t:\,h\in A_t^-}\pr{\tau_{\mathcal A}=t,\ X_t=v_h\mid H=h} \leq \frac{L}{n}\ . \label{ineq:h-neg}
\end{align}
to $\sum_hP_h$. Next, whenever $h\in A_t^+$, we have
\begin{align} 
    \pr{\tau_{\mathcal A}=t,\ X_t=v_h\mid H=h}
    =\frac1{n^2}\pr{\tau_{\mathcal A}\ge t\mid H=h} \notag \\
    \le \frac eL\pr{\tau_{\mathcal A}=t\mid H=h}, \label{ineq:L-bound}
\end{align}
where the equality follows because $v_h$ occurs with probability $1/n^2$ independently of everything else. For the inequality step, conditional on reaching $t$, the algorithm would also accept the $L$ smaller values indexed by $A_t^-$, whose total probability is at least $L/(en^2)$ by \eqref{eq:history-mass}. This gives
$$\pr{\tau_{\mathcal A}=t\mid H=h}
 \ge\frac{L}{en^2}\pr{\tau_{\mathcal A}\ge t\mid H=h},$$ and rearranging yields \eqref{ineq:L-bound}. Then, summing this inequality over all times $t \in [n]$ with $h\in A_t^+$ gives
\begin{align}
    \sum_{t:\,h\in A_t^+}\pr{\tau_{\mathcal A}=t,\ X_t=v_h\mid H=h}
    &\le\frac eL\sum_{t:\,h\in A_t^+}\pr{\tau_{\mathcal A}=t\mid H=h} \le\frac eL, \label{ineq:h-plus}
\end{align}
where the last inequality follows because the stopping events at distinct times are mutually exclusive.

Combining \eqref{ineq:h-neg} and \eqref{ineq:h-plus}, we obtain
\(
    \sum_{h=1}^K P_h\le \frac Ln+\frac{Ke}{L}.
\)
Substituting into \eqref{eq:history-reduction}, we infer
\begin{equation}
    \frac{r_u}{\off}
    \le\left(1+O\!\left(\frac1n\right)\right)
    \left(\frac LK+\frac{en}{L}+\frac1{n^2}\right).
    \label{eq:history-explicit}
\end{equation}
Choose $L=\lceil n^{3/2}\rceil\le K$. Since $K=n^2$, both $L/K$ and $en/L$ are $O(n^{-1/2})$. Thus, \eqref{eq:history-explicit} gives
\begin{equation*}
    \frac{r_u}{\off}=O(n^{-1/2}),
    \label{eq:history-rate}
\end{equation*}
for every initial seed $U = u$.

Finally, the observations are exchangeable. The classical secretary algorithm of \cite{dynkin1963} therefore guarantees an expected reward of at least $\off/e$. Hence, $\on\ge\off/e$, and we conclude
\[
    \frac{r(\mathcal A)}{\on}
    \le e\,\frac{r(\mathcal A)}{\off}
    =O(n^{-1/2})=O(K^{-1/4}).
\]
\end{proof}
}

\subsection{Constructing Value Oracle from PMFs in \iid and \noniid Models} \label{sssec:exact-off}
We explain how the value oracle for the conditional expected reward of the prophet given the information at time $t$, i.e., $Z_t^{(0)} = \expar{\max_{t' \in [n]}X_{t'} \mid \cal F_{t}}$, can be computed exactly only from the PMFs when the observations are conditionally independent given $H$. Indeed, we have $$\zz = \expar{M_0 \mid \cal F_t} = \expar{\expar{M_0 \mid H, \cal F_t} \mid \cal F_t}.$$ Thus, we first compute $\expar{M_0 \mid H = k, \cal F_t}$ for every $k \in [K]$, and then, we average these values by the posterior distribution, which satisfies $\prpar{H=k\mid\cal F_t}\propto\pi_k\prod_{s=1}^t\prpar{X_s=x_s\mid H=k}$ for the observed history $h_t=(x_1,\ldots,x_t)$. Suppose that $v_{q^*}$ is the maximum value observed so far in history $h_t$. Given that we have access to the distributions, and that the future samples are independent conditionally on $H$, we can write
\begin{align*}
    \ex{M_0 \mid H = k, \cal F_t} &= v_{q^*} + \sum_{q=q^*+1}^m (v_q - v_{q-1}) \pr{M_0 \ge v_q \mid H = k, \cal F_t} 
    \\ & = v_{q^*} + \sum_{q=q^*+1}^m (v_q - v_{q-1}) \pr{\max \{X_{t+1}, \cdots, X_n \} \ge v_q \mid H = k}  
    \\ & = v_{q^*} + \sum_{q=q^*+1}^m (v_q - v_{q-1}) \left (1- \prod_{s=t+1}^n \pr{X_s < v_q \mid H = k} \right ) \ .
\end{align*}
The right-hand side depends on the history only through $q^*$. Its values, over all $k\in[K]$, $t\in[n]$, and $q^*\in[m]$, can be tabulated once using $O(Kmn)$ reads from the PMF and $O(Kmn)$ arithmetic operations: the products $\prod_{s=t+1}^n\prpar{X_s<v_q\mid H=k}$ follow from a backward recursion over $t$, and the sums over $q$ from a backward recursion over $q^*$. Given this table, and since the posterior weights can be updated in $O(K)$ time per observation, we can compute $\zz$ in $O(K)$ time. %

\subsection{Hardness of Approximation with Sampling Oracles}\label{app:conditional-sampling-hardness}

We now consider an information model different from the one used throughout the paper: given any feasible history $h_t=(x_1,\ldots,x_t)$, we assume access to a sample oracle that returns an independent sample from $\mathcal L(X_{t+1}\mid X_{1:t}=h_t)$. This is the information model used by \cite{chen-goldberg2026}.

\begin{proposition}
    For every $\eps > 0$ and every polynomial-time algorithm $\cal A$ with a sample-access oracle, there exists an instance of the Bayesian \iid optimal stopping problem in which we have
    \[
        \frac{r(\cal A)}{\textup{\on}} \le \frac67 + \eps \ ,
    \]
    where \on{} is the Bayes-optimal algorithm that has full information of the distributions without any constraints on its runtime. 
\end{proposition}
\begin{proof} 
Fix an algorithm $\cal A$. We construct two simple instances of \iid prophet inequality problem, i.e., $K=1$. We show that an algorithm that has only sample access cannot obtain substantially different rewards between these two instances. From this, we infer that $\cal A$ cannot obtain reward more than $\frac67 \on$ in at least one of these instances which proves the proposition. 

In both instances, we set $n=2$ and $K=1$. 
Consider an integer $\lambda\ge2$, and let
\[
    L_\lambda=2^{2\lambda},\qquad
    d_0=2^{-3\lambda},\qquad d_1=2^{-\lambda}.
\]
In instance $\mathcal I_{i,\lambda}$, the observations $X_1,X_2$ are independent draws from $D_i$, where
\[
    D_i=
    \begin{cases}
        0,&\text{with probability }\frac12-d_i,\\
        1,&\text{with probability }\frac12,\\
        L_\lambda,&\text{with probability }d_i.
    \end{cases}
\]
Notice that the index $i$ labels the instance, not a hidden world; both instances have $K=1$ and the same support $\{0,1,L_\lambda\}$. The algorithm is not given the probabilities; rather, it can request a fresh sample from the conditional law given any history.

The \iid structure implies that every conditional sampling query returns a fresh draw from $D_i$, regardless of the supplied history. In particular, querying simulated histories does not provide any additional information beyond independent samples. Define
\[
    \mu_i=\extwo{i}{X_2}=\frac12+d_iL_\lambda, \quad \text{i.e.,} \quad
    \qquad
    \mu_0=\frac12+2^{-\lambda}<1,
    \qquad
    \mu_1=\frac12+2^\lambda>1.
\]
In words, $\mu_i$ is the expected value of the second observation in instance $\cal I_{i, \lambda}$. 
When $X_1=1$, the optimal policy accepts it in instance $0$ whereas it rejects $X_1$ and continues in instance $1$. In either instance, $\on$ accepts $L_\lambda$, rejects $0$, and accepts the second observation if it has not stopped earlier.

We claim that the polynomial-time algorithm $\cal A$ cannot distinguish these instances, implying almost identical rewards across the two. 
We may assume that $\mathcal A$ also accepts $L_\lambda$ and rejects $0$ at time 1, and always accepts the second observation if it reaches it. Let
\[
    a_i=\prtwo{i}{\tau_{\mathcal A}=1\mid X_1=1}
\]
be the probability of accepting the first observation $X_1 = 1$ when the instance is $\cal I_{i,\lambda}$. 

Let $\cal O_i$ be the set of samples collected by $\cal A$ when ran on instance $\cal I_{i,\lambda}$. Because $\cal A$ is a polytime algorithm, we have $|\cal O_i| = \poly(\lambda)$. 
The total variation distance between $D_0$ and $D_1$ is $d_1-d_0\le2^{-\lambda}$. With $|\cal O_i|$ many calls to the sample oracle, we have $\text{TV}(\cal O_1, \cal O_2) \le \poly(\lambda) \cdot 2^{-\lambda}$. Given that $a_i \in [0,1]$ is a function of $\cal O_i$, we infer that
\begin{equation}
    |a_0-a_1|\le \poly(\lambda)2^{-\lambda}.
    \label{eq:conditional-sampling-coupling}
\end{equation}

We now compare the reward of $\cal A$ in the two instances. 
Let $r_i(\mathcal A)=\extwo{i}{X_{\tau_{\mathcal A}}}$. Conditional on reaching the second time, the actual observation $X_2$ still has mean $\mu_i$. Evaluating $r_i{\cal A}$ by conditioning on $X_1$ gives
\begin{align}
    r_i(\mathcal A)
    &=d_iL_\lambda+\left(\frac12-d_i\right)\mu_i
      +\frac12\bigl(a_i+(1-a_i)\mu_i\bigr) \nonumber\\
    &=d_iL_\lambda+(1-d_i)\mu_i+\frac{a_i}{2}(1-\mu_i).
    \label{eq:conditional-sampling-reward}
\end{align}
Conversely, \on{} knows the distributions and can compute $\mu_0<1<\mu_1$. Then, \on{} sets $a_0=1$ and $a_1=0$. Plugging in this optimal choice in the reward formulae yields
\[
\begin{aligned}
    \on(\mathcal I_{0,\lambda})&=\frac34+O(2^{-\lambda}),
    &r_0(\mathcal A)&=\frac12+\frac{a_0}{4}+O(2^{-\lambda}),\\
    \on(\mathcal I_{1,\lambda})&=2^{\lambda+1}+O(1),
    &r_1(\mathcal A)&=2^\lambda\left(2-\frac{a_1}{2}\right)+O(1).
\end{aligned}
\]
We can therefore argue
\[
    \frac{r_0(\mathcal A)}{\on(\mathcal I_{0,\lambda})}
    =\frac23+\frac{a_0}{3}+O(2^{-\lambda}),
    \qquad
    \frac{r_1(\mathcal A)}{\on(\mathcal I_{1,\lambda})}
    =1-\frac{a_1}{4}+O(2^{-\lambda}).
\]
Using \eqref{eq:conditional-sampling-coupling} to replace $a_1$ by $a_0$ gives
\[
    \min_{i\in\{0,1\}}
    \frac{r_i(\mathcal A)}{\on(\mathcal I_{i,\lambda})}
    \le
    \min\left\{\frac23+\frac{a_0}{3},\,1-\frac{a_0}{4}\right\}
    + \poly(\lambda) \cdot 2^{-\lambda}.
\]
The term involving $a_0$ is maximized at $a_0=4/7$ with a value of 6/7. Thus, we have
\[
    \min_{i\in\{0,1\}}
    \frac{r_i(\mathcal A)}{\on(\mathcal I_{i,\lambda})} \le \frac6 7 + o(1) \quad \text{as } \lambda \to \infty.
\]
The proof is now complete. 
\end{proof}

Notice that the above proof shows hardness for a classic prophet inequality problem with $K=1$. Thus, such sample access creates a barrier for approximating the online optimum even in the classical setting of prophet inequality. Another implication of $K=1$ is that this proposition shows the same hardness for a different sample oracle that, given a world $h$, it returns a fresh sample from $D^{(h)}$. In general, this sample oracle is not comparable to one that only returns fresh samples conditoinal on the history, without per world knowledge. Nonetheless, the example above shows that both information access models create a barrier for developing (near) polynomial-time approximation schemes.

\section{Omitted Proofs from \Cref{sec:approximation}}
\subsection{Proof of \Cref{clm:z-alter-def}}\label{prf:z-alter-def}

For a sample path $\omega \in \Omega$, recall that $\omega_t$ is its $t$-step prefix. By the definition of the biased probability measure $\prtwo{*}{\cdot \mid \cal F_t}$ in \eqref{expr:biased-prob}, we have
\begin{align*}
    \extwo{*}{W_j\mid h_t}
    &=\sum_{\substack{\omega\in\Omega:\,\omega_t=h_t}}
      W_j(\omega)\prtwo{*}{\omega\mid h_t}\\
    &=\sum_{\substack{\omega\in\Omega:\,\omega_t=h_t}}
      W_j(\omega)\pr{\omega\mid h_t}
      \frac{M_0(\omega)}{Z_t^{(0)}(h_t)}\\
    &=\frac{\expar{M_0W_j\mid h_t}}{Z_t^{(0)}(h_t)}
     =\frac{Z_t^{(j)}(h_t)}{Z_t^{(0)}(h_t)}.
\end{align*}
The last equality uses $M_j=M_0W_j$. Multiplying by $Z_t^{(0)}(h_t)$ proves the claim. \hfill \qedsymbol

\subsection{Sampling from the Biased Distribution}\label{app:star-sampling}
In this subsection, we describe how to exactly and efficiently sample a future trajectory from the biased probability measure $\prtwo{*}{\omega \mid {\cal F}_t} = \pr{\omega \mid \cal F_t} \cdot \frac{M_0(\omega)}{Z_t^{(0)}}$ at time $t$, having observed a history $h_t$ with $Z_t^{(0)}(h_t)>0$. 

Let $h_s$ be the history until time $s$, and recall that $\cal F_s = \sigma(X_1,\ldots,X_s)$ is the natural filtration generated by the history. Also, recall that $\omega_s$ is the $s$-prefix of $\omega$, i.e., the first $s$ samples under $\omega$, and let $X_s(\omega)$ be exactly the $s$-th observation under $\omega$. We thus have
\begin{align*}
    \prtwo{*}{\omega \mid {\cal F}_t} &= \pr{\omega \mid h_t = \omega_t} \cdot \frac{M_0(\omega)}{Z_t^{(0)}} \\
    &= \prod_{s = t+1}^n \pr{\omega_s \mid h_{s-1} = \omega_{s-1}} \cdot \frac{\ex{M_0 \mid h_s = \omega_s}}{\ex{M_0 \mid h_{s-1} = \omega_{s-1}}} \ ,
\end{align*}
where the last equality uses chain rule $\prod_{s = t+1}^n \prpar{\omega_s \mid h_{s-1} = \omega_{s-1}} = \prpar{\omega \mid h_t = \omega_t}$. Consequently, to sample $\omega$ from $\prtwo{*}{\cdot \mid \cal F_t}$, it suffices to iteratively, for $s = t+1$ until $s = n$, sample the $s$-th observation $X_s(\omega)$ with probability $$\mu_s(\omega_s) = \pr{\omega_s \mid h_{s-1} = \omega_{s-1}} \cdot {\ex{M_0 \mid h_s = \omega_s}}/{\ex{M_0 \mid h_{s-1} = \omega_{s-1}}},$$ where we now use the fact that $\expar{M_0 \mid h_s = \omega_s}$ can be computed exactly. A value with $\expar{M_0 \mid h_s = \omega_s}=0$ has weight zero, so every sampled prefix has positive conditional expectation and all ratios are well defined. Notice that $\mu_s$ is a valid probability measure because
\begin{align*}
   & \sum_{X_s(\omega) \in \{v_1, \cdots, v_m\}} \pr{\omega_s \mid h_{s-1} = \omega_{s-1}} \cdot \frac{\ex{M_0 \mid h_s = \omega_s}}{\ex{M_0 \mid h_{s-1} = \omega_{s-1}}} \\ & \; \qquad = \frac{1}{\ex{M_0 \mid h_{s-1} = \omega_{s-1}}} \cdot \sum_{X_s(\omega) \in \{v_1, \cdots, v_m\}} \pr{\omega_s \mid h_{s-1} = \omega_{s-1}} \cdot {\ex{M_0 \mid h_s = \omega_s}}
   \\ & \; \qquad = \frac{1}{\ex{M_0 \mid h_{s-1} = \omega_{s-1}}} \cdot \ex{M_0 \mid h_{s-1} = \omega_{s-1}}  = 1 \ . 
\end{align*}
At each step, this requires evaluating $\expar{M_0 \mid h_s = \omega_s}$ for each of the $m$ candidate values of the $s$-th observation, at cost $O(K)$ each. As a result, the overall sampling procedure has $C_\star=O(nKm)$ time complexity.

\subsection{Proof of \Cref{lem:accuracy-analysis}}\label{prf:accuracy-analysis}

We first state a result that says \Cref{alg:w-oracle} approximates $W_j(\omega)$ accurately and efficiently (proof in Appendix~\ref{prf:w-estimation}):
\begin{claim}\label{clm:w-estimation}
Suppose $0<\alpha_0\le1$ and define $\alpha_{\ell+1}=1/(1+\ln(1/\alpha_\ell))$ for every $\ell\ge0$. For every integer $j\ge0$, sample path $\omega \in \Omega$, and accuracy parameters $0<\eta,\kappa<1$, the recursive estimator ${\cal W}_j(\omega,\eta,\kappa)$ in \Cref{alg:w-oracle} returns $\widehat W_j(\omega)$ such that $$ \frac{\widehat W_j(\omega)}{W_j(\omega)}\in[(1-\eta),(1-\eta)^{-1}] \quad \text{with probability at least $1-\kappa$ in time } \poly\!\left(2^{j^2}\left(\frac{Kmn}{\alpha_0\eta\kappa}\right)^{j+1}\right) \ .$$ More precisely, the runtime is
    $$
        O\!\left(
    756^{j^2}K m\,n^{j+1}(\alpha_0\eta)^{-2j}
    \left(j+\ln\frac{n}{\alpha_0\kappa\eta}\right)^j
    \right) \ .
    $$
\end{claim}
Given this claim, we can estimate $Z_t^{(j)}(h_t)$ via Monte carlo simulation, as the next claim shows (proof in Appendix~\ref{prf:z-estimator}):
\begin{claim}\label{clm:z-estimator}
    Consider some $j\ge1$, time $t\in[n]$, a fixed history $h_t$, and let $G_{j-1} = \prod_{\ell=0}^{j-1}\alpha_\ell$. Suppose we are given $r$ independent samples $(\omega^{(i)})_{i\in[r]}$ for continuations $X_{t+1},\ldots,X_n$ from the biased distribution $\prtwo{*}{\cdot\mid h_t}$. Moreover, suppose that for each $i$, we have an estimate $\widehat W_j(\omega^{(i)})$ of $W_j(\omega^{(i)})$ such that, conditionally on every realized $\omega^{(i)}$ and for some $0<\eta,\kappa<1$, it holds
    \[
        \frac{\widehat W_{j}(\omega^{(i)})}{ W_j(\omega^{(i)})} \in [(1-\eta), (1-\eta)^{-1}] \quad \text{with probability at least $1-\kappa$.}
    \]
    Then, $\widehat Z_t^{(j)}(h_t) = Z_t^{(0)}(h_t) \cdot \frac{1}{r}\sum_{i=1}^r \widehat W_{j}(\omega^{(i)})$ satisfies 
    \[
        \frac{\widehat Z_t^{(j)}(h_t)}{Z_t^{(j)}(h_t)} \in [(1-\eta)^2, (1-\eta)^{-2}]  \quad \text{with probability at least } 1 - \kappa r - 2  \exp \left(-\frac{\eta^2 r G_{j-1}}{3} \right) \ .
    \]
\end{claim}

We now combine the recursive estimator of $W_j$ with the Monte Carlo estimate of $Z_t^{(j)}$ from the above claims. 

For $j=0$, we compute $L_t^{(0)}(h_t)=Z_t^{(0)}(h_t)$ exactly by a call to the value oracle. We henceforth suppose $j\ge1$ and $Z_t^{(0)}(h_t)>0$. Let $G_{j-1}=\prod_{\ell=0}^{j-1}\alpha_\ell$, and choose
\[
    \eta=\frac\eps4,
    \qquad
    r=\left\lceil\frac{3}{\eta^2G_{j-1}}\ln\frac4\delta\right\rceil,
    \qquad
    \kappa=\frac{\delta}{2r}.
\]
Draw $r$ independent continuations $\omega^{(1)},\ldots,\omega^{(r)}$ from $\prtwo{*}{\cdot\mid h_t}$. For each completed sample path, use fresh randomness to compute $\widehat W_j(\omega^{(i)})={\cal W}_j(\omega^{(i)},\eta,\kappa)$, and return
\[
    L_t^{(j)}(h_t)=(1-\eta)^2\widehat Z_t^{(j)}(h_t),
    \qquad
    \widehat Z_t^{(j)}(h_t)
    =Z_t^{(0)}(h_t)\cdot\frac1r\sum_{i=1}^r\widehat W_j(\omega^{(i)}).
\]
By \Cref{clm:w-estimation}, conditionally on each sampled trajectory $\omega^{(i)}$, we have $\widehat W_j(\omega^{(i)})/W_j(\omega^{(i)})\in[1-\eta,(1-\eta)^{-1}]$ with probability at least $1-\kappa$. Thus, the hypothesis of \Cref{clm:z-estimator} holds, which implies that $\widehat Z_t^{(j)}(h_t)/Z_t^{(j)}(h_t)\in[(1-\eta)^2,(1-\eta)^{-2}]$ with probability at least $1-\kappa r-2\exp(-\eta^2rG_{j-1}/3)$. We now show that this failure probability is at most~$\delta$:
\[
    \kappa r+2\exp\left(-\frac{\eta^2rG_{j-1}}3\right)
    \overset{(i)}{\le}\frac\delta2+2\exp\left(-\ln\frac4\delta\right)
    =\delta \ ,
\]
where $(i)$ uses $\kappa r=\delta/2$ by the choice $\kappa=\delta/(2r)$, and $\eta^2rG_{j-1}/3\ge\ln(4/\delta)$ by the choice $r\ge\frac{3}{\eta^2G_{j-1}}\ln\frac4\delta$.
Consequently, with probability at least $1-\delta$, the estimate $\widehat Z_t^{(j)}(h_t)$ lies between $(1-\eta)^2Z_t^{(j)}(h_t)$ and $(1-\eta)^{-2}Z_t^{(j)}(h_t)$. Multiplying by $(1-\eta)^2$ gives
\[
    (1-\eps)Z_t^{(j)}(h_t)
    \le(1-\eta)^4Z_t^{(j)}(h_t)
    \le L_t^{(j)}(h_t)
    \le Z_t^{(j)}(h_t),
\]
where the first inequality uses $(1-\eta)^4\ge1-4\eta=1-\eps$. This proves the accuracy guarantee.

As for the runtime anlaysis, our estimation procedure consists of (i) drawing $r$ continuations from the biased distribution $\prtwo{*}{\cdot\mid h_t}$, each at cost $c_{\rm smpl}=O(nKm)$ (see Appendix~\ref{app:star-sampling}); and (ii) making $r$ calls to the recursive estimator ${\cal W}_j(\cdot,\eta,\kappa)$. Denoting the runtime of one such call by $T_j(\eta,\kappa)$, the overall runtime is
\begin{align}
    O(Kmn)+r\cdot\left(c_{\rm smpl}+T_j(\eta,\kappa)\right) \ . \label{expr:z-runtime}
\end{align}
It remains to bound $r$ and $T_j(\eta,\kappa)$. Let $\Lambda=j+\ln(n/(\alpha_0\delta\eps))$. Because $n\ge1$ and $0<\alpha_0,\eps,\delta<1$, we have
\begin{align}
    \Lambda\ge j\ge1 \ ,
    \qquad
    \ln\frac1\delta\le\Lambda-j \ ,
    \qquad \text{and} \qquad
    \ln\frac1{\alpha_0\eps}\le\Lambda-j \ . \label{ineq:lambda-facts}
\end{align}

We first bound $r$. Recall that $\eta=\eps/4$. We have
\begin{align}
    r\le\frac{3}{\eta^2G_{j-1}}\ln\frac4\delta+1
    \overset{(i)}{\le}\frac{12\cdot4^{j}}{(\alpha_0\eps)^2}\ln\frac4\delta+1
    \overset{(ii)}{\le}\frac{13\cdot4^{j}}{(\alpha_0\eps)^2}\ln\frac4\delta
    \overset{(iii)}{\le}\frac{26\cdot4^{j}}{(\alpha_0\eps)^2}\Lambda \ , \label{ineq:r-bound}
\end{align}
where $(i)$ uses $G_{j-1}^{-1}\le4^{j-1}\alpha_0^{-2}$ (from \Cref{clm:g-ub} in Appendix~\ref{prf:g-ub}), $(ii)$ uses $\alpha_0\eps<1$ and $\ln(4/\delta)>1$, and $(iii)$ uses $\ln(4/\delta)\le\ln4+\Lambda-j\le2\Lambda$, which follows from \eqref{ineq:lambda-facts} and $j\ge1$.

Next, we bound $T_j(\eta,\kappa)$. By \Cref{clm:w-estimation}, we have
\[
    T_j(\eta,\kappa)
    =O\!\left(756^{j^2} \cdot Km\,n^{j+1} \cdot (\alpha_0\eta)^{-2j} \cdot \left(j+\ln\frac{n}{\alpha_0\kappa\eta}\right)^j\right) \ .
\]
For the third factor, $\eta=\eps/4$ gives $(\alpha_0\eta)^{-2j}=16^j(\alpha_0\eps)^{-2j}$. For the last factor, $\kappa=\delta/(2r)$ gives
\begin{align*}
    j+\ln\frac{n}{\alpha_0\kappa\eta}
    =\Lambda+\ln(8r)
    &\overset{(i)}{\le}\Lambda+\ln208+j\ln4+2\ln\frac1{\alpha_0\eps}+\ln\Lambda
    \\ & \overset{(ii)}{\le}\Lambda+6+2j+2(\Lambda-j)+(\Lambda-1)
    \overset{(iii)}{\le}9\Lambda \ ,
\end{align*}
where $(i)$ uses the bound on $r$ in \eqref{ineq:r-bound}, $(ii)$ uses $\ln208\le6$, $\ln4\le2$, \eqref{ineq:lambda-facts}, and $\ln\Lambda\le\Lambda-1$, and $(iii)$ uses $\Lambda\ge1$. Since $16\cdot9=144$, we obtain
\[
    T_j(\eta,\kappa)
    =O\!\left(144^j\cdot756^{j^2}Km\,n^{j+1}(\alpha_0\eps)^{-2j}\Lambda^j\right) \ .
\]

We now combine the bounds. Since $c_{\rm smpl}=O(nKm)$ is dominated by the above bound on $T_j(\eta,\kappa)$, plugging the bounds on $r$ and $T_j(\eta,\kappa)$ into \eqref{expr:z-runtime} and using $4\cdot144=576$ gives the overall runtime
\[
    O\!\left(576^j\cdot756^{j^2}Km\,n^{j+1}(\alpha_0\eps)^{-(2j+2)}\Lambda^{j+1}\right) \ ,
\]
where the $O(Kmn)$ term is dominated because all other factors are at least $1$. Since $j\ge1$, we have
\[
    576^j\cdot756^{j^2}\le(576\cdot756)^{j^2}=435{,}456^{j^2}\le\left(2^{19}\right)^{j^2}=\zeta^{j^2} \ ,
\]
which proves the precise runtime bound in \Cref{lem:accuracy-analysis} with $\zeta=2^{19}$.

Finally, we show that this runtime is polynomial in the sense stated in the lemma's statement. Let $N=Kmn/(\alpha_0\eps\delta)\ge1$. We have $Km\,n^{j+1}\le N^{j+1}$ and $(\alpha_0\eps)^{-(2j+2)}\le N^{2j+2}$. Moreover, using $j\le2^j$ and $\ln y\le y$ for $y>0$, we get $\Lambda\le2^j+N\le2^{j+1}N$, and hence $\Lambda^{j+1}\le2^{(j+1)^2}N^{j+1}$. Combining these bounds with $(j+1)^2\le4j^2$ for $j\ge1$, the runtime is at most
\[
    O\!\left(2^{19j^2}\cdot2^{(j+1)^2}\cdot N^{4(j+1)}\right)
    \le O\!\left(2^{23j^2}N^{4(j+1)}\right)
    \le O\!\left(\left(2^{j^2}N^{j+1}\right)^{23}\right) \ ,
\]
which is polynomial in $2^{j^2}N^{j+1}$, as claimed. This completes the proof. \hfill \qedsymbol

\subsection{Proof of \Cref{clm:w-estimation}}\label{prf:w-estimation}
We split the proof in two parts for accuracy and runtime analysis. 
\paragraph{Accuracy analysis.} 
We prove the correctness of Algorithm~\ref{alg:w-oracle} by induction on $j$. The base case $j=0$ is immediate because $W_0=1$. For the inductive step, fix $j\ge1$ and assume that the lemma holds at level $j-1$. Recall the parameter choices in Algorithm~\ref{alg:w-oracle} with input parameters $(\eta,\kappa)$:
\[
    \eta'=1-(1-\eta)^{1/3}, \qquad
    I_j=\left\lceil\frac{3}{\eta'^2G_{j-1}}\ln\frac{4n}{\kappa}\right\rceil,
    \qquad
    \kappa'=\frac{\kappa}{2(nI_j+1)}.
\]
Since $0<\eta',\kappa'<1$, the induction hypothesis applies to every recursive call ${\cal W}_{j-1}(\cdot,\eta',\kappa')$.

We first analyze the estimates $\widehat Z_t^{(j-1)}$ in Line~\ref{line:prev-z}. Fix a time $t\in[n]$, and let $h_t$ be $t$-step prefix of $\omega$. We show that the ratio bound
\begin{align}
    \frac{\widehat Z_t^{(j-1)}}{Z_t^{(j-1)}(h_t)}
    \in[(1-\eta')^2,(1-\eta')^{-2}] \label{expr:z-good-event}
\end{align}
fails with probability at most $\kappa'I_j+\kappa/(2n)$. If $j=1$, every recursive call returns $\widehat W_0=W_0=1$, so $\widehat Z_t^{(0)}=Z_t^{(0)}(h_t)$ and the ratio bound always holds. Suppose instead that $j\ge2$. 
The continuations of completed sample paths $(\tilde\omega^{(t,i)})_{i\in[I_j]}$ from time $t$ are independent samples from $\prtwo{*}{\cdot\mid h_t}$. Moreover, each recursive call in Line~\ref{line:sample-w} uses fresh randomness. Thus, conditionally on every realized $\tilde\omega^{(t,i)}$, the induction hypothesis gives
\[
    \frac{\widehat W_{j-1}^{(t,i)}}{W_{j-1}(\tilde\omega^{(t,i)})}
    \in[(1-\eta'),(1-\eta')^{-1}]
    \quad\text{with probability at least $1-\kappa'$.}
\]
We can therefore apply Claim~\ref{clm:z-estimator} at level $j-1$ with $r=I_j$ and accuracy parameters $(\eta',\kappa')$. This shows that the ratio bound \eqref{expr:z-good-event} fails with probability at most
\begin{align*}
    \kappa'I_j
    +2\exp\left(-\frac{\eta'^2I_jG_{j-2}}3\right)
    &\le\kappa'I_j
    +2\exp\left(-\frac{\eta'^2I_jG_{j-1}}3\right)\\
    &\le\kappa'I_j+2\exp\left(-\ln\frac{4n}{\kappa}\right)
    =\kappa'I_j+\frac{\kappa}{2n},
\end{align*}
where the first inequality uses $G_{j-1}=\alpha_{j-1}G_{j-2}\le G_{j-2}$ and the second uses the choice of~$I_j$.

By the induction hypothesis, with probability at least $1-\kappa'$, the additional recursive call on $\omega$ in Line~\ref{line:self-w} satisfies
\[
    \frac{\widehat W_{j-1}(\omega)}{W_{j-1}(\omega)}
    \in[(1-\eta'),(1-\eta')^{-1}].
\]
A union bound over $t\in[n]$ shows that this estimate and all the preceding $Z$-ratio estimates in \eqref{expr:z-good-event} hold simultaneously with probability at least
\begin{align*}
    1-n\left(\kappa'I_j+\frac{\kappa}{2n}\right)-\kappa'
    &=1-\kappa'(nI_j+1)-\frac\kappa2\\
    &=1-\kappa.
\end{align*}
On the event that all estimates hold simultaneously, we have
\[
    \widehat \rho_{j-1}
    =\max_{t\in[n]}\frac{X_t(\omega)}{\widehat Z_t^{(j-1)}}
    \in[(1-\eta')^2\rho_{j-1}(\omega),(1-\eta')^{-2}\rho_{j-1}(\omega)].
\]
Recall the exact and estimated recursions
\begin{align*}
    W_j(\omega)
    &=W_{j-1}(\omega)
      \max\{\alpha_{j-1},\rho_{j-1}(\omega)\},\\
    \widehat W_j(\omega)
    &=\widehat W_{j-1}(\omega)
      \max\{\alpha_{j-1},\widehat\rho_{j-1}\}.
\end{align*}
Since $(1-\eta')^2\le1\le(1-\eta')^{-2}$, the multiplicative bounds for $\widehat\rho_{j-1}$ carry over to $\max\{\alpha_{j-1},\widehat\rho_{j-1}\}$. Combining them with the bounds for $\widehat W_{j-1}(\omega)$ in these recursions therefore gives
\[
    \widehat W_j(\omega)
    \in[(1-\eta')^3W_j(\omega),(1-\eta')^{-3}W_j(\omega)].
\]
Since $(1-\eta')^3=1-\eta$, the proof of the correctness is complete.

\paragraph{Runtime analysis.} Define $\eta_j=\eta$ and $\kappa_j=\kappa$. For every $\ell\in[j]$, a level-$\ell$ call uses
\begin{align*}
    \eta_{\ell-1}&=1-(1-\eta_\ell)^{1/3},
    &
    I_\ell&=\left\lceil
    \frac{3}{\eta_{\ell-1}^2G_{\ell-1}}
    \ln\frac{4n}{\kappa_\ell}
    \right\rceil, &
    B_\ell&=nI_\ell+1,
    &
    \kappa_{\ell-1}&=\frac{\kappa_\ell}{2B_\ell}.
\end{align*}
Here $B_\ell$ bounds the number of recursive level-$(\ell-1)$ calls. Let $T_\ell$ denote the running time of a level-$\ell$ call. Then the $nI_\ell$ biased-trajectory samples and the remaining arithmetic give, for a universal constant $c>0$,
\[
    T_\ell\le B_\ell T_{\ell-1}+c \cdot c_{\rm smpl} \cdot nI_\ell,
    \qquad T_0=O(1),
\]
where $c_{\rm smpl} = O(nKm)$ is the cost of one exact biased-trajectory sample from Appendix~\ref{app:star-sampling}. 
Unrolling the recurrence and using $nI_\ell\le B_\ell$ gives
\begin{align}
    T_j
    &\le T_0\prod_{\ell=1}^jB_\ell
    +c \cdot c_{\rm smpl} \cdot \sum_{q=1}^j nI_q\prod_{\ell=q+1}^jB_\ell \notag \\
    & = 
    T_0\prod_{\ell=1}^jB_\ell
    +c \cdot c_{\rm smpl} \cdot \sum_{q=1}^j (B_q - 1) \prod_{\ell=q+1}^jB_\ell \notag \\
    & = 
    T_0\prod_{\ell=1}^jB_\ell
    +c \cdot c_{\rm smpl} \cdot \left ( \prod_{\ell=1}^jB_\ell - 1 \right )
    =O\!\left(c_{\rm smpl}  \prod_{\ell=1}^jB_\ell\right). \label{ineq:tj-ub}
\end{align}

To bound this product in \eqref{ineq:tj-ub}, it remains to control the three quantities appearing in $B_\ell$, which are $G_{\ell-1}$, the propagated accuracy $\eta_{\ell-1}$, and the propagated failure probability $\kappa_\ell$. The next three claims do so, with their proofs delegated to the end of this subsection.

\begin{claim}\label{clm:g-ub}
Suppose $0<\alpha_0\le1$ and $\alpha_{q+1}=1/(1+\ln(1/\alpha_q))$ for every $q\ge0$. The following bounds hold for every $\ell\ge1$ and $j\ge1$, respectively:
\[
    \frac1{G_{\ell-1}}\le\frac{4^{\ell-1}}{\alpha_0^2}
    \qquad \text{and} \qquad
    \prod_{\ell=1}^j\frac1{G_{\ell-1}}
    \le\frac{4^{j(j-1)/2}}{\alpha_0^{2j}} \ .
\]
\end{claim}

\begin{claim}\label{clm:eps-ub}
For every $\ell\in[j]$, we have $\eta_{\ell-1}\ge\eta/3^{j-\ell+1}$, and consequently $\prod_{\ell=1}^j\eta_{\ell-1}^{-2}\le9^{j(j+1)/2}\eta^{-2j}$.
\end{claim}
    
\begin{claim}\label{clm:delta-ub}
For $0\le\ell\le j$, let $p_\ell = \ln(4n/\kappa_\ell)$ and $L = \ln(96\cdot9^j n^2/(\alpha_0^2\kappa\eta^2))$. These quantities satisfy
\[
    p_\ell\le2^{j-\ell}L
    \quad\text{and}\quad
    \prod_{\ell=1}^jp_\ell\le2^{j(j-1)/2}L^j \ .
\]
\end{claim}

Combining the three claims with $B_\ell = nI_\ell + 1 \le12np_\ell/(\eta_{\ell-1}^2G_{\ell-1})$ gives
\[
    \prod_{\ell=1}^jB_\ell
    \le 12^j\,8^{j(j-1)/2}\,9^{j(j+1)/2}\,
        n^j(\alpha_0\eta)^{-2j}L^j
    \le \zeta_0^{j^2}n^j(\alpha_0\eta)^{-2j}L^j,
\]
where we take $\zeta_0=108$, since $12^j8^{j(j-1)/2}9^{j(j+1)/2}=108^j72^{j(j-1)/2}\le108^{j^2}$ for $j\ge1$.
Let $R = \ln(n/(\alpha_0\kappa\eta))$. Using $j\ge1$ and $R\ge0$, we obtain
\[
    L\le\ln96+j\ln9+2R
    \le7(j+R).
\]
Since $7^j\zeta_0^{j^2}\le(7\zeta_0)^{j^2}=756^{j^2}$, we obtain
\[
    T_j
    =O\!\left(
      c_{\rm smpl}\cdot756^{j^2}n^{j}(\alpha_0\eta)^{-2j}
      \left(j+\ln\frac{n}{\alpha_0\kappa\eta}\right)^j
    \right).
\]
Substituting $c_{\rm smpl}=O(nKm)$ gives the precise runtime bound in \Cref{clm:w-estimation}.

We now show that this runtime is polynomial in the sense of \Cref{clm:w-estimation}. Let $N=Kmn/(\alpha_0\eta\kappa)\ge1$. We have $Km\,n^{j+1}\le N^{j+1}$ and $(\alpha_0\eta)^{-2j}\le N^{2j}$. Moreover, using $j\le2^j$ and $\ln y\le y$ for $y>0$, we get $j+\ln(n/(\alpha_0\kappa\eta))\le2^j+N\le2^{j+1}N$, and hence
\[
    \left(j+\ln\frac{n}{\alpha_0\kappa\eta}\right)^j\le2^{j(j+1)}N^j \ .
\]
Combining these bounds with $756\le2^{10}$ and $j(j+1)\le2j^2$ for $j\ge1$, the runtime is at most
\[
    O\!\left(756^{j^2}\cdot2^{j(j+1)}\cdot N^{4j+1}\right)
    \le O\!\left(2^{12j^2}N^{4(j+1)}\right)
    \le O\!\left(\left(2^{j^2}N^{j+1}\right)^{12}\right) \ ,
\]
which is polynomial in $2^{j^2}N^{j+1}$, as claimed. Finally, the case $j=0$ takes constant time.

Thus, both accuracy and runtime have been analyzed, showing the desired bounds. The proof of \Cref{clm:w-estimation} is now complete. \hfill \qedsymbol

\subsubsection{Proof of \Cref{clm:g-ub}.}\label{prf:g-ub}
Set $a_q=1/\alpha_q$. Then $a_0\ge1$ and $a_{q+1}=1+\ln a_q$. Since $1+\ln x\le2\sqrt{x}$ for $x\ge1$, induction gives $a_q\le4a_0^{1/2^q}$ for $q\ge1$.
Substitution and rearrangement of the factors now give both bounds:
\begin{align*}
    \frac1{G_{\ell-1}}
    &=a_0\prod_{q=1}^{\ell-1}a_q \le a_0\,4^{\ell-1}a_0^{\sum_{q=1}^{\ell-1}2^{-q}}
    \le4^{\ell-1}a_0^2,\\
    \prod_{\ell=1}^j\frac1{G_{\ell-1}}
    &=a_0^j\prod_{q=1}^{j-1}a_q^{j-q} \le a_0^j4^{\sum_{q=1}^{j-1}(j-q)}
       a_0^{\sum_{q=1}^{j-1}(j-q)/2^q}
    \le4^{j(j-1)/2}a_0^{2j}.
\end{align*}
\hfill \qedsymbol
\subsubsection{Proof of \Cref{clm:eps-ub}.}
Unrolling the recurrence gives $\eta_{\ell-1}=1-(1-\eta)^{1/3^{j-\ell+1}}$. For $0<r\le1$, concavity of $x\mapsto x^r$ gives $(1-\eta)^r\le1-r\eta$.
Taking $r=3^{-(j-\ell+1)}$ proves the first bound. Multiplying these bounds over $\ell\in[j]$ gives
\[
    \prod_{\ell=1}^j\frac1{\eta_{\ell-1}^2}
    \le\frac{9^{j(j+1)/2}}{\eta^{2j}} \ ,
\]
as desired. \hfill \qedsymbol

\subsubsection{Proof of \Cref{clm:delta-ub}.}\label{prf:delta-ub}
Note that we have $p_\ell=\ln(4n/\kappa_\ell)\ge\ln 4 >1$ which also gives $I_\ell \ge 1$. Together with Claims~\ref{clm:g-ub} and \ref{clm:eps-ub}, this gives
\begin{align*}
    I_\ell& = \left \lceil \frac{3}{\eta_{\ell-1}^2G_{\ell-1}}
    \ln\frac{4n}{\kappa_\ell} \right \rceil \le \frac{6p_\ell}{\eta_{\ell-1}^2G_{\ell-1}}
    \quad \text{and thus} 
    ,\\
    B_\ell & \le2nI_\ell
    \le\frac{12np_\ell}{\eta_{\ell-1}^2G_{\ell-1}} \le 12 np_\ell\cdot\frac{4^{\ell-1}}{\alpha_0^2}
       \cdot\frac{9^{j-\ell+1}}{\eta^2}
    \le\frac{12\cdot9^j n}{\alpha_0^2\eta^2}p_\ell \ .
\end{align*}
Using $\kappa_{\ell-1}=\kappa_\ell/(2B_\ell)$ and $p_\ell\ge1$, we obtain
\begin{align*}
    p_{\ell-1}
    &=p_\ell+\ln(2B_\ell) \le2p_\ell+
      \ln\left(\frac{24\cdot9^j n}{\alpha_0^2\eta^2}\right).
\end{align*}
Starting from $p_j=\ln(4n/\kappa)$ and iterating this recurrence yields
\[
    p_\ell \le2^{j-\ell}L \quad\text{and}\quad
    \prod_{\ell=1}^jp_\ell
    \le2^{j(j-1)/2}L^j \ ,
\]
as desired. The proof of the claim is now complete.  \hfill \qedsymbol

\subsection{Proof of \Cref{clm:z-estimator}}\label{prf:z-estimator}
Define the empirical averages of the exact and the estimated values by
\[
    \bar{\bs w}_t^{(j)}
    =\frac1r\sum_{i=1}^r W_j(\omega^{(i)}),
    \qquad
    \widehat{\bs w}_t^{(j)}
    =\frac1r\sum_{i=1}^r\widehat W_j(\omega^{(i)}).
\]
Recall that $W_j=\prod_{\ell=0}^{j-1}\max\{\rho_\ell,\alpha_\ell\} \le 1$. Together with $0<\alpha_\ell\le1$, this implies $G_{j-1}\le W_j\le1$ almost surely. By independence of the sampled trajectories, the variables $W_j(\omega^{(i)})$ are therefore independent and take values in $[0,1]$, and their average has mean $\expar{\bar{\bs w}_t^{(j)}}=\extwo{*}{W_j\mid h_t}\ge G_{j-1}$. Hence, a Chernoff Bound (see Appendix~\ref{app:chernoff}), applied to these $r$ variables with parameter $\eta$, gives
\[
    \pr{\left|\bar{\bs w}_t^{(j)}-\ex{\bar{\bs w}_t^{(j)}}\right|>\eta\ex{\bar{\bs w}_t^{(j)}}}
    \le 2\exp\left(-\frac{\eta^2r\ex{\bar{\bs w}_t^{(j)}}}3\right)
    \le 2\exp\left(-\frac{\eta^2rG_{j-1}}3\right).
\]
Moreover, a union bound over $i\in[r]$ shows that, with probability at least $1-\kappa r$,
\[
    \widehat W_j(\omega^{(i)})
    \in[(1-\eta)W_j(\omega^{(i)}),
        (1-\eta)^{-1}W_j(\omega^{(i)})]
    \qquad\text{for every }i\in[r].
\]
On this event, averaging over $i$ gives $\widehat{\bs w}_t^{(j)}\in[(1-\eta)\bar{\bs w}_t^{(j)},(1-\eta)^{-1}\bar{\bs w}_t^{(j)}]$.
Combining this event with the concentration event $|\bar{\bs w}_t^{(j)}-\expar{\bar{\bs w}_t^{(j)}}|\le\eta\expar{\bar{\bs w}_t^{(j)}}$ and using $1+\eta\le(1-\eta)^{-1}$, we conclude that, with probability at least $1-\kappa r-2\exp(-\eta^2rG_{j-1}/3)$,
\[
    \widehat{\bs w}_t^{(j)}
    \in\left[(1-\eta)^2\ex{\bar{\bs w}_t^{(j)}},
        (1-\eta)^{-2}\ex{\bar{\bs w}_t^{(j)}}\right].
\]
Finally, Claim~\ref{clm:z-alter-def} gives $Z_t^{(j)}(h_t)=Z_t^{(0)}(h_t)\expartwo{*}{W_j \mid h_t} = Z_t^{(0)}(h_t)\expartwo{*}{\bar{\bs w}_t^{(j)}}$, while $\widehat Z_t^{(j)}(h_t)=Z_t^{(0)}(h_t)\widehat{\bs w}_t^{(j)}$ by definition. 
Therefore, we have ${\widehat Z_t^{(j)}(h_t)}/{Z_t^{(j)}(h_t)} \in [(1-\eta)^2, (1-\eta)^{-2}]$ with probability at least $1-\kappa r-2\exp(-\eta^2rG_{j-1}/3)$ which completes the proof. \hfill \qedsymbol

\subsection{Proof of \Cref{lem:alpha-convergence}}\label{app:alpha-convergence}

Fix $0<\eps\le1$ and set $a_j=1/\alpha_j$. Since $0<\alpha_0\le1$, we have $a_0\ge1$ and $a_{j+1}=1+\ln a_j$. Let $f(x)=1+\ln x$ and define $k=\min\{\ell\ge0:\log_2^{\circ\ell}(a_0)\le16\}$.

The proof has two phases. Iterated logarithms first bring $a_j$ below a constant; we then show that $a_j-1$ decreases at rate $O(1/j)$ and $k=O(\log^*a_0)$. We claim that
\[
    f^{\circ\ell}(a_0)\le\log_2^{\circ\ell}(a_0)
    \qquad(0\le\ell\le k).
\]
For $\ell=0$, this is equality. If it holds for some $\ell<k$, then $\log_2^{\circ\ell}(a_0)>16$. Because $f$ is increasing and $f(x)\le\log_2x$ for every $x\ge16$, it follows that
\[
    f^{\circ(\ell+1)}(a_0)
    \le f\!\left(\log_2^{\circ\ell}(a_0)\right)
    \le\log_2^{\circ(\ell+1)}(a_0).
\]
Thus, we have $a_k=f^{\circ k}(a_0)\le16$.

Define $b_s=a_{k+s}-1$. Then $0\le b_0\le15$ and $b_{s+1}=\ln(1+b_s)$.
If $b_0=0$, then $b_s=0$ for every $s$; choosing $J=k$ gives $\alpha_J=1$, so the desired conclusion follows. Assume $b_0>0$. Then $0<b_s\le b_0\le15$ for every $s$. For every $b>0$, the integral representation of the logarithm gives
\begin{align*}
    \ln(1+b)
    &=b-\int_0^b\frac{x}{1+x}\,dx\\
    &\le b-\int_0^b\frac{x}{1+b}\,dx\\
    &=b-\frac{b^2}{2(1+b)}
      =\frac{b(2+b)}{2(1+b)}.
\end{align*}
Applying this estimate with $b=b_s$ yields
\[
    \frac1{b_{s+1}}
    \ge\frac{2(1+b_s)}{b_s(2+b_s)}
    =\frac1{b_s}+\frac1{2+b_s}
    \ge\frac1{b_s}+\frac1{17}.
\]
Iterating gives $1/b_s\ge s/17$ and, hence, $b_s\le17/s$ for $s\ge1$. Let $S=\lceil34/\eps\rceil$ and $J=k+S$.
For this choice of $J$, we have
\[
    \alpha_J
    =\frac1{1+b_S}
    \ge1-b_S
    \ge1-\frac\eps2.
\]
Moreover, we have $J=O(\log^*a_0+1/\eps)=O(\log^*(1/\alpha_0)+1/\eps)$.
\hfill \qedsymbol

\section{Omitted Proofs from \Cref{sec:competitive-algs}}

\subsection{Proofs of the claims used in \Cref{lem:main-hardness}}\label{prf:hard-instance}

\subsubsection{Posterior calculation.}\label{prf:posterior} Suppose that $M_t = n^{3j}$ which implies that the observed history has zero likelihood under \(D^{(h)}\) for \(h<j\). Now, fix \(j \le h < K\). For every observed sample \(X_s\), its probability under \(D^{(h+1)}\) is \(1-1/n^2\) times its probability under \(D^{(h)}\) by definition of \(D^{(h)}\). The likelihood ratio of the observed history is therefore
\[
\frac{
\pr{X_1,\ldots,X_t\mid H=h+1}
}{
\pr{X_1,\ldots,X_t\mid H=h}
}
=
\left(1-\frac1{n^2}\right)^t.
\]
Moreover, the prior weights satisfy
\(
\frac{\pi_{h+1}}{\pi_h}=\frac1{n^3}.
\)
Therefore, Bayes' rule gives
\[
\frac{\pr{H=h+1\mid\mathcal{F}_t}}
     {\pr{H=h\mid\mathcal{F}_t}}
=
\frac{(1-1/n^2)^t}{n^3}.
\]
Thus, the posterior weights on \(h=j,\ldots,K\) form a truncated geometric sequence satisfying
\[
\pr{H=h\mid\mathcal{F}_t}
\propto
\left(\frac{(1-1/n^2)^t}{n^3}\right)^{h-j}.
\]
for \(j\le h\le K\). Normalizing this sequence gives the stated formula. \hfill \qedsymbol

\subsubsection{Proof of \Cref{clm:mstar}.}\label{prf:mstar}
Recall that $p=1/n^2$ and $q=1-p$. Since $t\ge2$ and $0<q<1$, we have
\begin{align*}
    \frac{1-q^t}{tp}
    &=\frac1t\sum_{r=0}^{t-1}q^r
    \le\frac{1+(t-1)q}{t}\\
    &=1-\frac{t-1}{t}p
    \le1-\frac p2
    \le1-\frac{q^t}{n^3} \ ,
\end{align*}
where the last inequality uses $n\ge2$ and $q^t\le1$. Dividing by $1-q^t/n^3>0$ gives
\[
    m_t^*=\frac{1-q^t}{p(1-q^t/n^3)}\le t \ ,
\]
as desired. \hfill \qedsymbol

\subsection{Proof of \Cref{lem:iid-feasibility}}
\phantomsection\label{prf:huang-suff}
    
We begin by recalling the results of \cite{bai2025optimal}:
    
\begin{theorem}[c.f. \cite{bai2025optimal}, Theorem 1]
    For any $\beta \in [0, 1/e]$, suppose $\lambda_1, \lambda_2$ are two roots of $-x\ln x = \beta$ with $\lambda_1 \le \lambda_2$. If there exists a left-continuous non-increasing function $\theta:[\lambda_1,\lambda_2]\to[0,1]$ satisfying $\theta(\lambda_1) = 1$ and  
\begin{equation}
 \int_z^1\int_0^t
 \frac{1}{t}\,\theta\!\left(\max\{s,z\}\right)^t
 d s\,d t
 \ge \alpha\,\theta(z) \ , \quad \forall z \in [\lambda_1, \lambda_2] \ , \label{ineq:huang-suff}
\end{equation}
then, for every horizon $n \in {\bb N}$, there exists a policy that is $\alpha$-consistent and $\beta$-robust. 
\end{theorem}

Now, for every $\beta \in (0,1/e), q \in (0,1)$, define the function $\bar \theta_{\beta,q}(t) = q$ for all $t \in [\lambda_1, \lambda_2]$. Moreover, let $\bar \theta_{\beta,q}(t) = 1 $ for $t \in [0, \lambda_1]$ and $\bar \theta_{\beta,q}(t) = 0$ for $t \in (\lambda_2, 1]$. In what follows, we will prove the following lemma:
\begin{lemma}\label{lem:huang-suff} Setting $\alpha = 1/e + 1/K, \beta = 1/e - 2/(eK^2), q = \exp(-4e/K)$, and $\theta = \bar \theta_{\beta, q}$, inequality \eqref{ineq:huang-suff} is satisfied for every $K \ge 39$. 
\end{lemma}

It is immediate that this lemma implies \Cref{lem:iid-feasibility}.

{\it Proof of \Cref{lem:huang-suff}.}
We drop the subscript $K$ from variables to ease the notation, i.e., we have $\alpha = 1/e + 1/K$, $\beta = 1/e - 2/(eK^2), q = \exp(-4e/K)$. Let $\lambda_1 \le \lambda_2$ denote the two roots of $-x\ln x = \beta$, and $L(z)$ be the left-hand side of
\eqref{ineq:huang-suff}. 

For $z=\lambda_1$, the inner integral in
\eqref{ineq:huang-suff} equals
\[
 \int_0^t\theta(\max\{s,\lambda_1\})^t\,{d}s
 =
 \begin{cases}
  \lambda_1+(t-\lambda_1)q^t,
      &\lambda_1\le t\le\lambda_2,\\
  \lambda_1+(\lambda_2-\lambda_1)q^t,
      &\lambda_2<t\le1.
 \end{cases}
\]
Since $-\lambda_1\ln\lambda_1=\beta$, it follows that
\begin{align}
 L(\lambda_1)
 &=
 \beta+
 \int_{\lambda_1}^{\lambda_2}
 \left(1-\frac{\lambda_1}{t}\right)q^t\,{d}t
 +(\lambda_2-\lambda_1)
 \int_{\lambda_2}^1\frac{q^t}{t}\,{d}t
 =:A_1.
 \label{eq:q-step-L-a}
\end{align}
For $\lambda_1<z\le\lambda_2$, the inner integral equals $tq^t$
when $z\le t\le\lambda_2$ and $\lambda_2q^t$ when
$\lambda_2<t\le1$ and, hence,
\begin{align}
 \frac{L(z)}q
 &=
 \int_z^{\lambda_2}q^{t-1}\,{d}t
 +\frac{\lambda_2}{q}
 \int_{\lambda_2}^1\frac{q^t}{t}\,{d}t \notag\\
 &\ge
 \frac{\lambda_2}{q}
 \int_{\lambda_2}^1\frac{q^t}{t}\,{d}t
 =:A_2.
 \label{eq:q-step-L-z}
\end{align}
Since $\theta(\lambda_1)=1$ and $\theta(z)=q$ for
$\lambda_1<z\le\lambda_2$, it suffices to prove
$A_1,A_2\ge\alpha$.

Define $r_- = 1- e\lambda_1$ and $r_+ = e\lambda_2 - 1$ so that $\lambda_1 = \frac{1-r_-}{e}$ and $\lambda_2 = \frac{1 + r_+}{e}$. We have $r_-, r_+ > 0$ because $\lambda_1 < 1/e$. Since $\lambda_2=(1+r_+)/e$ and
$-\lambda_2\ln\lambda_2=\beta=1/e-2/(eK^2)$, we have
\[
 (1+r_+)\ln(1+r_+)-r_+= e\lambda_2 \ln (e\lambda_2) - e\lambda_2 + 1 = -e\beta + e\lambda_2 - e\lambda_2 + 1 = 1-e\beta =\frac{2}{K^2} \ .
\]
Therefore, we have
\begin{align}
 \int_0^{r_+}\ln(1+s)\,{d}s = (1+r_+)\ln(1+r_+)-r_+ = \frac{2}{K^2} \ . \label{eq:int1}
\end{align}
Because $\log(1+s)\le s$, \eqref{eq:int1} gives
\[
    \frac{2}{K^2} = \int_0^{r_+}\ln(1+s)\,{d}s \le \int_0^{r_+}s\,{d}s = \frac{r_+^2}{2} \ , 
\]
implying $r_+\ge2/K$. Moreover, since
$\ln(1+s)\ge s/(1+s)$, we can argue
\[
    \frac{2}{K^2} = \int_0^{r_+}\ln(1+s)\,{d}s \ge \int_0^{r_+} \frac{s}{1+s}\,{d}s \ge r_+ \cdot \frac{r_+}{2(1+r_+)}  \ ,
\] and, thus, $r_+ \le 2(1+\sqrt{K^2 + 1})/K^2 \le 4/K$. 

Similarly, since $\lambda_1=(1-r_-)/e$ and
$-\lambda_1\ln\lambda_1=\beta=1/e-2/(eK^2)$, we have
\[
 r_-+(1-r_-)\ln(1-r_-)=\frac{2}{K^2} \ ,
\]
which implies
\[
 \int_0^{r_-}-\ln(1-s)\,{d}s= r_-+(1-r_-)\ln(1-r_-) = \frac{2}{K^2} \ .
\]
For every $s\in[0,r_-]$, we have
\[
 -\ln(1-s)
 =\int_0^s\frac{{d}u}{1-u}
 \le \int_0^{s}\frac{{d}u}{1-r_-} = \frac{s}{1-r_-} \ .
\]
Therefore, we conclude
\[
 \frac{2}{K^2} = \int_0^{r_-}-\ln(1-s)\,{d}s
 \le
 \int_0^{r_-}\frac{s}{1-r_-}\,\mathrm{d}s
 =
 \frac{r_-^2}{2(1-r_-)} \ .
\]
Rearranging gives
\[
 K^2r_-^2+4r_--4\ge0.
\]
The roots of the corresponding quadratic equation are
$
 \frac{2(-1\pm\sqrt{K^2+1})}{K^2}$.
Since $r_->0$, it follows that
\[
 r_-
 \ge
 \frac{2(\sqrt{K^2+1}-1)}{K^2}
 =
 \frac{2}{\sqrt{K^2+1}+1}
 \ge
 \frac{2}{K+2} \ ,
\]
where the last inequality uses $\sqrt{K^2+1}\le K+1$. Combining this with $2/K\le r_+\le4/K$, we obtain
\begin{equation}
 \lambda_2-\lambda_1
 =
 \frac{r_++r_-}{e}
 \ge
 \frac1e\left(\frac2K+\frac{2}{K+2}\right)
 \ge
 \frac4{eK}-\frac4{eK^2} \ ,
 \label{eq:q-step-root-bounds}
\end{equation}
Here, the second inequality follows from
\( \frac{1}{K+2}\ge \frac1K-\frac{2}{K^2} = \frac{K-2}{K^2} \). Moreover, we have 
\begin{align}
    \lambda_2-1/e
 =
 {r_+}/{e}
 \le
 \frac4{eK}. \label{ineq:l2-ub}    
\end{align}

We are now ready to show that $A_1$ and $A_2$ are at least $\alpha$, starting with the former. Recall
\[
A_1 = \beta+
 \int_{\lambda_1}^{\lambda_2}
 \left(1-\frac{\lambda_1}{t}\right)q^t\,{d}t
 +(\lambda_2-\lambda_1)
 \int_{\lambda_2}^1\frac{q^t}{t}\,{d}t .
\]
Since $q^t\ge q$ for $t\in[0,1]$ and the weights in the definition of
$A_1-\beta$ are nonnegative, we have
\begin{align*}
 A_1-\beta
 &\ge
 q\left(
 \int_{\lambda_1}^{\lambda_2}
 \left(1-\frac{\lambda_1}{t}\right){d}t
 +(\lambda_2-\lambda_1)
 \int_{\lambda_2}^1\frac{{d}t}{t}
 \right) \\
 &=q\left(
 \lambda_2-\lambda_1
 +\lambda_1\ln\lambda_1
 -\lambda_2\ln\lambda_2
 \right)\\
 &\overset{(i)}{=}q(\lambda_2-\lambda_1)\\
 &\ge
 e^{-4e/K}\frac{4}{eK}\left(1-\frac1K\right)\\
 &\overset{(ii)}{\ge}
 \frac{4}{eK}
 \left(1-\frac{4e}{K}\right)
 \left(1-\frac1K\right)\\
 &\ge
 \frac{4}{eK}
 \left(1-\frac{1+4e}{K}\right),
\end{align*}
where $(i)$ uses that $\lambda_1,\lambda_2$ are roots of
$-x\ln x=\beta$, and $(ii)$ follows from $e^{-x}\ge1-x$. Because
\(
 \alpha-\beta = \frac1K+\frac{2}{eK^2}
\), we obtain
\begin{align*}
 A_1-\alpha
 =(A_1-\beta)-(\alpha-\beta) \ge
 \frac{4}{eK}
 \left(1-\frac{1+4e}{K}\right)
 -\frac1K-\frac{2}{eK^2} =
 \frac{(4-e)K-(16e+6)}{eK^2}.
\end{align*}
For every $K \ge 39$, the numerator is positive, implying that $A_1 \ge \alpha$, as desired.

We now move on to proving $A_2 \ge \alpha$. Because
\(
 \beta
 =
 -\lambda_2\ln\lambda_2
 =
 \lambda_2\int_{\lambda_2}^1\frac{{d}t}{t}
\), recalling the value of $q = \exp(-4e/K)$ and definition of $A_2$ in \eqref{eq:q-step-L-z}, we obtain
\[
 A_2-\beta
  = {\lambda_2}
 \int_{\lambda_2}^1\frac{q^{t-1}}{t}\,{d}t - \lambda_2 \int_{\lambda_2}^1 \frac{dt}{t} =
 \lambda_2\int_{\lambda_2}^1
 \frac{
  \exp\!\left(\frac{4e}{K}(1-t)\right)-1
 }{t}\,{d}t \overset{(iii)}{\ge} \lambda_2\int_{\lambda_2}^1
 \frac{
  \frac{4e}{K}(1-t)
 }{t}\,{d}t \ ,
\]
where step $(iii)$ uses $e^x-1\ge x$. We, thus, have \(A_2-\beta
 \ge\frac{4e}{K}J(\lambda_2) \), where we define 
\[J(x)=
 x\int_x^1\frac{1-t}{t}\,{d}t
 =
 -x\ln x-x+x^2 \ .
\]
We have $J(1/e)=1/e^2$. Furthermore, on $[1/e,1/2]$,
\[
 J'(x)+1=-\ln x-1+2x\ge \ln 2 > 0 \ ,
\]
because the derivative of the left-hand side is $2-1/x\le0$, so its
minimum on $[1/e, 1/2]$ is attained at $x=1/2$ with a value of $\ln 2$.

We now recall \eqref{ineq:l2-ub} that shows 
\(
 \lambda_2
 \le\frac1e+\frac4{eK}<\frac12
\) for every $K \ge 39$. Because $\lambda_2 \ge 1/e$, we have
\[
\begin{aligned}
 J(\lambda_2)
 =J(1/e)+\int_{1/e}^{\lambda_2}J'(x)\,{d}x \ge
 \frac1{e^2}
 -(1-\ln2)\left(\lambda_2-\frac1e\right) \ge
 \frac1{e^2}-\frac{4(1-\ln2)}{eK}.
\end{aligned}
\]
Consequently, we have
\[
 A_2-\beta
 \ge
 \frac{4e}{K}J(\lambda_2)
 \ge
 \frac4{eK}
 -\frac{16(1-\ln2)}{K^2}.
\]
Recalling
\(
 \alpha-\beta=\frac1K+\frac{2}{eK^2},
\) we obtain
\[
 A_2-\alpha = (A_2 - \beta) - (\alpha - \beta) 
 \ge
 \frac{4/e-1}{K}
 -
 \frac{16(1-\ln2)+2/e}{K^2} \ .
\]
In particular, the right-hand side is positive for every $K\ge39$.

Thus, we have proved $A_1,A_2\ge\alpha$. Since
$L(\lambda_1)=A_1$ and $\theta(\lambda_1)=1$,
\eqref{ineq:huang-suff} holds at $z=\lambda_1$. For
$\lambda_1<z\le\lambda_2$, \eqref{eq:q-step-L-z} gives
\[
 L(z)\ge qA_2\ge q\alpha=\alpha\theta(z) \ .
\]
Hence \eqref{ineq:huang-suff} holds for every
$z\in[\lambda_1,\lambda_2]$ and the proof is complete. \hfill \qedsymbol

\subsection{Proof of \Cref{prop:noniid-cr}}\phantomsection\label{prf:noniid-cr}
    The lower bound follows from the algorithm used for \Cref{cor:baseline-approximations}: find $h^\star = \arg\max_{h \in [K]} \pi_h \ex{\off | H = h}$ and use the classic prophet inequality of \cite{samuel1984comparison} for distribution $D_{h^\star}$. When $H = h^{\star}$, our algorithm is 1/2-competitive which yields the overall $1/(2K)$-competitiveness.  

    Conversely, the upper bound is from a concrete construction. Fix $\rho \in (0, 1)$ and write
\[
    C_\rho =\sum_{\ell=0}^{K-1}\rho^\ell \ ,
    \quad \text{and} \quad
    \pi_j = \frac{\rho^{j-1}}{C_\rho}
    \quad \text{for every } j \in [K] \ .
\] 
Let $n=K$. Conditional on \(\{H=j\}\), let the observations be deterministic:
\begin{equation}\label{eq:instance}
    X_t=
    \begin{cases}
        \rho^{-(t-1)}, & t\le j,\\
        0,             & t>j.
    \end{cases}
\end{equation}
In world \(j\), the
largest observation is \(\rho^{-(j-1)}\), and thus,
\begin{equation}\label{eq:prophet}
    \off  = \sum_{j=1}^K \frac{\rho^{j-1}}{C_\rho}\rho^{-(j-1)}
    =\frac{K}{C_\rho} \ .
\end{equation}
Now take an arbitrary online algorithm ${\cal A}$ that is potentially randomized, and let $\tau_{\cal A}$ be the corresponding stopping time. Represent all its internal randomness by a seed \(U\), independent of \(H\), and fix a
realization \(U=u\). The resulting rule is deterministic and non-anticipative. Run $K$ parallel instances of the algorithm with seed $u$, one for each value of $H \in [K]$.  Let \(T_u\in[K]\cup\{\infty\}\) be its stopping time when run in world $K$. Indeed, this stopping time corresponds to either a time in $1, \cdots, K$, or the algorithm never stops (if $T_u= \infty$). 

Suppose first that \(T_u\le K\). The algorithm would stop at time $t$ for all the runs corresponding to worlds $T_u, T_u+1, \cdots, K$, earning reward $\rho^{-(T_u - 1)}$. This is because the algorithm observes the exact sequence of values for all those worlds. By contrast, for worlds $1, \cdots, T_u-1$, the algorithm rejects all positive observations, since its history through time \(j\) agrees with that in world \(K\). All subsequent observations are zero, so its reward is zero. Hence, for a fixed seed $u$, the reward of the algorithm is $\rho^{-(T_u-1)} \{ H \ge T_u\}$. Taking expectation only over \(H\) gives
\begin{align}
    \ex{X_{\tau_{\cal A}} \mid U=u}
    &=\rho^{-(T_u-1)} \cdot \pr{H\ge T} = \rho^{-(T_u-1)} \cdot 
      \frac{\sum_{j=T_u}^{K}\rho^{j-1}}{C_\rho} = \frac{\sum_{\ell=0}^{K-T_u}\rho^\ell}{C_\rho}
      \le 1 \ . \label{ineq:fixed-seed-ub}
\end{align} 
If \(T_u=\infty\), the algorithm ${\cal A}$ passes every positive reward in every world and can
receive only zero, so the same bound holds. Since the bound \eqref{ineq:fixed-seed-ub} holds for every fixed $U = u$, averaging over \(U\) yields
\begin{equation}\label{eq:fixed-seed-online}
    \ex{X_{\tau_{\cal A}}} \le 1 \ .
\end{equation}
Combining \eqref{eq:prophet} and \eqref{eq:fixed-seed-online}, we infer
\[
    \frac{r({\cal A})}{\off} = \frac{\ex{X_{\tau_{\cal A}}}}{\off}
    \le \frac{C_\rho}{K}
    =\frac{1+\rho+\cdots+\rho^{K-1}}{K} \ .
\]
The right-hand side converges to \(1/K\) as \(\rho \to 0\). Therefore,
for any \(\eta>0\), taking \(\rho\) sufficiently small makes the competitive ratio of any online algorithm to be it at most
\(1/K+\eta\). Letting $\eta \to 0$ establishes the result.
\hfill \qedsymbol

\subsection{Proof of \Cref{prop:gen-cr}}\label{prf:gen-cr}
    The algorithm simply chooses one time $t$ uniformly at random from $\{1,\cdots, n\}$ and accepts it. With probability $1/n$, it accepts the item with the maximum value and thus its competitive ratio is at least $1/n$.

    We now construct a hard instance in which no algorithm can obtain a better competitive ratio, even when $K=1$. Fix some $B > 1$ and let $K=1$. Initially, $X_1 = 1$ with probability 1. For every $t < n$, generate $X_{t+1}$ conditionally on $X_t$ according to 
\[
X_{t+1}=
\begin{cases}
BX_t,&\text{with probability }1/B,\\
0,&\text{with probability }1-1/B,
\end{cases}
\]
where we note that when $X_t = 0$ for some $t$, we have $X_{t'} = 0$ for all $t'> t$.

Notice that \(
    \ex{X_{t+1} \mid \cal F_t} = \frac{1}{B}(B X_t) = X_t
\), i.e., the process is a martingale. Thus, by Doob's optional sampling (\Cref{thm:optional-sampling}), we have $\ex{X_{\thr}} = \on = 1$. Conversely, the prophet obtains a reward
\[
\begin{aligned}
\mathrm{OFF}
&=\sum_{t=1}^{n-1}
B^{t-1}B^{-(t-1)}
\left(1-\frac1B\right)
+B^{n-1}B^{-(n-1)}
\\
&=(n-1)\left(1-\frac1B\right)+1\\
&={n-\frac{n-1}{B}}.
\end{aligned}
\]
Now, taking $B \to \infty$ establishes $\on / \off \le 1/n$, which is the desired result. \hfill \qedsymbol

\section{Auxiliary lemmas}

\subsection{Chernoff Bound}\label{app:chernoff}

We use the following concentration inequality of \cite{chernoff1952measure}:
\begin{lemma}[Multiplicative Chernoff bound]
    Suppose $Y_1, \cdots, Y_n$ are independent random variables taking values in $[0,1]$ and let $\bar Y = \frac{1}{n} \sum_{i=1}^n Y_i$. Then for every $\theta \in [0,1]$, the following bound holds
    \[
        \pr{|\bar Y - \ex{\bar Y}| > \theta \ex{\bar Y}} \le  2 \exp \left (- \theta^2 n \ex{\bar Y}/3\right) \ .
    \]
\end{lemma}

\section{Figures}\label{app:figures}

\begin{figure}[H]
    \centering
    \begin{tikzpicture}
        \begin{axis}[
            width=0.60\textwidth,
            height=0.42\textwidth,
            xmin=0, xmax=1,
            ymin=0, ymax=1,
            xlabel={Baseline factor $\alpha$},
            ylabel={Approximation factor},
            xtick={0,0.2,0.4,0.6,0.8,1},
            ytick={0,0.2,0.4,0.6,0.8,1},
            grid=major,
            grid style={gray!20},
            tick align=outside,
            axis line style={gray!60},
            legend pos=north west,
            legend cell align=left,
            legend style={draw=none, fill=white, font=\small},
        ]
            \addplot[BrickRed, thick, dashed, domain=0:1] {x};
            \addlegendentry{Baseline: $\alpha$}

            \addplot[navyblue, thick, domain=0.001:1, samples=400]
                ({exp(1-1/x)}, {x});
            \addlegendentry{Boosted: $\frac{1}{1+\ln(1/\alpha)}$}
        \end{axis}
    \end{tikzpicture}
    \caption{Baseline and boosted approximation factors.
    The boosted factor is strictly larger for $0<\alpha<1$,
    and both equal $1$ at $\alpha=1$.
    The multiplicative improvement grows as $\alpha$ decreases.}
    \label{fig:boost}
\end{figure}
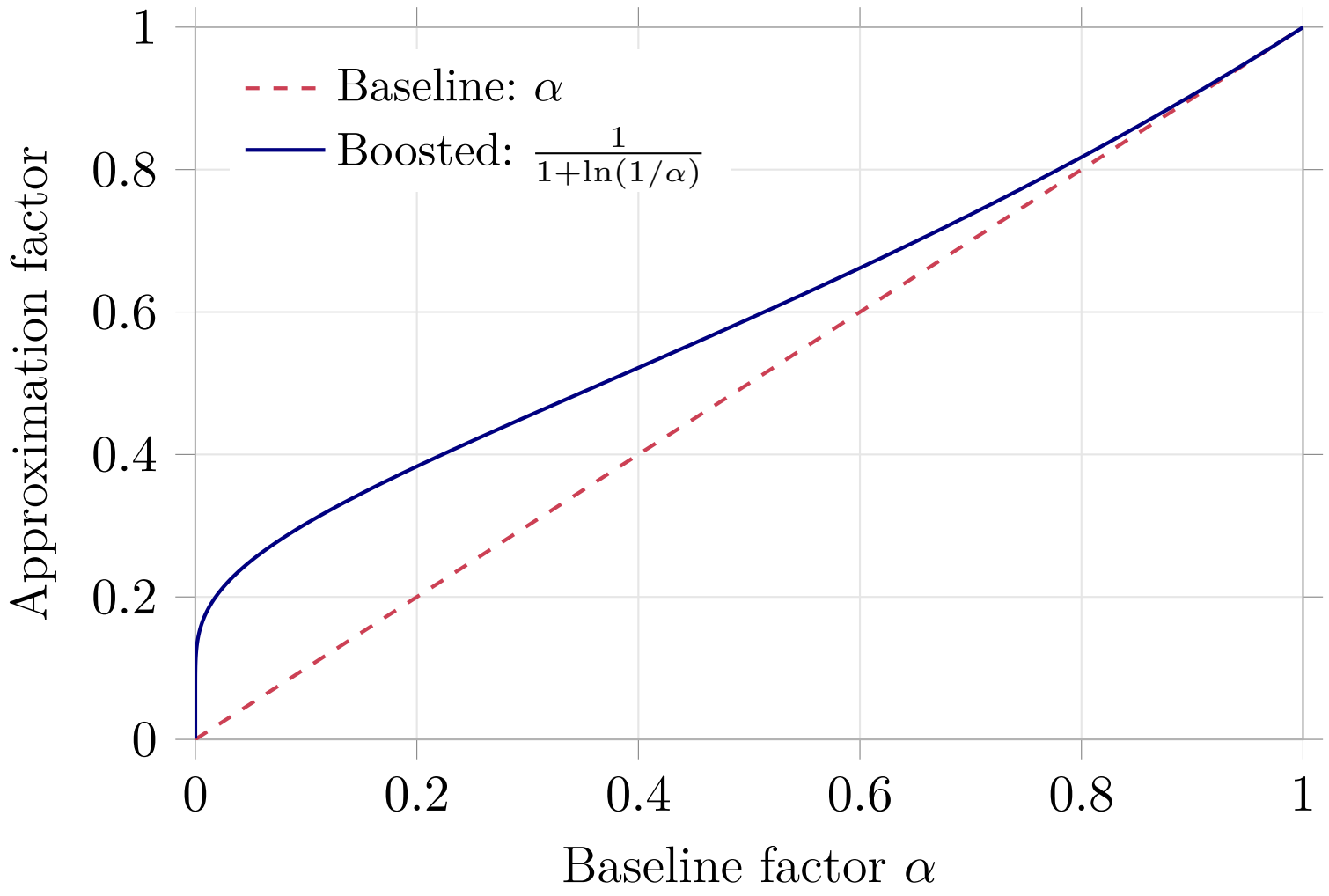

\end{APPENDICES}

\end{document}